\documentstyle[12pt,epsfig,graphics]{report}
\newcommand{\bfm}[1]{\mbox{\boldmath${#1}$}}
\begin{document}
\renewcommand{\thepage}{\Roman{page}}
\begin{center}
\Large \bf QUANTUM SYSTEMS OBEYING A GENERALIZED
EXCLUSION-INCLUSION PRINCIPLE\\ \vspace{10mm}
 \rm A.M. Scarfone\footnote{\small
e-mail: scarfone@polito.it}\\ \vspace{5mm}\small  Dipartimento di
Fisica - Politecnico di Torino\\ Corso Duca degli Abruzzi 24,
10129 Torino, Italy; \\ Istituto Nazionale di Fisica della Materia
- Unit\'a del
 Politecnico di Torino, \\
 Corso Duca degli Abruzzi 24,
10129 Torino, Italy.
\\ \normalsize \vspace{5mm}29, february 2000\\

\vspace{15mm} \bf Abstract
\end{center}
\vspace{5mm} \rm The aim of this work is to describe, at the
quantum level, a many body system obeying to a generalized
Exclusion-Inclusion Principle (EIP) originated by collective
effects, the dynamics, in mean field approximation, being ruled by
a nonlinear Schr\"odinger equation. The EIP takes its origin from
a nonlinear kinetic equation where the nonlinearities describe
interactions of different physical nature.\\ The method starts
from the study of the kinetic behavior of many particle system
which results to be nonlinear because of the interaction among the
particles and introduces an effective nonlinear potential $U_{\rm
EIP}$ which permits us to simulate the true interactions governing
the dynamics of the system. The power of the method is tested in
the case of spatially homogeneous classical $N$-particles system.
Its kinetic in the momenta space is described by a Markoffian
process. A judicious generalization of the particle current,
permits us to obtain at the equilibrium, a statistical
distribution interpolating in a continuous way the well known
quantum statistics (Fermi-Dirac or Bose-Einstein). \\ Systems with
statistical behavior interpolating between the Bose-Einstein and
the Fermi-Dirac were introduced fifty years ago. Up to now many
have been the attempts to generalize quantum statistics. As we
will discuss in the first chapter many of these generalized
statistics can be obtained by means of EIP by an appropriate
generalization of the many particle current expression. The result
of this approach is a nonlinear Schr\"odinger equation with a
complex nonlinearity. Only the imaginary part of this nonlinearity
is fixed by EIP, while the real one, is obtained by the
requirement that the system is canonical. This permits us to
include together within EIP other interactions of different
physical origin.\\ Systems obeying to EIP can be used to describe
a wide range of physical situations. In condensed matter we can
find many topics where EIP concepts can be applied like, for
instance, in superconductivity to describe the formation of
vortex-like excitations or in superfluidity to describe the
formation of Bose-Einstein condensates.

The plane of the work is the following.  In the first chapter,
after an introduction, we explain the origin of EIP. In the same
chapter we summarize the fundamental mathematical tools used in
the thesis: variational principle, Lagrangian and Hamiltonian
formalism for systems with infinity degrees of freedom, symmetries
and conservation quantities. In chapter II, EIP is obtained from
an appropriate deformation of the quantum current. From all the
possible deformations of the current we chose the simplest one to
study the mean properties of systems obeying to EIP. By means of
variational principle we obtain a canonical nonlinear
Schr\"odinger equation. We study the Lagrangian and Hamiltonian
formulation both in the standard $\psi$ representation than in the
hydrodynamic one. In chapter III we look at the symmetries of the
nonlinear Schr\"odinger equation obeying to EIP. By means of
N\"other theorem we obtain and discuss conserved quantities
associated with the symmetries of EIP potential and show that the
systems obeying to EIP are conservative. Only internal forces are
introduced in the system by the presence of EIP potential. In the
same chapter we consider an important class of nonlinear gauge
transformations. These transformations are obtained in order to
linearize the continuity equation. As a consequence, the
transformed evolution equation will be again a nonlinear
Schr\"odinger equation containing a purely real nonlinearity. In
chapter IV EIP-Schr\"odinger equation is coupled in a minimal way
to an abelian gauge field.  The dynamics of the gauge fields is
described by the Maxwell-Chern-Simons Lagrangian. In presence of
Chern-Simons coupling the evolution of the system is developed in
(2+1) space-time dimension so that the model can be used to
describe planar phenomena.  This situation, for instance, is
realized in the layer between two semiconductors or between
superconductor and normal conductor and in particular in
High-$T_c$ superconductors where Chern-Simons coupling is used to
describe the anyonic behavior of the excitations.  We show that in
this case, the anyonic statistic behavior ascribed to the system
by the presence of Chern-Simons Lagrangian is not destroyed by the
presence of EIP potential. The following two chapters are devoted
to the study of special solutions of evolution equations of the
system. In chapter V we study the solitary waves in neutral
systems. Here we obtain the differential equation for the shape of
the solitons, and study its mean properties. Using the nonlinear
gauge transformations obtained in chapter III we deduce the
effective potential responsible of the formation of the solitons.
Applications on the Bose-Einstein condensation are given. In
chapter VI we consider charge, static, self-dual vortex-like
solutions where the dynamics of gauge fields is ruled by the
Chern-Simons Lagrangian. The shapes of the vortices as well as its
electric and magnetic fields are obtained by numerical integration
of appropriate differential equations. The physical properties of
this solution are matched to the corresponding solutions known in
literature where EIP is absent. Conclusions are reported in the
last chapter.

\vspace{5mm} \noindent PACS: 03.65.-w, 05.20.Dd, 11.15.-q\\

\chapter*{Contents}
 \markright{Contents} \contentsline
{chapter}{\numberline {I} \ \ Introduction to EIP}{1}
\contentsline {section}{\numberline {1.1}Considerations on
quantum\\ many body systems}{1} \contentsline
{section}{\numberline {1.2}What is EIP}{7} \contentsline
{section}{\numberline {1.3}Mathematical background}{13}
\contentsline {chapter}{\numberline {II} \ \ Canonical System
Obeying to EIP}{18} \contentsline {section}{\numberline
{2.1}Canonical systems}{18} \contentsline {section}{\numberline
{2.2}Hydrodynamic formulation}{24} \contentsline
{section}{\numberline {2.3}Physical observable}{27} \contentsline
{subsection}{\numberline {2.3.1}Ehrenfest relations}{28}
\contentsline {subsection}{\numberline {2.3.2}Mathematical
proofs}{31} \contentsline {chapter}{\numberline {III} \ \
Symmetries and Conservation Laws}{36} \contentsline
{section}{\numberline {3.1}Lie Symmetries}{37} \contentsline
{section}{\numberline {3.2}Conserved quantities}{41} \contentsline
{section}{\numberline {3.3}Nonlinear gauge transformations}{46}
\contentsline {chapter}{\numberline {IV} \ \ EIP-Gauged
Schr\"odinger Model}{52} \contentsline {section}{\numberline
{4.1}MCS model with EIP}{53} \contentsline {section}{\numberline
{4.2}Hamiltonian formulation}{57} \contentsline
{subsection}{\numberline {4.2.1}Derivation of Ehrenfest
relations}{62} \contentsline {section}{\numberline
{4.3}Conservation laws}{64} \contentsline {chapter}{\numberline
{V} \ \ Canonical Systems Obeying to EIP: Solitons}{72}
\contentsline {section}{\numberline {5.1}Solitons}{73}
\contentsline {section}{\numberline {5.2}Effective potential}{78}
\contentsline {section}{\numberline {5.3}Applications}{82}
\contentsline {chapter}{\numberline {VI} \ \ EIP-Gauged
Schr\"odinger Model: Chern-Simons vortices}{84} \contentsline
{section}{\numberline {6.1}Static solutions}{86} \contentsline
{section}{\numberline {6.2}Vortex like solutions}{92}
\contentsline {section}{\numberline {6.3}Numerical analysis}{95}
\contentsline {chapter}{\numberline {VII} \ \ Conclusions}{98}
\contentsline {chapter}{\numberline {} \ \ Bibliografy}{101}
\contentsline {chapter}{\numberline {} \ \ Figure}{110}



\setcounter{chapter}{1}
\setcounter{equation}{0}
\chapter*{Chapter I\\
\vspace{10mm}Introduction to EIP}
\renewcommand{\thepage}{\arabic{page}}
\markright{Chap. I - Introduction to EIP} \setcounter{page}{1}
\section{Considerations on quantum\\
many body systems}

Macroscopic systems are characterized by a high number of degrees
of freedom that makes impossible to determine exactly their
evolution. This is due to many reasons like, for instance, the
large number of evolution equations that must be solved (one for
each degree of freedom), the incomplete knowledge of the initial
conditions, the approximate knowledge of the physical interaction
and so on. It also true that for complex systems it is more
important to have information about physical observable that
represent mean quantities characterizing the system as a whole.\\
We deal with the evolution equations describing the dynamics of
the mean values of these observable. Because of the high number of
degrees of freedom, it is more convenient to study complex systems
in the context of field theories. Fields carry infinite degrees of
freedom and their dynamics is ruled out by partial differential
equations. The self-interactions introduced by the collective
effects of the many particle systems are in general described by
nonlinear terms in the evolution equations.\\ An important
question in the topics of many body system is the statistic
behavior of its constituents. This question is even more important
for a quantum system where concept of statistics plays an
important role.\\ In this thesis we deal with a nonlinear theory
describing in the mean field approximation an interacting many
body system where the nonlinear interactions are introduced
starting from the kinetics of the system. Therefore the topics of
the nonlinear partial differential equations and the statistic
behavior of a quantum many body system are important subjects of
this thesis. Let us clarify these two arguments.

We begin with a review about quantum statistics. Since the early
days of quantum mechanics, it was clear that, from the principle
of undistinguishability, the statistical behavior of a collection
of identical particles must be different from the classical one.
The particle statistics determines the structure of the many body
wave functions, that turn out to be completely symmetric under
permutations of identical particles named bosons which obey to the
Bose-Einstein statistics or completely antisymmetric under
permutation of identical particles, now called fermions and
obeying to the Fermi-Dirac statistics.\\ The statistics obeyed by
fermions and bosons have many important implications in quantum
mechanics. For instance, the Pauli exclusion principle, which
gives null probability for two fermions to be in the same quantum
state, takes its origin from the antisymmetric fermionic wave
functions; it has consequences like the quality of emitted
spectrum of the atoms or the stability of compact objects likes
white dwarfs or neutron stars. On the contrary, the totally
symmetric bosonic wave function has the effect of enhancing the
probability for the bosons to occupy a quantum state if this one
is already occupied. This effect is responsible of phenomena like
superconductivity or superfluidity.\\ In the early fifty Green
\cite{Green,Chnuki} found that the principles of quantum mechanics
allow two kinds of statistics called {\sl para-Bose} and {\sl
para-Fermi} statistics \cite{Dellantonio}. The parastatistics of
order $p$ are defined as the identical particles statistics in
three dimensional space under the restriction of a possible number
of particles in the symmetric or antisymmetric state for the
para-Fermi and para-Bose respectively. The case with order $p=1$
corresponds to the ordinary Fermi and Bose statistics. The
different cases can be described by trilinear commutation
relations among the creation and annihilation operators
\cite{Govorkov}. The parastatistics was applied to subnuclear
components like quarks \cite{Greenberg} to solve, for example, the
puzzle in the quantum number of the barionic resonance
$\Delta^{++}$.\\ The case of a particle described by a value of
$p$ not an integer has been studied in Ref.
\cite{Ignatev,Biedenharn}, in order to take into account small
violations of Pauli exclusion principle or Bose statistics. In Ref
\cite{Greenberg1} this particle was called {\sl paronic}. However,
the corresponding quantum field theories for such a particle turns
out to have negative norm states and, as a consequence, are not
acceptable \cite{Doplicher,Fredenhagen}. This saga culminates with
a recent study of infinite statistics without assumptions on the
parameter $p$ \cite{Greenberg2,Mohapatra}, all representations of
the symmetric group can occur. The particles obeying this type of
statistics are called {\sl quons}
\cite{Greenberg2,Greenberg3,Greenberg4}. The quonic statistics are
described by the $q$-deformed bilinear commutation relations:
\begin{eqnarray}
a\,a^\dag-q\,a^\dag\,a=1 \ ,\label{quonic}
\end{eqnarray}
where $q$ is a $C$-number with $|q|\leq1$ and according to the
Fredenhagen theorem \cite{Fredenhagen}, cannot be embedded in the
local algebra of observable. Differently from the paronic
statistics the quonic one has positive definite squared norms for
state vectors but notwithstanding Greenberg succeeded only in the
nonrelativistic quantum theory due to the locality problem for the
infinity statistics.\\ Differently from the quonic statististics,
the concept of $q$-deformed oscillators or $q$-{\sl oscillators}
\cite{Chaichian,He,Ubriaco} takes its origin from the concept of
quantum groups \cite{Faddeev}. $q$-oscillator, in spite of classic
oscillators obeys to the $q$-deformed unitary algebra $SU_q(N)$:
\begin{eqnarray}
a\,a^\dag\mp q\,a^\dag\,a=q^{\mp N} \ ,\label{qoscillator}
\end{eqnarray}
with $q\in C$ and $N$ is the number operator functions of the
annihilation and creation operator $a$ and $a^\dag$. In
(\ref{qoscillator}) the sign $-$ or $+$ is reported to the
$q$-bosons or $q$-fermions, respectively. The $q$-oscillators can
formally be defined in any dimensional space but then violate the
fundamental axioms of quantum field theories in terms of the
relation between spin and statistics.

The situation changes radically in many body systems, the dynamics
being confined in two spatial dimensions. As it was discussed by
Wilczek \cite{Wilczek,Wilczek1}, the statistic behavior of a
system depends on the property of interchange of identical
particles and is related to the topological property of the
configuration space of a collection of identical particles. In
more than two dimensions only two possibilities are present. Here
the fundamental group is that of permutation which has two
one-dimensional representations corresponding to completely
symmetric (Bosons) or antisymmetric (Fermions) wave functions.
Differently, in two spatial dimensions the permutation group is
replaced by the braid group \cite{Artin}: the spin is not
quantized in integer or half integer value and particles obey any
statistics interpolating between the Bose and Fermi ones. These
particles are called {\sl anyons}. Successively in Ref.
\cite{Goldin} were achieved similar conclusions using a completely
different method based on the study of the unitary representation
of the current algebra and diffeomorfism group. Anyons can occur
in many physical applications in those condensed matter systems
that can be effectively regarded as two dimensional. For example,
anyons can occur in fractional quantum Hall effect where
collective excitations have been identified as localized
quasi-particles of fractional charges, fractional spin and
fractional statistics \cite{Langhulin,Arovas,Halperin} and also
they can occur in high temperature superconductors recently
discovered \cite{Chen}.\\ Is not too difficult to show in a
heuristic way how anyonic statistics born in two spatial
dimensions. Let $\psi(1,\,2)$ be the function describing two
identical particles and assume that when we move particle 2 around
particle 1 by an angle $\Delta\,\varphi$ the wave function changes
as:
\begin{eqnarray}
\psi(1,\,2)\rightarrow\psi^\prime(1,\,2)=e^{i\,\nu\,\Delta\varphi}\,\psi(1,\,2)
 \ ,
\end{eqnarray}
acquiring a phase factor depending on a statistical parameter
$\nu$. Now we look at the exchange of the two particles. This can
be realized in two ways (see figure 1.1): moves particle 2 around
particle 1 by an angle $\Delta\varphi=\pi$ or at the opposite
side, by an angle $\Delta\varphi=-\pi$. In the two cases the wave
function acquires an extra phase $exp(\pm\pi\,\nu)$. It is easy to
recognize that in more than two spatial dimensions the two paths
are topologically equivalent, so identifying the two
transformations we have the relation:
\begin{eqnarray}
e^{i\,\pi\,\nu}=e^{-i\,\pi\,\nu} \ ,\label{anyonic}
\end{eqnarray}
which is true only if $\nu=0,\,1$ modulo 2, corresponding,
respectively, to the well known bosonic and fermionc statistics.
In more than two dimensions there are no other possibilities.
Differently, in two dimensions we can not deform with continuity
the paths, one into the other. They are topologically and
physically distinct operations. The Eq. (\ref{anyonic}) does not
necessarily hold any more and the statistical  parameter $\nu$ can
be not shrunken to take the value 0 or 1. The anyonic statistics
was obtained in a more rigorous way by Y. -S. Wu \cite{Wu}. For a
review on anyons see for example Ref. \cite{Lerda}.\\ We have
emphasized the anyonic statistics because in chapter IV and VI we
discuss a particle system obeying an exclusion-inclusion
principle, where the matter field is coupled to an abelian gauge
field whose dynamics is described by means of Chern-Simons
Lagrangian. As it was discussed in Ref.
\cite{Wilczek,Arovas1,Hansson} the presence of the Chern-Simons
gauge field confers the anyonic behavior to the system.

Another definition of generalized quantum statistics has been formulated
by Haldane \cite{Haldane,Wu1}, is based on the rate of the number
of the available states in a system of fixed size decreasing as
more particles are added to it. This statistics is called {\sl exclusion statistics}.
The statistics of Haldane is formulated without any reference to spatial dimensions
of the system. In his formulation of exclusion statistics, Haldane
defines a generalized Pauli exclusion principle introducing the
dimension $d_N$ of the Hilbert space for single particle states as a
finite and extensive quantity that depends on the number $N$ of
particles contained in the system. The exclusion principle implies that
the number of available single particle states decreases as the
occupational number increases
\begin{eqnarray}
\Delta\,d_N=-g\,\Delta\,N \ ,\label{Haldane}
\end{eqnarray}
where $g$ is the parameter that characterizes the complete or partial
action of the exclusion principle and makes possible the interpolation
between the Bose-Einstein ($g=0$) and the Fermi-Dirac ($g=1$)
statistics. The relevance of the Haldane statistics is in its implication
in the fractional quantum Hall effect and anyonic physics, in the
Calogero-Sutherland model \cite{Calogero,Sutherland,Sen} and in the Luttinger model
\cite{Murthy,Veigy}.

It is known \cite{Uhlenbeck} that the effects due to the
statistics are imposed by the Pauli exclusion principle to a system of
free fermions and can be simulated by a repulsive potential in the
coordinate space. Analogously, free bosons can be submitted to an
attractive potential. We refer to it as a {\sl statistical potential}.
The statistical potential will be a nonlinear function of the fields
describing the system and its spatial derivatives.

Several nonlinear Schr\"odinger equations (NLSEs) have been
studied in the past and recently, they are commonly used in  many
different fields of research in physics. The cubic equation
\cite{Gross}, for instance, with the nonlinearity proportional to
$\pm|\psi|^2$, has been used to study the dynamical evolution of a
boson gas with $\delta$-function pairwise repulsion or attraction,
responsible of its anyonic-like behavior \cite{Barashenkov}.
Recently, this equation has been used to describe the
Bose-Einstein condensation
\cite{Stringari,Holland,Edwards,Ruprecht,Smerzi} and the dynamics
of two-dimensional radiating vortices \cite{Ivonin}. The nonlinear
term $|\psi|^2$ appears also in the Ginzburg-Landau model of the
superconductivity \cite{Papanicolaou}, a phenomenon investigated
also by means of the Eckhaus equation which is a NLSE with a
nonlinearity of the type $|\psi|^2+\alpha\,|\psi|^4$ \cite{5gk}.
The same equation appears in superfluidity, where the properties
of a gas of bosons interacting via a two-body attractive and
three-body repulsive $\delta$ function inter-particle potential
are investigated \cite{3gk,4gk}. The Eckhaus equation can describe
nonlinear waves in optical fibers with a "normal" dependence of
the refractive index on the light intensity \cite{Ivonin}.\\
Another important example where nonlinearities in the
Schr\"odinger equation induce a statistical behavior is given by
Schr\"odinger-Chern-Simons theory. In Ref. \cite{Jackiw} it was
shown that the gauge fields can be expressed as functions of the
matter fields and therefore can be eliminated from the initial
equation by transforming it into a highly nonlinear Schr\"odinger
equation which describe the same anyonic system.\\ In literature
we can find NLSEs with complex and derivative type nonlinearities
involving the quantities
$({\bfm\nabla}\rho)^2,\,\Delta\rho,\,{\bfm
j}\cdot{\bfm\nabla}\rho,\,{\bfm\nabla}\cdot{\bfm j}$
\cite{8gk,Doebner1,Doebner2} as, for instance, in the
Doebner-Goldin equation associated with a certain unitary group
representation and describing irreversible and dissipative quantum
systems. NLSEs with nonlinearities involving the quantity $\bf j$
have been also introduced to study planar systems of particles
with anyon statistics \cite{Aglietti}.
\\ We will see in chapter II that the potential introduced by the
EIP is complex and derivative in the field $\psi$ and $\psi^\ast$.

Solution spectrum of nonlinear partial differential equations is much
rich than the linear one.
Generally, the solutions of a nonlinear PDE can be split in two distinct classes:
in the first we have the solutions that
can be obtained with the perturbative method. This is possible if the
coupling constant of the nonlinearity is small.
In the other class we find the nonperturbative solutions, obtained
integrating directly the nonlinear PDE. This solutions depend on the coupling
constant of the nonlinearity and are divergent in the zero limit. So, it is not possible
to go continuously from the solutions of perturbative class to the
nonperturbative class.
In this one we find the soliton solutions. They are wave packets that
propagate freely asymptotically not changing their shape and velocity
also after a collision.
These were discovered in the last century by Russel, but their interest
in physics was emphasized only thirty years ago.
Solitons are solutions of nonlinear partial differential equation.
For instance, if we take into account the dispersion relation of the linear Schr\"odinger
equation we have:
\begin{eqnarray}
\hbar\,\omega=\frac{\hbar^2\,{\bfm k}^2}{2\,m} \ ,
\end{eqnarray}
and looking for the group velocity ${\bfm v}_g=d{\bfm k}/d\omega$:
\begin{eqnarray}
{\bfm v}_g=\frac{\hbar\,{\bfm k}}{m} \ ,
\end{eqnarray}
which depends on the number wave $\bfm k$. This means that each
component in the wave packet propagates with a different velocity
and therefore the modulation of the shape of the packet is not
maintained. Responsible of this dispersion is the Laplacian
operator appearing in the kinetic term. To avoid this effect a
confining potential is required which makes the evolution equation
nonlinear. \\ Soliton solutions were found in many fields, for
example light impulse in the wave guide
\cite{Chiao,Talanov,Kelley,Shabat}, propagation in electric
circuits and plasma waves \cite{Scott}. Their particle behavior
makes them interesting in the physics of elementary particles
\cite{Rajaraman} like for instance the t'Hooft-Polyakov monopole
\cite{tHooft,Polyakov}.\\ The question of the research of the
solutions of nonlinear differential equations is one of the most
important topics of mathematical physics of the last years. One of
the most powerful is the inverse scattering method \cite{Kruskal}.
It is a canonical transformation in the action-angle fields in
which the Hamiltonian of the system appears to be diagonalized.
This transformation is a Fourier transformation plus a Laplace
one. When an evolution equation is solved with this method, the
Hamiltonian spectrum results decomposed in a continuous part given
by the perturbative solutions plus a discrete contribution given
by the soliton solutions. Equations solved with the inverse
scattering method are named $S$-integrable because their solutions
are expressed as function on a spectral parameter. The more easy
$S$-integrable equation is of course the Schr\"odinger equation
which is solved by means of Fourier transformation.\\ Finally a
new class of integrable equations have made their appearance
recently. These are called $C$-integrable because can be
linearized by means of change of dependent variables. In this
category we can find for example the Burger equation \cite{Olver}
or the Ekhaus equation \cite{Calogero1}.

\setcounter{equation}{0}
\section{What is EIP}

We present the generalized Exclusion-Inclusion Principle (EIP) in
the configuration space. For a rigorous introduction of the EIP we
remind to the reference
\cite{Quarati,Quarati1,Kaniadakis,Kaniadakis1,Kaniadakis2,Kaniadakis3}.
In the first part of this section the discussion is keep at a
classic level.\\ The EIP takes its origin from a classical
nonlinear kinetic that takes into account inhibition or
enhancement of the particle transition probabilities in the phase
space.\\ We start by considering a Markoffian process in a
$D$-dimensional phase space. If we identify the state of the
system by a vector in the phase space and setting $\pi(t,\,{\bfm
u}\rightarrow{\bfm v})$ the transition probability from the state
$\bfm u$ to the state $\bfm v$, the evolution equation for the
distribution function $n(t,\,{\bfm v})$ can be written as
\begin{eqnarray}
\frac{\partial\,n(t,\,{\bfm v})}{\partial\,t}=\int\left[\pi(t,\,{\bfm
u}\rightarrow{\bfm v})-\pi(t,\,{\bfm v}\rightarrow{\bfm u})\right]\,d^D{\bfm u}
\ .\label{fokker}
\end{eqnarray}
The exclusion-inclusion principle is introduced into the classical
transition probabilities by means of an inhibition or an enhancement
factor.\\
In Ref. \cite{Quarati1} it was postulated the following expression of the transition
probability:
\begin{eqnarray}
\pi(t,\,{\bfm v}\rightarrow{\bfm u})=r(t,\,{\bfm v},\,{\bfm v}-{\bfm
u})\,\phi[n(t,\,{\bfm v})]\,\psi[n(t,\,{\bfm u})] \ ,
\end{eqnarray}
where $r(t,\,{\bfm v},\,{\bfm v}-{\bfm u})$ is the transition rate,
$\phi[n(t,\,{\bfm v})]$ is a function depending on the
occupational distribution at the initial state $\bfm v$ and
$\psi[n(t,\,{\bfm u})]$ depends on the arrival state.
The function $\phi(n)$ must obey the condition $\phi(0)=0$ because the transition
probability is equal to zero if the initial state is empty. Furthermore, the
function $\psi(n)$ must obey the condition $\psi(0)=1$ because,
if the arrival state is empty, the transition probability is not modified.
The classical linear case is obtained if we chose $\phi(n)=n$ and
$\psi(n)=1$.\\
In the following, for reason of simplicity, we consider an one-dimensional
space, in the first neighbor interaction approximation.\\
If we consider an infinitesimal transition from the state $v$ to the
state $v+dv$, the transition rate can be defined as \cite{Quarati1}
\begin{eqnarray}
r(t,\,v,\,\pm dv)\,dv^2=D(t,\,v)\pm{1\over2}J(t,\,v)\,dv \
,\label{rate}
\end{eqnarray}
where $J(t,\,v)$ and $D(t,\,v)$ are the drift and diffusion coefficients, respectively.
In the simple case in which in a time interval $dt$ only one transition
is allowed, Eq. (\ref{fokker}) becomes
\begin{eqnarray}
\nonumber
\frac{\partial\,n(t,\,v)}{\partial\,t}&=&\pi(t,\,v-dv\rightarrow v)+\pi(t,\,v+dv\rightarrow
v)\\
&-&\pi(t,\,v\rightarrow v-dv)-\pi(t,\,v\rightarrow v+dv) \ .\label{bilancio}
\end{eqnarray}
The (\ref{bilancio}) is a balance equation between the particle
coming in $v\pm dv\rightarrow v$ in the site $v$ with respect to
that coming out $v\rightarrow v\pm dv$. It describes in the
continuum limit the discrete Markoffian process reported in figure
1.2.\\
Expanding the r.h.s. of Eq. (\ref{bilancio}) in powers of
$dv$, up to second order, in the limit $dv\rightarrow0$ and taking
into account the transition rate defined in Eq. (\ref{rate}), we
obtain the following generalized, nonlinear Fokker-Planck
equation:
\begin{eqnarray}
\nonumber
\frac{\partial\,n(t,\,v)}{\partial\,t}&=&\frac{\partial}
{\partial\,v}\left[\left(J(t,\,v)+\frac{\partial\,D(t,\,v)}
{\partial\,v}\right)\,\phi(n)\,\psi(n)\right.\\
&+&\left.D(t,\,v)\,\left(\psi(n)\frac{\partial\,\phi(n)}{\partial\,v}
-\phi(n)\frac{\partial\,\psi(n)}{\partial\,v}\right)\right] \ .
\end{eqnarray}
This is a continuity equation for the distribution function $n=n(t,\,v)$
\begin{eqnarray}
\frac{\partial\,n(t,\,v)}{\partial\,t}-\frac{\partial\,j(t,\,v,\,n)}{\partial\,v}=0
 \ ,
\end{eqnarray}
where the particle current $j=j(t,\,v,\,n)$ is given by:
\begin{eqnarray}
j=\left(J(t,\,v)+\frac{\partial\,D(t,\,v)}{\partial\,v}\right)\,\phi(n)\,\psi(n)
+D(t,\,v)\,\left(\psi(n)\frac{\partial\,\phi(n)}{\partial\,v}
-\phi(n)\frac{\partial\,\psi(n)}{\partial\,v}\right) \ .\label{corrente}
\end{eqnarray}
When the system reaches the equilibrium configuration, the current
(\ref{corrente}) must be equal to zero. In particular, if we
consider Brownian particles, the drift and the diffusion
coefficients are given by:
\begin{eqnarray}
J=\gamma\,v \ ,\hspace{30mm} D=\frac{\gamma}{\beta\,m} \ ,
\end{eqnarray}
with $\gamma$ a dimensional constant and $\beta=1/k_B\,T$ where
$k_B$ is the Boltzmann constant.\\
In this situation, when $j=0$ Eq. (\ref{corrente}) can be viewed as a first
order differential equation with solution:
\begin{eqnarray}
\frac{\phi(n)}{\psi(n)}=e^{-\epsilon} \ ,\label{equilibrium}
\end{eqnarray}
where $\epsilon=\beta\,(E-\mu)$ and $E=m\,v^2/2$ is the kinetic energy.
The integration constant $\mu$ is
the chemical potential which can be evaluated by fixing the number of particles of the
system.

We show now how we can select the functions $\phi(n)$ and
$\psi(n)$ in order to obtain the equilibrium distribution of some
of the statistical distributions described in the previous
section.\\ If we make the choice:
\begin{eqnarray}
\phi(n)=n \ ,\hspace{30mm}\psi(n)=1+\kappa\,n \ ,
\end{eqnarray}
from (\ref{equilibrium}) we obtain the following stationary distribution:
\begin{eqnarray}
n=\frac{1}{e^\epsilon-\kappa} \ .
\end{eqnarray}
Here we recognize the well know Maxwell-Boltzmann (MB),
Bose-Einstein (BE) and Fermi-Dirac (FD) distributions when we fix the value of the statistic
parameter $\kappa=0,\,\pm1$ respectively. Moreover for
$-1<\kappa<1$ we have fractional distributions interpolating between the BE and
FD ones.\\
As a second example we pose:
\begin{eqnarray}
\nonumber
&&\phi(n)=n\,(1-g\,n)^\frac{1-g}{2}\,[1+(1-g)\,n]^\frac{g}{2} \
,\\
&&\\
&&\psi(n)=(1-g\,n)^\frac{1+g}{2}\,[1+(1-g)\,n]^{1-\frac{g}{2}} \ ,
\end{eqnarray}
and obtain the distribution of Haldane's statistics:
\begin{eqnarray}
n\,e^\epsilon=(1-g\,n)^g\,[1+(1-g)\,n]^{1-g} \ ,
\end{eqnarray}
obtained in Ref. \cite{Wu1}. The statistical parameter $g$ is defined in (\ref{Haldane}).
Finally if we pose:
\begin{eqnarray}
\phi(n)=[n]_q \ ,\hspace{30mm}\psi(n)=[1+\sigma\,n]_q \ ,
\end{eqnarray}
where $[x]_q$ is defined as:
\begin{eqnarray}
[x]_q=\frac{q^x-q^{-x}}{q-q^{-1}} \ ,
\end{eqnarray}
we obtain the $q$-oscillator distribution:
\begin{eqnarray}
e^\epsilon=\frac{\sinh[\eta\,(1+\sigma\,n)]}{\sinh(\eta\,n)} \ ,
\end{eqnarray}
with $\eta=\log\,q$. The parameter $\sigma=\pm1$ permits us to
describe bosonic or fermionic $q$-oscillator respectively.

We can give other examples but it is now clear how to obtain
different statistical distributions starting from a nonlinear
kinetic.\\ In this thesis {\sl we shall study, in mean field
approximation, a canonical quantum system obeying to an
exclusion-inclusion principle obtained by choosing $\phi(n)=n$ and
$\psi(n)=1+\kappa\,n$}.\\ This choice is made because it is the
most easy to treat and, as we have shown before, permits us to
simulate a system with an equilibrium distribution interpolating
between the BE and FD distribution.\\

Now we introduce EIP in a quantum system. \\  Let us start by
looking at an exclusion-inclusion principle in the configuration
space. We consider the classical stochastic Marcoffian process in
a 3-dimensional space described by the following forward nonlinear
Fokker-Planck equation (FPE):
\begin{eqnarray}
\frac{\partial\,\rho}{\partial\,t}+{\bfm\nabla}\cdot{\bfm
j}^{(+)}=0 \ ,\label{continuita}
\end{eqnarray}
with
\begin{eqnarray}
{\bfm j}^{(+)}={\bfm
u}^{(+)}\,\rho\,(1+\kappa\,\rho)-D\,{\bfm\nabla}\,\rho \
,\label{corrente1}
\end{eqnarray}
where $\rho=\rho(t,\,{\bfm x})$ is the occupational number or
particle distribution in the configuration space, ${\bfm u}^{(+)}$
is the forward velocity and $\kappa\in I\!\!R$.\\ Beside the
forward FPE (\ref{corrente1}) we must also consider the backward
one with current:
\begin{eqnarray}
{\bfm j}^{(-)}={\bfm
u}^{(-)}\,\rho\,(1+\kappa\,\rho)+D\,{\bfm\nabla}\,\rho \ .
\end{eqnarray}
The semisum of the forward and backward FPEs will give us
\begin{eqnarray}
\frac{\partial\,\rho}{\partial\,t}+{\bfm\nabla}\left[{\bfm
v}\,\rho\,(1+\kappa\,\rho)\right]=0 \ ,\label{cccc}
\end{eqnarray}
where $\bfm v$ is the current velocity given by:
\begin{eqnarray}
{\bfm v}={1\over2}\,\left[{\bfm u}^{(+)}+{\bfm u}^{(-)}\right] \ .
\end{eqnarray}
 We stress one more on the meaning of the current.
We define the transition probability from the site $\bfm x$ to
${\bfm x}^\prime$ as $\pi(t,\,{\bfm x}\rightarrow{\bfm
x}^\prime)=r(t,\,{\bfm x},\,{\bfm x}^\prime)\,\rho(t,\,{\bfm
x})\,[1+\kappa\,\rho(t,\,{\bfm x}^\prime)]$ with $r(t,\,{\bfm
x},\,{\bfm x}^\prime)$ the transition rate and the choice
$\phi(\rho)=\rho$, $\psi=1+\kappa\,\rho$. The transition
probability depends on the particle population $\rho(t,\,{\bfm
x})$ of the starting point $\bfm x$ and also on the population
$\rho(t,\,{\bfm x}^\prime)$ of the arrival point ${\bfm
x}^\prime$. If $\kappa>0$ the $\pi(t,\,{\bfm x}\rightarrow{\bfm
x}^\prime)$ introduces an inclusion principle. In fact the
population at the arrival point ${\bfm x}^\prime$ stimulates the
transition and the transition probability increases linearly with
$\rho(t,\,{\bfm x}^\prime)$. In the case $\kappa<0$ the
$\pi(t,\,{\bfm x}\rightarrow{\bfm x}^\prime)$ takes into account
the Pauli exclusion principle. If the arrival point ${\bfm
x}^\prime$ is empty $\rho(t,\,{\bfm x}^\prime)=0$, the
$\pi(t,\,{\bfm x}\rightarrow{\bfm x}^\prime)$ depends only on the
population of the starting point. If the arrival site is populated
$0<\rho(t,\,{\bfm x}^\prime)\leq\rho_{\rm max}$ the transition is
inhibited. The range of values the parameter $\kappa$ can assume
is bounded by the condition that $\pi(t,\,{\bfm x}\rightarrow{\bfm
x}^\prime)$ be real and positive as the $r(t,\,{\bfm x},\,{\bfm
x}^\prime)$. Then we conclude that $\kappa$ is limited from below
by the condition $\kappa\geq-1/\rho_{\rm max}$.

In the Nelson picture \cite{Nelson,Fenyes,Weizel}, a quantum system can be viewed as
a stochastic
process where the particles are subjected to a Brownian diffusion with a
coefficient $D=\hbar/2\,m$. The quantum system is described by the
complex wave function $\psi=\psi(t,\,{\bfm x})$ and the quantity
$\rho=|\psi|^2$ is interpreted as the particle probability density. We make
the ansatz:
\begin{eqnarray}
\psi=\rho^{1/2}\,\exp\left(\frac{i}{\hbar}S\right) \ ,\label{ansatz}
\end{eqnarray}
where the phase $S$ is related to the velocity $\bfm v$ as
\cite{Weizel,Kaniadakis5}:
\begin{eqnarray}
{\bfm v}=\frac{{\bfm\nabla}\,S}{m} \ .\label{fase}
\end{eqnarray}
Using Eqs. (\ref{ansatz}) and (\ref{fase}) it is immediate to
obtain from Eq. (\ref{cccc}) the expression for the quantum
current:
\begin{eqnarray}
{\bfm
j}=-\frac{i\,\hbar}{2\,m}\,(1+\kappa\,\rho)\,(\psi^\ast\,{\bfm\nabla}\psi-
\psi\,{\bfm\nabla}\psi^\ast) \ ,\label{EIPcorrente}
\end{eqnarray}
which we can rewrite also as:
\begin{eqnarray}
{\bfm j}=\frac{{\bfm\nabla}S}{m}\,\rho\,(1+\kappa\,\rho) \ ,\label{generalcorrente}
\end{eqnarray}
which is our starting assumption.\\
In Ref. \cite{Kaniadakis6}, by using Nelson stochastic quantization
method, it was obtained
the following NLSE:
\begin{eqnarray}
i\,\hbar\,\frac{\partial\,\psi}{\partial\,t}=-\frac{\hbar^2}{2\,m}\,\Delta\,\psi+\Lambda(\rho,\,{\bfm
 j})\,\psi+V\,\psi \ ,\label{schroedinger1}
\end{eqnarray}
where:
\begin{eqnarray}
\Lambda(\rho,\,{\bfm j})=W(\rho)+i\,{\cal W}(\rho,\,{\bfm j}) \ ,
\end{eqnarray}
is the complex nonlinearity introduced by the EIP, and the expression of
the imaginary part is:
\begin{eqnarray}
{\cal W}(\rho,\,{\bfm
j})=-\kappa\,\frac{\hbar}{2\,\rho}\,{\bfm\nabla}\left(\frac{{\bfm
j}\,\rho}{1+\kappa\,\rho}\right) \ .\label{immaginario1}
\end{eqnarray}
The real part $W(\rho)$ depends drastically on the quantization method.
In the picture of stochastic quantization it was obtained the form:
\begin{eqnarray}
W(\rho)=\kappa\,\frac{\hbar^2}{4\,m}\,\left[\frac{\Delta\,\rho}
{1+\kappa\,\rho}+\frac{2-\kappa\,\rho}{2\,\rho\,(1+\kappa\,\rho)^2}({\bfm
\nabla}\,\rho)^2\right] \ .\label{reale1}
\end{eqnarray}
It can be shown that the system described by Eqs. (\ref{schroedinger1}),
(\ref{immaginario1}) and (\ref{reale1}) admits a continuity
equation (\ref{continuita}) with the current given by Eq.
(\ref{EIPcorrente}).\\
We define in the following the {\sl quantum system obeying to EIP}
as a system whose dynamics is described by a NLSE admitting a
continuity equation with
the quantum current gave by Eq. (\ref{EIPcorrente}).

We conclude this section by considering some examples where EIP can be usefully applied. In nuclear
physics the correlation effects between pairs of nucleons, viewed as
fermions, are quite relevant in the interpretation of experimental results.
Similarly, the interactions among bosons are relevant in various nuclear
modes (superfluid model, interacting boson model, mean field boson
approximation) and allow the explanation of many collective nuclear properties.
The interaction among the fermionic valence nucleons outside the core produces
pairs of correlated nucleons that can be approximated as particles with a
behavior intermediate between fermionic and bosonic ones.  This nuclear
state (quasideuteron state) can be viewed as a particles system that obeys
to EIP.  Recently, it was studied a semiclassical model of photofission in the
quasideuteron energy region \cite{Quarati2}. We described the quasideuteron state as
a mixture of fermion and boson states, with a good agreement of our
calculated photofission cross sections of several heavy nuclei and
experimental results.\\
Another example is the Bose-Einstein condensation.
The condensation originates from an attraction of statistical nature
(Bose-Einstein statistics) among the particles.
In several papers the Bose-Einstein condensation
is studied by means of a cubic NLSE \cite{Stringari,Holland} which describes in mean field
approximation an attractive interaction between two bodies. In place of the cubic
and simplest interaction, other interactions can be considered as,
for instance, the one introduced by EIP to simulate an attraction
among the particles.\\
In condensed matter we can consider the problem of the hopping transport
on a lattice of ionic conductors: two ions having the same charge cannot
occupy the same site due to their natural electrostatic repulsion; also
the motion of the couple electron-hole in a semiconductor can be described
by means of EIP. In fact, while electrons and hole are fermions, together can
be considered excited states behaving differently from that of a
fermion or a boson.


\setcounter{equation}{0}
\section{Mathematical background}

In this section we summarize the canonical method used in the
study of systems with infinite degrees of freedom whose dynamics
is described by a partial differential equations.\\ Let us start
with some definitions and notations. Let $M$ to be a complex
smooth manifold with dimension $D$ mapped by the coordinates $x_i$
with $i=1,\,\cdots,\,D$. Let $\cal F$ be the algebra of the
functionals on $M\rightarrow I\!\!R$ of the type: ${\cal F}\ni
P=\int{\cal P}\,d^Dx$. We call the quantity $\cal P$ functional
density. Let $\psi(t,\,{\bfm x})$ and $\psi^\ast(t,\,{\bfm x})$ be
two fields on $M$\footnote{Because the theory developed in this
thesis is nonrelativistic, here and after $t$ play the role of a
parameter}, we set $\Psi\equiv(\psi,\,\psi^\ast)$ the complex
2-vector field on $M\times M$. We assume that the 2-vector field
$\Psi$ vanishes quickly on the boundary of $M$.\\ We consider now
a non relativistic canonical quantum system described by the
Lagrangian density ${\cal L}[\psi]\equiv{\cal L}(\partial_t
\psi^\ast,\,\partial_t\psi,\,[\psi^\ast],\,[\psi])\in{\cal F}$
depending on the scalar field $\psi\in M$ and its derivatives with
$\partial_t=\partial/\partial\,t$. Here and in the following we
adopt the notation in square bracket to indicate the dependence
from the spatial derivative of any order. Moreover, for our
purpose, we deal with the case in which the spatial derivative in
the Lagrangian density are introduced by $\bfm\nabla$ operator.
Thus, the following notations means:
\begin{eqnarray}
{\cal L}([a])\equiv{\cal
L}(a,\,{\bfm\nabla}a,\,{\bfm\nabla}^2a,\,{\bfm\nabla}^3a,\,\cdots)
\ .
\end{eqnarray}
Posing the action functional:
\begin{equation}
{\cal A}=\int{\cal L}\,d^Dx\,dt \ ,\label{deflagran}
\end{equation}
the evolution equations for the fields $\psi$ and $\psi^\ast$ can be obtained by
a variational principle \cite{Gelfand}:
\begin{equation}
\delta\,{\cal A}=0 \ ,\label{vara}
\end{equation}
where the variation of a functional $F\in {\cal F}$ is a 2-vector
defined as:
\begin{eqnarray}
\delta\,F=\left(\frac{\delta\,F}{\delta\,\psi},\,
\frac{\delta\,F}{\delta\,\psi^\ast}\right) \ .
\end{eqnarray}
The functional derivative can be defined by means of Euler operator:
\begin{eqnarray}
\frac{\delta\,F}{\delta\,\psi}\equiv{\rm E}_{_\psi}({\cal A}) \
,\hspace{30mm} \frac{\delta\,F}{\delta\,\psi^\ast}\equiv{\rm
E}_{_{\psi^\ast}}^\ast({\cal A}) \ ,
\end{eqnarray}
which it is given by:
\begin{eqnarray}
{\rm
E}_{_\psi}=\sum_{n=0}^\infty\,\sum_J\,(-D)^J\frac{\partial}{\partial_{_J}\,\psi}
\ ,\label{euler}
\end{eqnarray}
where the second sum is extended over multi-indices
$J=(j_t,\,j_1,\,\cdots,\,j_D)$ with $0\leq j_i\leq n$,
$i=t,\,1,\,\cdots,\,D$ and $j_t+\sum\,j_i=n$. Eq. (\ref{vara}) are
the Euler-Lagrange equations for the field $\psi^\ast$ and $\psi$
respectively. From Eq. (\ref{vara}) we obtain:
\begin{equation}
\sum_{n=0}^\infty(-1)^n\,{\bfm\nabla}^n\frac{\partial\,{\cal
 L}}{\partial({\bfm\nabla}^n\,\psi)}-\frac{d}{d\,
t}\frac{\partial\,{\cal L}}{\partial(\partial_t\,\psi)}=0 \
,\label{lagrange}
\end{equation}
and its conjugate equation.\\
It is possible to show that the Euler operator satisfies the following
property:
\begin{eqnarray}
{\rm E}(\frac{\partial\,B}{\partial\,t}+{\bfm\nabla}\,C)=0 \ ,
\end{eqnarray}
with $B,\,C\in{\cal F}$. Therefore, Lagrangian density which are
total derivatives does not gives contribute to the evolution
equations. It are named {\sl null Lagrangian}.\\ In this thesis,
we will consider canonical systems described by the following
class of Lagrangian density in (3+1) dimensions:
\begin{eqnarray}
{\cal
L}=i\,\frac{\hbar}{2}\left(\psi^\ast\,\frac{\partial\psi}{\partial\,t}-
\psi\,\frac{\partial\psi^\ast}{\partial\,t}\right)-\frac{\hbar^2}{2\,m}
|{\bfm\nabla}\psi|^2-U([\psi^\ast],\,[\psi])-V\,\psi^\ast\,\psi \ ,\label{derivata}
\end{eqnarray}
where $V$ describes an external potential and
$U([\psi^\ast],\,[\psi])$ is a real nonlinear potential.\\ Using
the Lagrangian (\ref{derivata}) we obtain the following NLSE for
the field $\psi$:
\begin{eqnarray}
i\,\hbar\,\frac{\partial\,\psi}{\partial\,t}=-\frac{\hbar^2}
{2\,m}\,\Delta\,\psi+\sum_{n=0}^\infty\,(-1)^n\,{\bfm\nabla}^n
\frac{\partial\,U}{\partial({\bfm\nabla}^n\,\psi)}+V\,\psi \ ,
\end{eqnarray}
which is a derivative NLSE.

In a canonical theory, we can express the motion equations also in
the Hamilton formalism. This requires that functional space $M$ is
a {\sl Poisson manifold}, i.e. a $D$-dimensional smooth manifold
equipped with the Poisson brackets. This one is defined as:
\begin{eqnarray}
\{F,\,G\}=\int(\delta F\,{\cal D}\,\delta G)\,d^Dx \ ,
\end{eqnarray}
where $F,\,G\in {\cal F}$. $\cal D$ is the Hamiltonian linear
operator acting on ${\cal F}$ which might depends by $\psi$ and
its derivative.\\ The Poisson brackets must satisfy the following
properties:
\begin{itemize}
\item Linear
\item Skew-symmetric
\item Jacobi identity
\end{itemize}
This property is reflected on the definition of $\cal D$. The
first property is evident from the linearity of $\cal D$ and the
definition of Poisson brackets, the second require $\cal D$ to be
skew-adjoint: ${\cal D}^\dag=-{\cal D}$. Finally, the third
condition is:
\begin{eqnarray}
\{\,\{P,\,Q\},\,R\}+\mbox{cyclic permutations}=0 \ .
\end{eqnarray}
This last relation is true if $\cal D$ is a constant quantity not
dependent on the fields $\psi$ and $\psi^\ast$.\\ Eqs.
(\ref{lagrange}) can be expressed in the Hamiltonian formalism if
a functional ${\cal F}\ni H=\int{\cal H}\,d^Dx$ called Hamiltonian
function exists ($\cal H$ is the Hamiltonia density) so that:
\begin{eqnarray}
\frac{\partial\,\Psi}{\partial\,t}={\cal D}\,\delta\, H \ ,\label{hamilton}
\end{eqnarray}
which is equivalent to the following expression in terms of
Poisson brackets:
\begin{eqnarray}
\frac{\partial\,\Psi}{\partial\,t}=\{\Psi,\,H\} \ .
\end{eqnarray}
In general, given a functional $F\in {\cal F}$ describing a
physical observable of a system with Hamiltonian $H$, its time
evolution can be written as:
\begin{eqnarray}
\frac{d\,F}{d\,t}=\{F,\,H\}+\frac{\partial\,F}{\partial\,t} \ .\label{poissonn}
\end{eqnarray}
We remark {\sl en passant} that for a system with finite
dimension, the Poisson brackets must satisfy one more condition:
it must be a derivative operator which means to satisfy the
Leibnitz rule $\{A\,B,\,C\}=A\,\{B,\,C\}+\{A,\,C\}\,B$. For
infinitely dimensional systems this condition is not imposed
because the multiplication between elements of ${\cal F}$ is not
well defined. In fact, it gives two functionals $A,\,B\in {\cal
F}$, their product is not expressible as integral of a density
functional i.e. the relation:
\begin{eqnarray}
\int{\cal A}\,d^Dx\,\int{\cal B}\,d^Dx=\int{\cal C}\,d^Dx \
,\hspace{20mm}\mbox{for same density ${\cal C}$}
\end{eqnarray}
is not generally satisfied.\\
Let us now introduce the fields $\pi_{_\psi}$ and $\pi_{_{\psi^\ast}}$, canonically conjugate
momenta of the fields $\psi$ and $\psi^\ast$, respectively and define:
\begin{equation}
\pi_{_\psi}=\frac{\partial\,{\cal L}}{\partial(\partial_t\,\psi)} \ ,\hspace{30mm}
\pi_{_{\psi^\ast}}=\frac{\partial\,{\cal L}}{\partial(\partial_t\,{\psi^\ast})} \ .
\label{defmomenti}
\end{equation}
We can write the Hamiltonian density
${\cal H}([\psi],\,[\pi_{_\psi}])$ related
to the Lagrangian density by the Legendre transformation \cite{Goldstein}):
\begin{equation}
{\cal H}=\pi_{_\psi}\,{\partial\,\psi\over\partial\,t}+\pi_{_{\psi^\ast}}\,{\partial\,\psi^\ast\over\partial\,t}-{\cal
L} \ ,\label{legendre}
\end{equation}
and the evolution of the system is described by the equations:
\begin{eqnarray}
\nonumber
&&\frac{\partial\,\psi}{\partial\,t}=\frac{\delta\,H}{\delta\,\pi_{_\psi}}
\ ,\\ \label{ham}\\ \nonumber
&&\frac{\partial\,\pi_{_\psi}}{\partial\,t}=-\frac{\delta
H}{\delta\,\psi} \ .
\end{eqnarray}
It is easy to see that Eqs. (\ref{ham}) are equivalent to Eq.
(\ref{hamilton}) if we make the choice for the Hamiltonian
operator:
\begin{eqnarray}
{\cal D}=\left(
\begin{array}{cc}
0&1\\
-1&0
\end{array}
\right) \ ,\label{hoperatore}
\end{eqnarray}
which is skew-hermitian and satisfies the Jacobi identity because it has
constant entry.\\
With this definition for the operator $\cal D$ the expression
(\ref{poissonn}) becomes:
\begin{eqnarray}
\nonumber \frac{d\,F}{d\,t}&=&\int\left[\frac{\delta\,F({\bf
x})}{\delta\,\psi({\bf z})} \frac{\delta\,H({\bf
y})}{\delta\,\pi_{_\psi}({\bf z})}-\frac{\delta\,H({\bf
y})}{\delta\,\psi({\bf z})}\frac{\delta\,F({\bf x})}
{\delta\,\pi_{_\psi}({\bf z})} \right]\,d^Dz\\
&+&\int\left[\frac{\delta\,F({\bf x})}{\delta\,\psi^\ast({\bf z})}
\frac{\delta\,H({\bf y})}{\delta\,\pi_{_{\psi^\ast}}({\bf
z})}-\frac{\delta\,H({\bf y})}{\delta\,\psi^\ast({\bf
z})}\frac{\delta\,F({\bf x})} {\delta\,\pi_{_{\psi^\ast}}({\bf
z})} \right]\,d^Dz+\frac{\partial\,F}{\partial\,t} \ ,
\label{poisson1}
\end{eqnarray}
which can be simplified following the Dirac procedure \cite{Dirac}
(see sections 2.1 and 4.2).\\ An important fact, in canonical
theories, follows from the N\"other theorem \cite{Noether}, which
states the link between symmetries for the Lagrangian (or
Hamiltonian) and the conserved physical quantities related to it.
We resume briefly the general method.\\ The N\"other theorem says
that for each one-parameter symmetry group of the system, there is
a physical observable related to it:
\begin{eqnarray}
Q=\int{\cal J}^0\,d^Dx \ ,
\end{eqnarray}
which is conserved as a consequence of the continuity equation:
\begin{eqnarray}
\frac{\partial\,{\cal J}^0}{\partial\,t}+\frac{\partial\,{\cal
J}^i}{\partial\,x_i}=0 \ ,\label{cn1}
\end{eqnarray}
where $i=1,\,\cdots,\,D$ (sum on $i$ is assumed) and ${\cal J}^i$
are the components of a $D$-vector describing the flux density
associated to the ${\cal J}^0$.\\ In fact, let $\delta\,\psi$ be
the variation produced by the action of a group transformation on
$\psi$, it generates a symmetry for the system if the action
(\ref{deflagran}) does not change. This implies for the
Lagrangian:
\begin{equation}
\delta {\cal L}=\partial_\nu f^\nu \ , \label{divergence}
\end{equation}
which says that it might change for a total divergence (null Lagrangian).
To give an example, we consider the easy case in which only the first
derivative of the fields are present in the Lagrangian density.
Computing the variation in $\cal L$ we obtain:
\begin{eqnarray}
\frac{\partial\,{\cal
L}}{\partial\,\psi}\,\delta\,\psi+\frac{\partial\,{\cal
L}}{\partial(\partial_\nu\psi)}\,\delta\,\partial_\nu\psi+c.c.=\partial_\nu
 f^\nu \ ,
\end{eqnarray}
and after integration by part, taking into account the motion
equations (\ref{cccc}) we arrive at Eq. (\ref{cn1}) where:
\begin{eqnarray}
{\cal J}^\nu=\frac{\partial\,{\cal L}}{\partial\,(\partial_\nu\psi)}
\,\delta\,\psi+\frac{\partial\,{\cal L}}{\partial\,(\partial_\nu\psi^\ast)}
\,\delta\,\psi^\ast-f^\nu \ . \label{ncurrent1}
\end{eqnarray}
We remember that the ${\cal J}^\nu$ is not univocally defined. In fact, we
can add to the expression (\ref{ncurrent1}) the gradient of a
skew-symmetric 2-tensor:
\begin{eqnarray}
{\cal J}^\nu\rightarrow\widetilde{\cal J}^\nu={\cal
J}^\nu+\partial_\mu\,T^{\mu\nu} \ ,\label{sostituzione}
\end{eqnarray}
with
\begin{eqnarray}
T^{\mu\nu}=-T^{\nu\mu} \ .
\end{eqnarray}
As a consequence of the skew-symmetry the new vector $\widetilde{\cal
J}^\nu$ satisfies the same continuity equation (\ref{cn1}) and the
conserved quantities are left unchanged by the substitution (\ref{sostituzione})
if the field goes to zero on the boundary of $M$.\\


\setcounter{chapter}{2}
\setcounter{section}{0}
\setcounter{equation}{0}
\chapter*{Chapter II\\
\vspace{10mm}Canonical Systems Obeying to the\\
Exclusion-Inclusion Principle} \markright{Chap. II - Canonical
Systems Obeying to EIP}

In this chapter we introduce a class of canonical nonlinear
Schr\"odinger equations obeying to a generalized
inclusion-exclusion principle (EIP). This is accomplished trough
an opportune deformation of the expression of the quantum current
of the matter field $\psi$. The Lagrangian and Hamiltonian
structure are studied both in the $\psi$-representation and in the
hydrodynamic one. The approach used to include the EIP in the
evolution equation does not determine in a unique way the form of
the nonlinear potential. In fact, we show that in the nonlinear
potential $U_{_{\rm EIP}}$, which is a complex functional, its
real part is not determinable from kinetic considerations. We
impose the canonicity of the system in order to determine the real
part of the nonlinear potential. The final result is different
from those obtained by performing other quantization method.
Moreover, the canonicity required leave us the possibility to
include an arbitrary real nonlinear potential $U[\rho]$. This
arbitrarity can be used to introduce other interactions acting on
the system simultaneously with the EIP. Finally we study the mean
property of the system. In particular we analyze the Eherenfest
relations for observables like energy, linear and angular
momentum.

\setcounter{equation}{0}
\section{Canonical systems}
We start from the general expression of a nonlinear Schr\"odinger
Lagrangian in $(D+1)-$ dimension of the form:
\begin{eqnarray}
{\cal L}=i\,\frac{\hbar}{2}\,\left(\psi^\ast\,\frac{\partial\,\psi}{\partial\,t}
-\psi\,\frac{\partial\,\psi^\ast}{\partial\,t}\right)
-\frac{\hbar^2}{2\,m}\,|{\bfm\nabla} \psi|^2-U([\psi^\ast],\,[\psi])
-V\,\psi^\ast\,\psi \ ,\label{lagrangiana}
\end{eqnarray}
where
$\bfm\nabla=(\partial/\partial\,x_1,\,\partial/\partial\,x_2,\,
\dots\partial/\partial\,x_D)$ is the gradient operator in
$D$-dimension, $\hbar$ is a constant with the dimension of an
action that we identify with the Planck constant and $m$ is the
mass parameter.\\ The first two terms in the Lagrangian density
are the same encountered in the standard linear quantum
description, the quantity $V$ is a potential describing external
interaction. $U([\psi^\ast],\,[\psi])$ is the nonlinear term which
we assume to be an analytic smooth functional of the fields
$\psi$, $\psi^\ast$ and their spatial derivatives. We assume $U$
real to make the system not dissipative.\\ Using the Lagrangian we
introduce the action:
\begin{eqnarray}
{\cal A}=\int{\cal L}\,d^Dx\,dt \ .\label{action}
\end{eqnarray}
Applying the Euler operator E$_{_{\psi^\ast}}$, defined in Eq.
(\ref{euler}), to Eq. (\ref{action}), we obtain the following NLSE
for the field $\psi$:
\begin{eqnarray}
i\,\hbar\,\frac{\partial\,\psi}{\partial\,t}=-\frac{\hbar^2}{2\,m}
\Delta\,\psi+\frac{\delta}{\delta\,\psi^\ast}\,U([\psi^\ast],\,[\psi])+V\,\psi \ ,\label{schroedinger2}
\end{eqnarray}
where $(\Delta=\partial^2/\partial\,x_1^{\,2}+
\partial^2/\partial\,x_2^{\,2}+\cdots+\partial^2/\partial\,x_D^{\,2})$
is the $D$-dimensional Laplacian operator.\\ To obtain the
expression of the nonlinear potential $U([\psi^\ast],\,[\psi])$ in
order to include the EIP in the system it is convenient to
consider the Bohm-Madelung representation for the wave function
$\psi$:
\begin{equation}
\psi(t,1\,{\bfm x})=\rho(t,\,{\bfm x})^{1/2}\,\exp\,
\left[{i\over\hbar}S(t,\,{\bfm x})\right] \ .\label{ansatz1}
\end{equation}
The hydrodynamic fields, density of particles $\rho$ and phase $S$ \cite{Bohm}
are related with $\psi$ by means of:
\begin{eqnarray}
&&\rho=|\psi|^2 \ ,\label{ro}\\
&&S=i\,\frac{\hbar}{2}\,\log\left(\frac{\psi^\ast}{\psi}\right) \ .\label{s}
\end{eqnarray}
The nonlinear
potential $U([\rho],\,[S])$ becomes now a real functional of the fields $\rho$, $S$ and their
spatial derivatives.\\
By taking into account the Leibnitz rule for the functional derivative:
\begin{eqnarray}
\frac{\delta}{\delta\,\psi^\ast}=\frac{\delta\,\rho}{\delta\,\psi^\ast}\,
\frac{\delta}{\delta\,\rho}+\frac{\delta\,S}{\delta\,\psi^\ast}\,
\frac{\delta}{\delta\,S}=\psi\,\frac{\delta}{\delta\,\rho}+
i\,\frac{\hbar}{2\,\rho}\,\psi\,\frac{\delta}{\delta\,S} \ ,
\end{eqnarray}
Eq. (\ref{schroedinger2}) is rewritten in the form:
\begin{eqnarray}
i\,\hbar\,\frac{\partial\,\psi}{\partial\,t}=-\frac{\hbar^2}{2\,m}
\Delta\,\psi+\left[\frac{\delta}{\delta\,\rho}\,U([\rho],\,[S])\right]
\,\psi+i\,\frac{\hbar}{2\,\rho}\left[\frac{\delta}{\delta\,S}\,U([\rho],\,[S])\right]
\,\psi+V\,\psi \ .\label{schroedinger3}
\end{eqnarray}
Introducing the nonlinear quantities:
\begin{eqnarray}
&&W([\rho],\,[S])=\frac{\delta}{\delta\,\rho}\,U([\rho],\,[S]) \ ,\label{re}\\
&&{\cal
W}([\rho],\,[S])=\frac{\hbar}{2\,\rho}\,\frac{\delta}{\delta\,S}\,U([\rho],\,[S])
\ ,\label{im}
\end{eqnarray}
the NLSE (\ref{schroedinger3}) becomes:
\begin{eqnarray}
i\,\hbar\,\frac{\partial\,\psi}{\partial\,t}=-\frac{\hbar^2}{2\,m}
\Delta\,\psi+W([\rho],\,[S])\,\psi+i\,{\cal W}([\rho],\,[S])\,\psi+V\,\psi \ .\label{schroedinger4}
\end{eqnarray}
Using this expression it is easy to obtain the continuity equation
of the system. By
taking the product of (\ref{schroedinger4}) times $\psi^\ast$ and subtract
the complex conjugate form, we arrive to the expression:
\begin{eqnarray}
\frac{\partial\,\rho}{\partial\,t}+{\bfm\nabla}\cdot{\bfm
j}_0=\frac{2}{\hbar}\,\rho\,{\cal W}([\rho],\,[S]) \ ,\label{continuity2}
\end{eqnarray}
where
\begin{eqnarray}
{\bfm j}_0=-\frac{i\,\hbar}{2\,m}\,(\psi^\ast\,{\bfm\nabla}\psi^\ast-
\psi\,{\bfm\nabla}\psi^\ast) \ , \label{corr}
\end{eqnarray}
is the standard quantum current density.\\
Following the discussion made in section 1.2, we require that the system
described by (\ref{schroedinger4}) admits the following expression for
the quantum current [cfr. Eq. (\ref{EIPcorrente})]:
\begin{eqnarray}
{\bfm j}=(1+\kappa\,\rho)\,{\bfm j}_0 \ ,\label{corrente2}
\end{eqnarray}
and that the continuity equation (\ref{continuity2}) becomes:
\begin{eqnarray}
\frac{\partial\,\rho}{\partial\,t}+{\bfm\nabla}\cdot{\bfm j}=0 \ .\
\label{continuity3}
\end{eqnarray}
This can be accomplished if,
taking into account expression (\ref{corrente2}) and
(\ref{continuity2}), we select for the term $\cal W$ the form:
\begin{eqnarray}
{\cal W}([\rho],\,[S])
=-\kappa\,\frac{\hbar}{2\,\rho}\,{\bfm\nabla}\cdot\left(
{\bfm j}_0\,\rho\right)
=-\kappa\,\frac{\hbar}{2\,\rho}\,{\bfm\nabla}\cdot\left(
\frac{{\bfm j}\,\rho}{1+\kappa\,\rho}\right) \ ,\label{re1}
\end{eqnarray}
which has the same form of Eq. (\ref{immaginario1}) obtained in Ref.
\cite{Kaniadakis6} using the stochastic quantization method.
Eq. (\ref{re1}) can be expressed with the hydrodynamic fields as:
\begin{eqnarray}
{\cal
W}([\rho],\,[S])=-\kappa\,\frac{\hbar}{2\,\rho}\,{\bfm\nabla}\left(
\frac{{\bfm\nabla}S}{m}\,\rho^2\right) \ ,
\end{eqnarray}
and taking into account the dependence of $\cal W$ from the nonlinear
potential $U([\rho],\,[S])$ given by Eq. (\ref{im}) we obtain:
\begin{eqnarray}
U([\rho],\,[S])=\kappa\,\frac{({\bfm\nabla}S)^2}{2\,m}\,\rho^2+\widetilde{U}([\rho])
 \ ,\label{bbbb}
\end{eqnarray}
which is defined modulo a real functional of the field $\rho$ and
its spatial derivative.\\ We note that starting from the
continuity equation only the dependence of $U([\rho],\,[S])$ from
the phase can be determinate whilst the dependence from the field
$\rho$ is not fixed which means that an arbitrary real quantity
$\widetilde{U}([\rho],\,[S])$ functional of $\rho$, can be added
compatibly with the EIP, without removing the canonicity of the
system. Hereinafter we call "{\sl EIP potential}" the quantity
$U_{\rm EIP}(\rho,\,S)$ given by:
\begin{eqnarray}
U_{\rm EIP}(\rho,\,S)=\kappa\,\frac{({\bfm\nabla}S)^2}{2\,m}\,\rho^2 \ .\label{ueip}
\end{eqnarray}
This potential depends on $\rho$ and on ${\bfm\nabla}\,S$.
Inserting Eq. (\ref{bbbb}) in Eq. (\ref{re}) we obtain for the
real part of the nonlinearity the expression:
\begin{eqnarray}
W(\rho,\,S)=\kappa\,\frac{m}{\rho}\left(\frac{\bfm
j}{1+\kappa\,\rho}\right)^2+F([\rho]) \ ,\label{re2}
\end{eqnarray}
with
\begin{eqnarray}
F([\rho])=\frac{\delta}{\delta\,\rho}\,\widetilde{U}([\rho]) \ .\label{f}
\end{eqnarray}
We have obtained the following result:\\
{\it Starting from the
expression of the quantum current $\bfm j$ appearing in the continuity equation,
it is possible to deduce a NLSE compatible
with it. This NLSE generally contains a complex nonlinearity. Only its imaginary
part is fixed while the real one required an additional constraints. If
we require that
the system obeying the EIP is canonical, we obtain for the real part
the quantity $W(\rho,\,S)$ gives by Eq. (\ref{re2}) that is defined modulo a real functional
$F([\rho])$}.\\
In Ref. \cite{Kaniadakis6} the iter of the stochastic
quantization leads to a NLSE with the same expression of $\cal W$ but a
different form for $W$ [see Eq. (\ref{reale1})].

Returning again to the fundamental fields $\psi$ and $\psi^\ast$, the expression
of the potential (\ref{ueip}) takes the form:
\begin{equation}
U_{_{\rm EIP}}([\psi],\,[\psi^\ast])
=-\kappa\frac{\hbar^2}{8\,m}(\psi^\ast\,{\bfm\nabla}\psi-\psi\,{\bfm\nabla}\psi^\ast)^2
\ ,\label{eip}
\end{equation}
so that we can write the most general form of the Lagrangian of a system obeying the
EIP as:
\begin{eqnarray}
\nonumber {\cal
L}&=&i\,\frac{\hbar}{2}\,\left(\psi^\ast\,\frac{\partial\,\psi}{\partial\,t}
-\psi\,\frac{\partial\,\psi^\ast}{\partial\,t}\right)
-\frac{\hbar^2}{2\,m}\,|{\bfm\nabla}\psi|^2
+\kappa\,\frac{\hbar^2}{8\,m}(\psi^\ast\,{\bfm\nabla}\psi-\psi\,{\bfm\nabla}\psi^\ast)^2\\
&-&\widetilde{U}([|\psi|^2])-V\,\psi^\ast\,\psi \
,\label{lagrangiana1}
\end{eqnarray}
where $\widetilde{U}([|\psi|^2])$ is a real arbitrary smooth
functional which might depend on $\rho$ and its spatial
derivatives.\\ In this thesis, if not otherwise specified, the
arbitrary potential $\widetilde{U}(\rho)$ is an analytic
nonderivative functional of $\rho$ only. This potential can be
used to describe other interactions in the system. By an
appropriately choice of its form, the Lagrangian
(\ref{lagrangiana}) can be used to describe different physical
systems in presence of collective interactions and obeying to the
EIP (see chapter V).

Using Eq.(\ref{schroedinger2}) we obtain the following NLSE:
\begin{eqnarray}
\nonumber
i\,\hbar\,\frac{\partial\,\psi}{\partial\,t}
=&-&\frac{\hbar^2}{2\,m}\Delta\,\psi
-\kappa\,\frac{\hbar^2}{2\,m}(\psi^\ast\,{\bfm\nabla}\psi-\psi\,{\bfm\nabla}\psi^\ast)\,\nabla\psi\\
&-&\kappa\,\frac{\hbar^2}{4\,m}\,{\bfm\nabla}(\psi^\ast\,{\bfm\nabla}\psi-\psi\,{\bfm\nabla}\psi^\ast)\,\psi
+F(\rho)\,\psi+V\,\psi \ ,\label{schroedinger5}
\end{eqnarray}
where $F(\rho)$ is now given by $F(\rho)=\partial\,\widetilde
{U}(\rho)/\partial\,\rho$.\\ Eq. (\ref{schroedinger5}) can also be
written using the expression of the current (\ref{corrente2}) and
the particle density $\rho$ in the form:
\begin{eqnarray}
i\,\hbar\,\frac{\partial\,\psi}{\partial\,t}
=-\frac{\hbar^2}{2\,m}\Delta\,\psi+\Lambda(\rho,\,{\bfm j})\,\psi
+F(\rho)\,\psi+V\,\psi \ ,\label{schroedinger6}
\end{eqnarray}
where the complex nonlinearity $\Lambda(\rho,\,{\bfm j})$ is given by:
\begin{eqnarray}
\Lambda(\rho,\,{\bfm j})=\kappa\,\frac{m}{\rho}\left(\frac{\bfm j}{1+\kappa\,\rho}\right)^2
-i\,\kappa\,\frac{\hbar}{2\,\rho}\,{\bfm\nabla}
\left(\frac{{\bfm j}\,\rho}{1+\kappa\,\rho}\right) \ .
\end{eqnarray}

The quantum system described by the Lagrangian density
(\ref{lagrangiana}) is canonical. This can be verified defining
the fields $\pi_{_\psi}$ and $\pi_{_{\psi^\ast}}$, canonically
conjugated to the field $\psi$ and $\psi^\ast$, by means of the
relations (\ref{defmomenti}):
\begin{eqnarray}
&&\pi_{_\psi}=i\,\frac{\hbar}{2}\,\psi^\ast \ ,\label{moment1}\\
&&\pi_{_\psi^\ast}=-i\,\frac{\hbar}{2}\,\psi \ .\label{moment2}
\end{eqnarray}
It is well known that $\pi_{_\psi}$ and $\pi_{_{\psi^\ast}}$ are
proportional to the fields $\psi^\ast$ and $\psi$, so that, while
in the Lagrangian formalism $\psi$ and $\psi^\ast$ are independent
fields, in the Hamiltonian formalism they are canonically
conjugated. Following Ref.\cite{Dirac}, Eqs. (\ref{moment1}) and
(\ref{moment2}) give rise to the primary constrains:
\begin{eqnarray}
&&\xi_1=\pi_{_\psi}-i\,\frac{\hbar}{2}\,\psi^\ast \
,\label{constr1}\\
&&\xi_2=\pi_{_{\psi^\ast}}+i\,\frac{\hbar}{2}\,\psi \
.\label{constr2}
\end{eqnarray}
Performing the Legendre transformation (\ref{legendre}), it is easy to see
that the Hamiltonian density can be written as:
\begin{eqnarray}
{\cal H}=\frac{\hbar^2}{2\,m}\,|{\bfm\nabla} \psi|^2
-\kappa\,\frac{\hbar^2}{8\,m}(\psi^\ast\,{\bfm\nabla}\psi-
\psi\,{\bfm\nabla}\psi^\ast)^2+\widetilde{U}(\psi^\ast\,\psi)+V\,\psi^\ast\,\psi
\ .\label{hamiltoniana}
\end{eqnarray}
Let us introduce now the Poisson brackets between two functionals:
\begin{eqnarray}
\nonumber \{f({\bfm x}),\,g({\bfm y})\}&=&\int
\left[\frac{\delta\,f( {\bfm x})}{\delta\,\psi({\bfm {z})}}
\frac{\delta\,g({\bfm y})}{ \delta\,\pi_{_{\psi}}({\bfm {z})}}-
\frac{\delta\,f({\bfm y})}{\delta\,\pi_{_{\psi}}({\bfm z)}}
\frac{\delta\,g({\bfm x})}{\delta\,\psi({\bfm
{z})}}\right]\,d^Dz\\
 &+&\int \left[\frac{\delta\,f( {\bfm
x})}{\delta\,\psi^\ast({\bfm {z})}} \frac{\delta\,g({\bfm y})}{
\delta\,\pi_{_{\psi^\ast}}({\bfm {z})}}- \frac{\delta\,f({\bfm
y})}{\delta\,\pi_{_{\psi^\ast}}({\bfm z)}} \frac{\delta\,g({\bfm
x})}{\delta\,\psi^\ast({\bfm {z})}}\right]\,d^Dz \ . \label{poiss}
\end{eqnarray}
The second class primary derivative (\ref{constr1}) and
(\ref{constr2}) satisfy the relation:
\begin{eqnarray}
\{\xi_1(t,\,{\bfm x}),\,\xi_2(t,\,{\bfm
y})\}=-i\,\hbar\,\delta^{(D)}({\bfm x}- {\bfm y}) \ ,
\end{eqnarray}
and can be accommodated by the introduction of the Dirac brackets:
\begin{eqnarray}
\nonumber \{f({\bfm x}),\,g({\bfm y})\}_D&=&\{f({\bfm
x}),\,g({\bfm y})\}+ \frac{i}{\hbar}\,\int\{f({\bfm
x}),\,\xi_1({\bfm z})\}\,\{ \xi_2({\bfm z}),\,g({\bfm
y})\}\,d^Dz\\ &-&\frac{i}{\hbar}\,\int\{f({\bfm x}),\,\xi_2({\bfm
z})\}\,\{ \xi_1({\bfm z}),\,g({\bfm y})\}\,d^Dz \ .
\end{eqnarray}
Expression (\ref{poiss}) can be simplified if one solves the
derivative (\ref{constr1}), (\ref{constr2}) for $\pi_{_{\psi}}$
and $\pi_{_{\psi^\ast}}$ and treats $f$ and $g$ as functionals of
$\psi$ and $\psi^\ast$ only. A straightforward calculation gives:
\begin{eqnarray}
\{f({\bfm x}),\,g({\bfm y})\}_D=&-&\frac{i}{\hbar}\,\int
\left[\frac{\delta\,f( {\bfm x})}{\delta\,\psi({\bfm {z})}}
\frac{\delta\,g({\bfm y})}{ \delta\,\psi^\ast({\bfm {z})}}-
\frac{\delta\,g({\bfm y})}{\delta\,\psi({\bfm z)}}
\frac{\delta\,f({\bfm x})}{\delta\,\psi^\ast({\bfm
{z})}}\right]\,d^Dz \ .\label{poisson2}
\end{eqnarray}
Using this expression we can obtain the evolution equations for the
fields $\psi$ and $\psi^\ast$ in the Poisson formalism:
\begin{eqnarray}
\frac{\partial\,\psi}{\partial\,t}&=&\{\psi,\,H\} \ ,\label{pua1}\\
\frac{\partial\,\psi^\ast}{\partial\,t}&=&\{\psi^\ast,\,H\} \ ,\label{pua2}
\end{eqnarray}
which give the Hamiltonian equations:
\begin{eqnarray}
i\,\hbar\,\frac{\partial\,\psi}{\partial\,t}
&=&\frac{\delta\,H}{\delta\,\psi^\ast} \ ,\label{ham1}\\
-i\,\hbar\,\frac{\partial\,\psi^\ast}{\partial\,t}
&=&\frac{\delta\,H}{\delta\,\psi} \ ,\label{ham2}
\end{eqnarray}
that are the Schr\"odinger equations for the fields $\psi$ and $\psi^\ast$,
respectively. It can be easily verified that the chose equations are
equal to Eq. (\ref{hamilton}) introduced in
section 1.3 with the Hamiltonian operator given by (\ref{hoperatore}).

\setcounter{equation}{0}
\section{Hydrodynamic formulation}
In this section we describe the NLSE with EIP in the hydrodynamic
representation. Its utility will be seen in chapter V, where we
study explicit solutions of the model. In particular we make use
of the hydrodynamic formulation in order to obtain the solitary
wave solutions of the system.

It is well known \cite{Bohm} that a quantum system can be seen as
a Madelung-like fluid \cite{Madelung} described by means of the
fields $\rho$ and $S$ trough a coupled system of differential
equations. The velocity field associated to the fluid is related
to the phase of $\psi$ from the relation ${\bfm
v}={\bfm\nabla}\,S/m$ and its dynamics is described by means of a
{\sl Hamilton-Jacobi} like equation in presence of a nonlinear
term function of $\rho$ which was called quantum potential. Let us
remember now that the dynamical equation for the phase is only
formally similar to the Hamilton-Jacobi equation. In fact, in the
classical mechanics, the Hamilton-Jacobi equation describes the
evolution of the Hamilton principal function $S$ that is the
generator of a canonical transformation and is identified {\sl a
posteriori} with the action of the system: $S=\int L\,dt$. On the
other hand, differently from the Bohm picture, the phase $S$ of
the field $\psi$ is a dynamical field canonically conjugate to the
density $\rho$ and is not a generator of canonical
transformations.

By introducing Eq. (\ref{ansatz1}) in Eq. (\ref{schroedinger5})
and separating the real part from the imaginary one, we obtain a
coupled system in $\rho$ and $S$:
\begin{eqnarray}
\frac{\partial\,S}{\partial\,t} +\frac{({\bfm\nabla}
S)^2}{2\,m}+\kappa\,\rho\,\frac{({\bfm\nabla} S)^2}{m}
-\frac{\hbar^2}{2\,m}
\frac{\Delta\sqrt{\rho}}{\sqrt{\rho}}+F(\rho)+V=0 \ ,\label{hj}
\end{eqnarray}
\begin{equation}
\frac{\partial\,\rho}{\partial\,t}+
{\bfm\nabla}\left[\frac{{\bfm\nabla} S}{m}\,\rho\,(1+\kappa\,\rho)\right]=0 \ .\label{cont}
\end{equation}
Equation (\ref{hj}) is a {\it Hamilton-Jacobi} type equation, where
the fourth term is the quantum
potential \cite{Bohm}. The third term
is the real part $W(\rho,\,{\bfm j})$ of the term introduced
by the EIP potential (\ref{ueip}),
as can be verified taking in mind the expression of
the current
$\bfm j$ given by Eq. (\ref{corrente2}), that in the new field becomes:
\begin{equation}
{\bfm j}=\frac{{\bfm\nabla} S}{m}\,\rho\,(1+\kappa\,\rho) \ .\label{corr1}
\end{equation}
Finally the last two terms in Eq. (\ref{hj}) are the extra nonlinearity
and the external potential.
We can see that
the imaginary part ${\cal W}(\rho,\,{\bfm j})$ of the term introduced
by the EIP potential does not appear in the {\it Hamilton-Jacobi} equation (\ref{hj}).\\
It is easy to recognize that Eq. (\ref{cont}) is the continuity equation in
the $\rho\,$-$S$ representation.

The {\it Hamilton-Jacobi} equation (\ref{hj}) and the continuity
equation (\ref{cont}) can be derived from a variational principle,
making use of the Euler operator (\ref{euler}) written in the
$\rho$ and $S$ representation, obtained from Eq. (\ref{euler})
with the substitution $\psi\rightarrow\rho$ and $\psi\rightarrow
S$ respectively. In fact, starting from the Lagrangian density
$\widetilde{\cal L}$ which is given by:
\begin{eqnarray}
\widetilde{\cal L}=-\frac{\partial\,S}{\partial\,t}\,\rho
-\frac{({\bfm\nabla}S)^2}{2\,m}\rho-\kappa\,\frac{({\bfm\nabla}S)^2}
{2\,m}\rho^2-\frac{\hbar^2}{8\,m}\frac{({\bfm\nabla}
\rho)^2}{\rho}-\widetilde{U}(\rho)-V\,\rho \ ,\label{lagrangiana2}
\end{eqnarray}
and applied to the action $\widetilde{\cal A}[\rho,\,S]$ defined
by (\ref{deflagran}) the Euler operator:
\begin{equation}
{\rm E}_\rho(\widetilde{\cal A})=0 \ ,\hspace{10mm} {\rm
E}_S(\widetilde{\cal A})=0 \ ,
\end{equation}
we obtain the Eqs. (\ref{hj}) and (\ref{cont}). In the Lagrangian
density (\ref{lagrangiana2}) the first three terms are the same as
occurring in the linear Schr\"odinger equation, while the
nonlinear contribution is given by the forth and six the term,
where the fourth is the potential introduced by the EIP. The term
$-\hbar^2\,({\bfm\nabla}\,\rho)^2/(8\,m\,\rho)$ in Eq.
(\ref{lagrangiana2}) is responsible of the presence of the quantum
potential in Eq. (\ref{hj}),
$U_q=-\hbar^2\,\Delta\rho^{1/2}/(2\,m\,\rho^{1/2})$.
Notwithstanding, in literature the quantum potential is referred
to the nonlinear term that appears in the {\sl Hamilton-Jacobi}
equation.

Now we introduce the Hamiltonian procedure. Let us show the
results without describing the Dirac procedure.\\ The momentum
$\pi_{_S}$, canonically conjugate to $S$, is given by Eq.
(\ref{defmomenti}), that now becomes:
\begin{eqnarray}
&&\pi_{_S}=-\rho \ .
\end{eqnarray}
Moreover, we have $\pi_{_\rho}=i\,\hbar/2$. Therefore, $\pi_{_S}$
is proportional to $\rho$, while $\pi_{\rho}$ is a constant; the
number of degrees of freedom is the same in both the Lagrangian
and Hamiltonian formalism.\\ The Hamiltonian density, function of
the canonically conjugate fields $S$ and $-\rho$, can be deduced
taking into account Eq. (\ref{legendre}) and (\ref{lagrangiana2}):
\begin{eqnarray}
\tilde{\cal H}&=&\frac{({\bfm\nabla}
S)^2}{2\,m}\,\rho+\kappa\,\frac{({\bfm\nabla} S)^2}{2\,m}\,\rho^2
+\frac{\hbar^2}{8\,m}\frac{({\bfm\nabla}\rho)^2}{\rho}+\widetilde
U(\rho)+V\,\rho \ .
\end{eqnarray}
The {\it Hamilton-Jacobi} and the continuity equations take the form:
\begin{eqnarray}
&&\frac{\partial\,S}{\partial\,t}
=-\frac{\delta\,\tilde H}{\delta\,\rho} \ ,\label{ham3}\\
&&\frac{\partial\,\rho}{\partial\,t}
=\frac{\delta\,\tilde H}{\delta\,S} \ .\label{ham4}
\end{eqnarray}
The same equations, in the Poisson formalism, can be rewritten as:
\begin{eqnarray}
&&\frac{\partial\,S}{\partial\,t}
=\{S,\,\tilde H\} \ ,\label{pua3}\\
&&\frac{\partial\,\rho}{\partial\,t}
=\{\rho,\,\tilde H\} \ .\label{pua4}
\end{eqnarray}
The evolution equations (\ref{ham1}), (\ref{ham2}) or
(\ref{pua1}), (\ref{pua2}) deduced from $H(\psi,\,\psi^\ast)$ have
the same form of the equations (\ref{ham3}), (\ref{ham4}) or
(\ref{pua3}), (\ref{pua4}) deduced from $\tilde H(\rho,\,S)$.
Then, according to a well-established procedure, we can relate the
fields $\psi$-$\psi^\ast$ to the fields $S$-$\rho$ by means of a
canonical transformation \cite{Goldstein}. The equations of motion
in the $S$-$\rho$ representation will be used in chapter V to
study particular soliton solutions of Eq. (\ref{schroedinger6})
that preserve their shapes in the time. As we will show in chapter
V, we are able to decouple the system of equations (\ref{ham3}),
(\ref{ham4}) or equivalently Eqs. (\ref{pua3}), (\ref{pua4})
obtaining a differential equation in the variable $\rho$ only,
whose solutions define the solitons of the systems with EIP.

In conclusion of this section we study the most simplest solutions
of the NLSE (\ref{schroedinger5}) in presence of the nonlinearity
introduced by EIP only, i.e. when we neglect the potentials
involving $\widetilde U$ and $V$. These are the planar waves:
\begin{eqnarray}
\psi(t,\,{\bf x})=A\,\exp\left[i({\bfm k}\cdot{\bfm
x}-\omega\,t)\right] \ ,\label{wave}
\end{eqnarray}
where ${\bfm k}$ is the wave number and $A$ a complex constant. After inserting
Eq. (\ref{wave}) in (\ref{schroedinger5}) we obtain the following
dispersion relation for the planar waves:
\begin{eqnarray}
\omega=\frac{\hbar\,{\bfm k}^2}{2\,m}\,(1+2\,\kappa\,|A|^2) \ ,
\end{eqnarray}
which reproduces the well known dispersion relation of the Schr\"odinger equation when the
EIP is switched off $(\kappa\rightarrow0)$.
Let us remark that in presence of EIP, the angular frequency depends also on the
amplitude of the wave. A this point it is important to remember that,
differently from the linear theory, in the NLSE the normalization
constant is not
arbitrary. Generally, the amplitude of a solution is related to the other
parameter of the system. For instance, the cubic NLSE \cite{Scott}
admits soliton solutions whose velocity is related to the amplitude.
This amplitude increases when the soliton velocity decrease.\\
We remember also that because of the nonlinear nature of the systems
the more general solution can not be  expressible as superposition of planar
waves how it is made in the linear case, by using the Fourier transformation method. More sophisticated
mathematical tools, like, for instance, the inverse scattering method
\cite{Kruskal} are needed to found the general solutions.
In this thesis we do not develop this procedure.\\
From Eq. (\ref{hamiltoniana}), it appears that the system may be
unstable in the case $\kappa<-1/|A|^2$. In fact, the energy of the modes with wave
number ${\bfm k}$ is:
\begin{eqnarray}
E=\frac{\hbar^2{\bfm k}^2}{2\,m}\,|A|^2\,(1+\kappa\,|A|^2) \ ,
\end{eqnarray}
which must be positive. This imposes the condition for the repulsive
systems
\begin{eqnarray}
|A|^2\leq\rho_{\rm max}\equiv 1/|\kappa|
\end{eqnarray}
according to the exclusion principle in the configuration space \cite{Kaniadakis6}

\setcounter{equation}{0}
\section{Physical observable}
Now we study the time evolution of the average of the most
important physical observable that describe the system in presence
of EIP. We will identify the motion constants of the system.  The
results of this section will be obtained again in the next chapter
from the analysis of the symmetries satisfied by the system. The
section is organized in to parts: in the first we describe the
general results. The mathematical proofs are collected in the
second part.

\subsection{Ehrenfest relations}
Let us assume that the nonlinear potential $\widetilde U(\rho)$
and the field $\psi$ vanish at infinity so that the surface terms
can be disregarded. Moreover we assume that the potential
$\widetilde U(\rho)$ depends on the space and on the time only
through the field $\rho$. In this section we assume $D=3$.\\  To
obtain the Ehrenfest relations of the system obeying to Eq.
(\ref{schroedinger6}) we start, at first, with the definition of
average of an Hermitian operator $\widehat{A}=\widehat{A}^\dag$:
\begin{eqnarray}
<A>=\int\psi^\ast\,\widehat{A}\,\psi\,d^3x
 \ .
\end{eqnarray}
Hereinafter we normalize the system as:
\begin{eqnarray}
N=\int\psi^\ast\,\psi\,d^3x \ ,\label{e1}
\end{eqnarray}
where $N$ is an integer, so that the field $\rho$ assumes the meaning of
a density
of probability of position of a $N$-body system. Of course the
normalization is maintained in the time thanks to the
continuity equation (\ref{continuity3}):
\begin{eqnarray}
\frac{d\,N}{d\,t}=0 \ .
\end{eqnarray}
This means that the number of particles of the system do not
change during the evolution. Using Eq. (\ref{poisson1}) it is easy
to obtain the following relationship for the time evolution of the
average of $\widehat{A}$:
\begin{eqnarray}
\frac{d}{d\,t}\,<\widehat{A}>=\frac{i}{\hbar}\int\left[\frac{\delta
H}{\delta\psi}\,\widehat{A}\,\psi-\psi^\ast\,\widehat{A}\,\frac{\delta
H}{\delta\psi^\ast}\right]\,d^3x
+<\frac{\partial\,\widehat{A}}{\partial\,t}>
 \ .\label{ehr2}
\end{eqnarray}
Here the last term takes eventually into account explicit time dependence
of the functional $A$.
Let us call $\widehat{\cal{O}}$ the operator in the r.h.s. of the NLSE
(\ref{schroedinger6}) that can be rewritten in the form:
\begin{equation}
i\,\hbar\frac{\partial\,\psi}{\partial\,t}=\widehat{\cal{O}}\,\psi \ ,
\end{equation}
with
\begin{equation}
\widehat{\cal{O}}=\widehat{H}_0+W(\rho,\,{\bfm j})+i\,{\cal
W}(\rho,\,{\bfm j})+F(\rho) \ ,\label{o}
\end{equation}
where $\widehat{H}_0=(-\hbar^2/2\,m)\,\Delta+V({\bfm x})$ is the
Hamiltonian operator of the linear theory. We can write the
relation (\ref{ehr2}) in the following form:
\begin{eqnarray}
\frac{d}{d\,t}\,<\widehat{A}>=
\frac{i}{\hbar}\,<[{\rm Re}\,\widehat{\cal{O}},\,\widehat{A}]>
+{1\over\hbar}<\{{\rm Im}\,\widehat{\cal{O}},\,\widehat{A}\}>
+<\frac{\partial\,\widehat{A}}{\partial\,t}> \ ,\label{ehrenfest}
\end{eqnarray}
where the symbols $[\cdot,\cdot]$ and $\{\cdot,\cdot\}$ stand for the
commutator and the anticommutator, respectively
(see also Ref. \cite{Doebner1} where an example of the Ehrenfest relations in
a nonlinear Schr\"odinger equation with complex potential is discussed).

After setting $\widehat{A}=\widehat{\bfm x}_c$ in Eq.
(\ref{ehrenfest}), where $\widehat{\bfm x}_c= \widehat{{\bfm
x}}/N$, we obtain the Ehrenfest relationship for the time
evolution of the center of mass frame:
\begin{eqnarray}
\frac{d}{d\,t}\,<\widehat{\bfm x}_c>=-i\,\frac{\hbar}{2\,m\,N}
\int(1+\kappa\,\rho)\,
\left(\psi^\ast\,{\bfm\nabla}\psi-\psi\,{\bfm\nabla}\psi^\ast\right)
\,d^3x \ .\label{e2}
\end{eqnarray}
It is worth remarking that the EIP introduces the additional
quantity $<\kappa\,\rho\,\widehat{\bfm P}>$ which is equal to
$\kappa\,m\,\rho\,{\bfm j}/(1+\kappa\,\rho)$. Note also that the
right hand side of Eq. (\ref{e2}) can be written as:
\begin{eqnarray}
\frac{d}{d\,t}\,<\widehat{\bfm x}_c>={1\over N}\int{\bfm j}\,d^3x
\ ,\label{xxx}
\end{eqnarray}
which appear formally the same as in the standard linear quantum mechanics.

A second relation is obtained setting $\widehat{A}=\widehat{\bfm P}$:
\begin{eqnarray}
\frac{d}{d\,t}\,<\widehat{\bfm
P}>=-\int\psi^\ast\,{\bfm\nabla}\,V\,\psi\,d^3x-\int\psi^\ast\,{\bfm\nabla}\,F\,\psi\,d^3x
\ .\label{e3}
\end{eqnarray}
Equation (\ref{e3}) can be regarded as the second law of the
dynamics \cite{Hasse1,Moura}. The dynamics of the mean value of
the momentum is governed by an {\it effective potential} given by
the sum of the external potential $V$ and of the nonlinearity
$F(\rho)$. The EIP potential does not affect, on the average, the
dynamics of the system because, due to their particular form, the
terms $W$ and $\cal W$ satisfy the relation
$<[W,\nabla]-i\,\{{\cal W},\nabla\}>=0$. For the most frequent
nonlinearity $F(\rho)$ generally appearing in the nonlinear
Schr\"odinger equations the last term can be dropped and the
Newtonian behavior is restored \cite{Hasse1} (see the next section
for the proof of this statement). On the contrary, other dynamical
equations, like the sine-Gordon equation, seem to show a different
behavior with respect to the Newtonian one. In this case the
reason is in the kink-like solution of this equation that do not
vanish on the boundary at infinity so that the surface terms can
not be neglected.

Than, setting $\widehat{A}=\widehat{\bfm L}$, where $\widehat{\bfm
L}$ is the angular momentum operator whose components are
$\widehat{L}_i=\varepsilon_{ijk}\, x_j\, \widehat {P}_k$, we
obtain:
\begin{eqnarray}
\frac{d}{d\,t}\,<\widehat{\bfm L}>=-\int\psi^\ast\,({\bfm
x}\wedge{\bfm\nabla}\,V)\,\psi\,d^3x
-\int\psi^\ast({\bfm x}\wedge{\bfm\nabla}\,F)\,\psi\,d^3x \ .\label{e4}
\end{eqnarray}
Like in the previous relation, the EIP potential does not
contribute to the average of the angular momentum. Again if the
nonlinear potential has a well behavior at infinity the last term is unremarkable
on the dynamics of the system.

We discuss now the Ehrenfest relation concerning the energy. The
energy of a canonical system, as we are considering here, is given
by $E=H$ where $H$ is the Hamiltonian given by Eq.
(\ref{hamiltoniana}). We can define a Hamiltonian operator
$\widehat{H}$ whose average value is $<\widehat{H}>=H$. It is
easily verified that:
\begin{equation}
\widehat{H}=-\frac{\hbar^2}{2\,m}\,\Delta+\frac{1}{\rho}\,U_{_{\rm
EIP}}(\rho,\,{\bfm j})+\frac{1}{\rho}\,\widetilde{U}(\rho)+V({\bfm
x},\,t) \ .\label{hh}
\end{equation}
If we compare this expression of $\widehat{H}$ with the expression
of the operator $\widehat{\cal{O}}$ given by Eq. (\ref{o}) we
find:
\begin{equation}
\widehat{H}\not=\widehat{\cal{O}} \ ,
\end{equation}
which means that the Hamiltonian operator of a nonlinear canonical
system does not coincide with the operator $\widehat{\cal{O}}$ of
the r.h.s. of the NLSE whilst, in the case of the linear theories,
we have $\widehat{\cal{O}}=\widehat{H}=\widehat{H}_0$.\\ Within
the definition $E=<\widehat{H}>$, we obtain the following
relationship:
\begin{eqnarray}
\frac{d\,E}{d\,t}=<\frac{\partial\,V}{\partial\,t}> \ ,\label{e5}
\end{eqnarray}
which means that when time dependent external potentials are
absent, the system is conservative being $d\,E/d\,t=0$. We may
conclude that the EIP does not introduce dissipative effects.\\ We
remark that for a noncanonical model, where no Hamiltonian is
present, the energy of the system is assumed generally as
$E=<\widehat{\cal O}>$ where $\widehat{\cal O}$ is the operator of
the r.h.s. of the corresponding NLSE. Several models with
nonlinearities in the r.h.s. of the Schr\"odinger equation,
characterized by time independent average values, have been
developed. For instance, in the Kostin NLSE \cite{Kostin,Schuch},
the operator $\widehat{\cal{O}}$ is defined as
$\widehat{\cal{O}}=\widehat{H}_0+(\hbar\,\gamma/2\,i)\,[\log(\psi/\psi^\ast)-<\log(\psi/\psi^\ast)>]$
being a real quantity, the energy of the system is defined as
$E=<\widehat{\cal{O}}>$. In this case the non conservation of
$<\widehat{\cal{O}}>$ implies energy dissipation of the system.

In conclusion, we have shown that the EIP potential (\ref{eip}) describes a conservative
system. For a free system ($V=0$) and when the non-linear
potential $U(\rho)$ has a good behavior at infinity, we are able to identify four
constants of motion:
\begin{eqnarray}
&&N=\int\rho\,d^3x \ ,\\ &&<\widehat{{\bfm
P}}>=-i\,\frac{\hbar}{2}\,\int\left(\psi^\ast\,{\bfm\nabla}\psi-
\psi\,{\bfm\nabla}\psi^\ast\right)\,d^3x \ ,\label{en1}\\
&&<\widehat{{\bfm L}}>=-i\,\frac{\hbar}{2}\,\int{\bfm
x}\times\left(\psi^\ast\,{\bfm\nabla}\psi-\psi\,{\bfm\nabla}\psi^\ast\right)\,d^3x
\ ,\\ &&E=\int{\cal H}\,d^3x \ ,\label{p1}
\end{eqnarray}
representing respectively the energy, the momentum, the angular momentum
and the number of particles, conserved in virtue of the
continuity equation (\ref{continuity3}).

\subsection{Mathematical proofs}

We deduce here the Ehrenfest relations for the observable
$N,\,<\widehat{\bfm x}_c>,\,<\widehat{\bfm P}>,\,<\widehat{\bfm
L}>,\,E$ given from Eqs. (\ref{e1}), (\ref{e2}), (\ref{e3}),
(\ref{e4}) and (\ref{e5}). In the following we make the hypothesis
that the fields vanish steeply at infinity and that we can neglect
surface terms.\\ The first relation, Eq. (\ref{e1}), is a trivial
consequence of the continuity equation, while the (\ref{e2}), can
be easily obtained from Eq. (\ref{continuity3}) in the following
way:
\begin{eqnarray}
\frac{d}{d\,t}<{\bfm x}_c>={1\over N}\int{\bfm
x}\,\frac{d\,\rho}{d\,t}\,d^3x=-{1\over N}\int{\bfm
x}\,{\bfm\nabla}\cdot{\bfm j}\,d^2x={1\over N}\int{\bfm j}\,d^3x \
,
\end{eqnarray}
where we have performed an integration by parts in the last
step.\\ Let us consider now Eq. (\ref{e3}) for the component $i$
of the momentum. Using Eq. (\ref{ehr2}) for
$<\widehat{P}_i>=i(\hbar/2)\int(\psi^\ast\,\partial_i\psi-\psi\,\partial_i\psi^\ast)\,d^3x$
and taking into account the following functional derivative:
\begin{eqnarray}
\frac{\delta\,<\widehat{P}_i>}{\delta\,\psi^\ast}&=&i\,\hbar\,\partial_i\psi
\ ,\\ \nonumber \frac{\delta
H}{\delta\,\psi^\ast}&=&\frac{\hbar^2}{2\,m}\,\Delta
\psi+\kappa\,\frac{\hbar^2}{2\,m}\left[\psi^\ast\,{\bfm\nabla}\psi-\psi\,{\bfm\nabla}
\psi^\ast\right]\,{\bfm\nabla}\psi\\
&+&\kappa\,\frac{\hbar^2}{4\,m}\,{\bfm\nabla}\left[\psi^\ast\,{\bfm\nabla}\psi-\psi\,
{\bfm\nabla}\psi^\ast\right]\,\psi+F(\rho)\,\psi+V\,\psi \
,\label{a5}
\end{eqnarray}
and their conjugate, we obtain:
\begin{eqnarray}
\nonumber
\frac{d}{d\,t}<P_i>&=&\int\Biggl\{-\frac{\hbar^2}{2\,m}
\left[\partial_i\psi^\ast\,\Delta\psi+\partial_i\psi\,\Delta\psi^\ast\right]\\
\nonumber
&+&\kappa\,\frac{\hbar^2}{2\,m}\left[\psi^\ast\,{\bfm\nabla}\psi-\psi\,{\bfm\nabla}\psi^\ast\right]
\left[{\bfm\nabla}\psi^\ast\,\partial_i\psi-{\bfm\nabla}\psi\,\partial_i\psi^\ast\right]\\
\nonumber
&+&\kappa\,\frac{\hbar^2}{4\,m}\,{\bfm\nabla}\left[\psi^\ast\,{\bfm\nabla}\psi-\psi\,{\bfm\nabla}\psi^\ast\right]
\left[\psi^\ast\,\partial_i\psi-\psi\,\partial_i\psi^\ast\right]\\
&-&\left[F(\rho)+V\right]\,\partial_i\,\rho\Biggr\}\,d^3x \ .\label{a7}
\end{eqnarray}
Integrating by parts twice the first term and one time the third in
the right hand side of (\ref{a7}) we obtain:
\begin{eqnarray}
\nonumber
\frac{d}{d\,t}<P_i>&=&\int\Biggl\{-\frac{\hbar^2}{2\,m}
\partial_i(\Delta\psi^\ast\,\psi)
-\kappa\,\frac{\hbar^2}{8\,m}\partial_i\left[\psi^\ast\,{\bfm\nabla}\psi
-\psi\,{\bfm\nabla}\psi^\ast\right]^2\\
&-&\left[F(\rho)+V\right]\,\partial_i\,\rho\Biggr\}\,d^3x \ ,
\end{eqnarray}
and neglecting surface terms we are left with:
\begin{eqnarray}
&&\frac{d}{d\,t}<P_i>=\int\rho\,\partial_i\,V\,d^2x+\int\rho\,\partial_i\,F(\rho)\,d^2x \ .\label{a13}
\end{eqnarray}
Taking into account the relation $F=d\widetilde U/d\rho$, the last
integral in (\ref{a13}) can be written as:
\begin{equation}
\int\partial_i\,(\rho\,F-\widetilde U)\,d^2x \ ,
\end{equation}
and therefore, if the potential and the field $\rho$ has a well
behavior at infinity, it can be ignored. Then Eq. (\ref{a13}) is
equal to Eq. (\ref{e3}).\\ Note that in Eq. (\ref{e3}) it does not
appears the contribution of the nonlinear potential $\widetilde
U(\rho)$. This statement is true also for nonlinear potential
dependent on $\rho$ and its spatial derivative $\widetilde
U([\rho])$. In fact, in this case we have $F=\delta\,\widetilde
U/\delta\,\rho$ and the last term in (\ref{a13}) is, as before, a
surface integral.\\ For Eq. (\ref{e4}) we can follow the same
procedure used for the relationship (\ref{e3}). We only give the
trace of the procedure.\\ Setting
$<\widehat{L}_i>=i(\hbar/2)\int\epsilon_{ijk}
\,x_j\,(\psi^\ast\,\partial_k\psi-\psi\,\partial_k\psi^\ast)\,d^3x$,
by using Eq. (\ref{ehr2}) we have:
\begin{eqnarray}
\nonumber
\frac{d}{d\,t}<\widehat{L}_i>&=&\int\epsilon_{ijk}\,x_j\Biggl\{-\frac{\hbar^2}{2\,m}
\left[\partial_k\psi^\ast\,\Delta\psi+\partial_k\psi\,\Delta\psi^\ast\right]\\
\nonumber
&+&\kappa\,\frac{\hbar^2}{2\,m}\left[\psi^\ast\,{\bfm\nabla}\psi-\psi\,{\bfm\nabla}\psi^\ast\right]
\left[{\bfm\nabla}\psi^\ast\,\partial_k\psi-{\bfm\nabla}\psi\,\partial_k\psi^\ast\right]\\
\nonumber
&+&\kappa\,\frac{\hbar^2}{4\,m}\,{\bfm\nabla}\left[\psi^\ast\,{\bfm\nabla}\psi-\psi\,{\bfm\nabla}\psi^\ast\right]
\left[\psi^\ast\,\partial_k\psi-\psi\,\partial_k\psi^\ast\right]\\
&-&\left[F(\rho)+V\right]\,\partial_k\,\rho\Biggr\}\,d^3x \
.\label{a14}
\end{eqnarray}
Now we have perform the partial integration, as it was done in Eq.
(\ref{a7}), taking into account that the quantity
$\epsilon_{ijk}\partial_k\,x_j$ vanishes because the skew-symmetry
of $\epsilon_{ijk}$ and disregarding the surface terms we obtain:
\begin{eqnarray}
&&\frac{d}{d\,t}<\widehat{L}_i>=\int\epsilon_{ijk}\,x_j\,\partial_k
\left[V+F(\rho)\right]\,\rho\,d^2x \ .\label{a15}
\end{eqnarray}
Computing as before the nonlinear quantity $F(\rho)$ we obtain Eq. (\ref{e4}).

Finally, we derive the relation (\ref{e5}). We begin writing the
EIP potential as a function of the density $\rho=\psi^\ast\,\psi$
and of the gradient of the phase
$S=(i\,\hbar/2)\,\log(\psi^\ast/\psi)$. Equation (\ref{ehr2})
becomes:
\begin{eqnarray}
\nonumber
\frac{d\,E}{d\,t}&=&\int\left[-\frac{\hbar^2}{2\,m}\,\frac{\partial\,\psi^\ast}{\partial\,t}\,\Delta\,
\psi-\frac{\hbar^2}{2\,m}\,\psi^\ast\,\Delta\,\frac{\partial\,\psi}{\partial\,t}\right.\\
&+&\left.\frac{\partial}{\partial\,t}\left(\widetilde{U}+U_{_{\rm
EIP}}\right)+
V\,\frac{\partial\,\rho}{\partial\,t}+\rho\,\frac{\partial\,V}{\partial\,t}\right]\,d^3x
\ .\label{aa3}
\end{eqnarray}
By using the equations of motion of the fields $\psi$ and
$\psi^\ast$:
\begin{eqnarray}
i\,\hbar\,\frac{\partial\,\psi}{\partial\,t}&=&\left[-\frac{\hbar^2}{2\,m}
\,\Delta+V+F+W+i\,{\cal W}\right]\,\psi \ ,\\
-i\,\hbar\,\frac{\partial\,\psi^\ast}{\partial\,t}&=&\left[-\frac{\hbar^2}{2\,m}
\,\Delta+V+F+W-i\,{\cal W}\right]\,\psi^\ast \ ,
\end{eqnarray}
with ${\cal W}$, $W$ and $F$ given by (\ref{re1}), (\ref{re2}) and
(\ref{f}) respectively, the Eq. (\ref{aa3}) becomes:
\begin{eqnarray}
\nonumber \frac{d\,E}{d\,t}&=&\int\left[
i\,\frac{\hbar^3}{4\,m^2}\,\Delta\,\psi^\ast\,\Delta\,\psi
-i\,\frac{\hbar}{2\,m}\,\psi^\ast\,\left(V+F+W-i\,{\cal
W}\right)\,\Delta\,\psi\right]\,d^3x\\ \nonumber
&-&\int\left[i\,\frac{\hbar^3}{4\,m^2}\,\psi^\ast\,\Delta^2\,\psi
+i\,\frac{\hbar}{2\,m}\,\psi^\ast\,\Delta\,\left[\left(V+F+W+i\,{\cal
W}\right)\,\psi\right]\right]\,d^3x\\ \nonumber
&+&\int\left[\left(V+\frac{\partial\,\widetilde
U}{\partial\,\rho}+\frac{\partial\,U_{_{\rm
EIP}}}{\partial\,\rho}\right)
\,\frac{\partial\,\rho}{\partial\,t}+\frac{\partial\,U_{_{\rm
EIP}}}{\partial({\bfm\nabla}\,S)}\,
\frac{\partial\,({\bfm\nabla}\,S)}{\partial\,t}\right]\,d^3x\\
&+&<\frac{\partial\,V}{\partial\,t}>
 \ ,\label{aa6}
\end{eqnarray}
where we have used the relationship:
\begin{eqnarray}
\frac{\partial}{\partial\,t}\,\left(\widetilde U+U_{_{\rm
EIP}}\right)= \frac{\partial\,\widetilde
U}{\partial\,\rho}\,\frac{\partial\,\rho}{\partial\,t}
+\frac{\partial\,U_{_{\rm EIP}}}{\partial\,\rho}
\,\frac{\partial\,\rho}{\partial\,t} +\frac{\partial\,U_{_{\rm
EIP}}}{\partial({\bfm\nabla}\,S)}\,
\frac{\partial\,({\bfm\nabla}\,S)}{\partial\,t} \ .
\end{eqnarray}
In Eq. (\ref{aa6}), integrating by parts and neglecting the surface terms, we obtain:
\begin{eqnarray}
\nonumber \frac{d\,E}{d\,t}&=&-\frac{i\,\hbar}{2\,m}\int(V+F+W)\,
\left(\psi^\ast\,\Delta\,\psi-\psi\,\Delta\,\psi^\ast\right)\,d^3x\\
\nonumber &-&\frac{\hbar}{2\,m}\int{\cal W}\,
\left(\psi^\ast\,\Delta\,\psi+\psi\,\Delta\,\psi^\ast\right)\,d^3x\\
\nonumber &+&\int\left\{\left(V+\frac{\partial\,\widetilde
U}{\partial\,\rho}+\frac{\partial\,U_{_{\rm
EIP}}}{\partial\,\rho}\right) \,\frac{\partial\,\rho}{\partial\,t}
-{\bfm\nabla}\left[\frac{\partial\,U_{_{\rm
EIP}}}{\partial({\bfm\nabla}\,S)}\right]\,
\frac{\partial\,S}{\partial\,t}\right\}\,d^3x\\
&+&<\frac{\partial\,V}{\partial\,t}>
 \ .\label{aa10}
\end{eqnarray}
Using Eqs. (\ref{hj}) and (\ref{cont}), that we can rewrite in the
form:
\begin{equation}
\frac{\partial\,\rho}{\partial\,t}=-{\bfm\nabla}\left(\frac{{\bfm\nabla}
\,S}{m}\,\rho\right)+\frac{2}{\hbar}\,\rho\,{\cal W} \ ,
\end{equation}
\begin{equation}
\frac{\partial\,S}{\partial\,t}=
\frac{\hbar^2}{2\,m}\frac{\Delta\,\sqrt{\rho}}{\sqrt{\rho}}-\frac{({\bfm\nabla}\,S)^2}{2\,m}-V-F-W
 \ ,
\end{equation}
and taking into account the relations:
\begin{equation}
\frac{\partial\,\widetilde
U}{\partial\,\rho}+\frac{\partial\,U_{_{\rm
EIP}}}{\partial\,\rho}=F+W
\end{equation}
\begin{equation}
{\bfm\nabla}\left[\frac{\partial\,U_{_{\rm
EIP}}}{\partial({\bfm\nabla}\,S)}\right]=-\frac{2}{\hbar}\,\rho\,{\cal
 W} \ ,
\end{equation}
\begin{equation}
-\frac{i\,\hbar}{2\,m}\,\left(\psi^\ast\,\Delta\,
\psi-\psi\,\Delta\,\psi^\ast\right)={\bfm\nabla}\,\left(\frac{{\bfm\nabla}\,S}{m}\,\rho\right)
 \ ,
\end{equation}
\begin{equation}
\left(\psi^\ast\,\Delta\,\psi+\psi\,\Delta\,\psi^\ast\right)=
2\,\rho\,\left[\frac{\Delta\,\sqrt{\rho}}{\sqrt{\rho}}-\left(\frac{{\bfm\nabla}\,S}{\hbar}\right)^2\right]
 \ ,
\end{equation}
Eq. (\ref{aa10}) becomes:
\begin{eqnarray}
\nonumber
\frac{d\,E}{d\,t}&=&\int(V+F+W)\,{\bfm\nabla}\,\left(\frac{{\bfm\nabla}\,S}{m}\,\rho\right)\,d^3x
-\frac{\hbar}{m}\int\rho\,{\cal W}\,\left[\frac{\Delta\,\sqrt{\rho}}
{\sqrt{\rho}}-\left(\frac{{\bfm\nabla}\,S}{\hbar}\right)^2\right]\,d^3x\\
\nonumber
&+&\int(V+F+W)\left[-{\bfm\nabla}\left(\frac{{\bfm\nabla}\,S}{m}\,\rho\right)+\frac{2}{\hbar}\,\rho\,{\cal
 W}\right]\,d^3x\\
\nonumber
&+&\frac{2}{\hbar}\,\int\rho\,{\cal
W}\,\left[\frac{\hbar^2}{2\,m}\frac{\Delta\,\sqrt{\rho}}
{\sqrt{\rho}}-\frac{({\bfm\nabla}\,S)^2}{2\,m}-V-F-W\right]\,d^3x\\
&+&<\frac{\partial\,V}{\partial\,t}>
 \ .
\end{eqnarray}
We can immediately obtain:
\begin{equation}
\frac{d\,E}{d\,t}=<\frac{\partial\,V}{\partial\,t}> \ ,\label{eee}
\end{equation}
that is the relation (\ref{e5}). We note the absence of
contribution of the nonlinear potentials $U_{\rm EIP}(\rho,\,{\bfm
j})$ and $\widetilde U(\rho)$ to the average of the energy of the
system. We conclude by seeing that relation (\ref{eee}) can be
obtained straightforwardly by using the following property of the
Poisson brackets:
\begin{equation}
\{f,\,f\}=0 \ ,
\end{equation}
valid for every functional $f$ and therefore, we have:
\begin{equation}
\frac{d\,E}{d\,t}=\langle\frac{\partial\,V}{\partial\,t}\rangle \ ,
\end{equation}
if we take into account that explicit time dependence in the
Hamiltonian can occur only in the external potential $V(t,\,{\bfm
x})$. Notwithstanding we have preferred to obtain it in a more
rigorous fashion because of its importance.

\setcounter{chapter}{3} \setcounter{section}{0}
\setcounter{equation}{0}
\chapter*{Chapter III\\
\vspace{10mm}Symmetries and Conservation Laws}
\markright{Chap. III - Symmetries and Conservation Laws}

The concept of symmetry plays an important role in the search of
the solutions of dynamical equations. In the case of dynamical
systems with infinite degrees of freedom described by partial
differential equations (PDE), the Liouville integrability requires
the knowledge of infinite symmetries. In fact many integrable PDE
show this property like, for instance, the linear Schr\"odinger
equation $i\,\psi_t+\psi_{xx}=0$ \cite{Calogero1}, the Korteweg-de
Vries (KdV) $u_t-6\,u\,u_x+u_{xxx}=0$ and the modified KdV (mKdV)
$u_t-6u^2\,u_x+u_{xxx}=0$ \cite{Miura}, the cubic NLSE
$i\,\psi_t+\psi_{xx}+|\psi|^2\,\psi=0$ \cite{Zakharov,Ablowitz},
the Kaup-Newell equation
$i\,\psi_t+\psi_{xx}-i(\psi\,\psi^\ast\,\psi)_x=0$ \cite{Kaup},
the Chen, Lee and Liu equation
$i\,\psi_t+\psi_{xx}-i\,|\psi|^2\,\psi_x=0$ \cite{Chen} and
others.\\ When the evolution equation can be obtained from a
variational principle, the powerful N\"other theorem
\cite{Noether} gives the possibility to compute directly the
associate conserved quantities satisfying appropriate continuity
equations. The symmetries associated to geometrical
transformations are very importants. In fact, as it is known,
these are related to conserved physical quantities like the total
energy, the linear and angular momentum of the system and so on.\\
In this chapter we study the symmetries described by the connected
local Lie groups $\cal G$ depending on $r$ parameters
$\xi_i(t,\,{\bfm x},\,\psi,\,\psi^\ast)$ which are functions of
the independent variables $t,\,{\bfm x}$ and the dependent ones
$\psi$ and $\psi^\ast$. They are continuous transformations
mapping solutions of the evolution equations obeying the EIP in
other solutions. We show that in presence of the potential
introduced by EIP some geometrical symmetries of the linear
Schr\"odinger equation are lost.\\ When the full group of symmetry
$\cal G$ is known, we can study the set of solutions which are
invariant under the action of symmetries belonging to the coset
${\cal G}/{\cal K}$ where $\cal K$ is the set of subgroup of $\cal
G$ in $r-1$ parameters. In fact in this case, the original PDE is
reduced to an ordinary differential equation (ODE) which can be
solved by quadrature. We perform this computation in chapter V
where soliton solutions obeying the EIP are obtained with this
procedure.\\ At the end of this chapter we will discuss in some
detail a particular class of nonlinear gauged symmetries which can
be considered as a generalization of the gauge transformation
introduced recently by Doebner and Goldin \cite{Doebner2}. Because
it can be applied to a wide class of NLSE, we will study this
transformation in a general contest. The importance of this
transformation is that in order to make real the complex
nonlinearity that occurs in the NLSE and as a consequence it
linearizes the continuity equation. In some cases, known in the
literature \cite{Doebner2,Calogero1,Calogero2,Calogero3,Doebner3}
it is also a powerful tool to linearize the NLSE at whole.

\setcounter{equation}{0}
\section{Lie Symmetries}
Let us rewrite the EIP-Schr\"odinger equation
(\ref{schroedinger5}), after an appropriate rescaling of the
space-time coordinates, in the form:
\begin{eqnarray}
i\,\psi_t+\psi_{_{\bfm{xx}}}+\kappa\,(\psi^\ast\,\psi_{_{\bfm
x}}-\psi\,\psi^\ast_{_{\bfm x}})\,\psi_{_{\bfm x}}+
{\kappa\over2}\,(\psi^\ast\,\psi_{_{\bfm{xx}}}
-\psi\,\psi_{_{\bfm{xx}}}^\ast)\,\psi=0 \ ,\label{schroedinger7}
\end{eqnarray}
where we use of the stenography notation: $\psi_{_{\bfm
x}}\equiv{\bfm\nabla}\,\psi,\,\,\psi_{_{\bfm{xx}}}\equiv\Delta\,\psi$.\\
We are concerned with the study of the symmetries of the EIP
potential and thus, we neglect in this chapter both the nonlinear
and the external potentials $\widetilde{U}(\rho)$ and $V$. The
results obtained can be generalized when these quantities are
present.\\ Being Eq. (\ref{schroedinger7}) canonical, we can
perform the study of the symmetries both at the level of the
Lagrangian and directly on the evolution equation. The study of
the symmetries from the PDE is more general, because as we will
show, not all the symmetries of the PDE are also symmetries of its
Lagrangian. Therefore we start by studying the symmetries of Eq.
(\ref{schroedinger7}) and after in the next section, starting from
its Lagrangian, we derive, by means of N\"other theorem, the
appropriate conserved quantities (when it occurs).\\ To find the
Lie symmetries of Eq. (\ref{schroedinger7}) we adopt the geometric
method presented in the Olver text-book \cite{Olver} to which we
send the reader for a complete treatment. We make a summary of the
general concepts.\\ The first step in the research of the
symmetries for a PDE is to look up it as an algebraic equation.
For this purpose, we introduce the {\sl jet}-space
$\widetilde{F}\equiv M\times F\times F^{^{\bfm x}}\times
F^{^{\bfm{xx}}}\times F^t$ which is an extension of the
configuration space. Here $M$ is the configuration space mapped by
the space-time coordinates, $F$ is the space mapped by the
function $\psi$ and $\psi^\ast$ when are considered as independent
fields. In the same fashion $F^{^{\bfm x}}$ is mapped from the
function $\psi_{_{\bfm x}}$ and $\psi_{_{\bfm x}}^\ast$ and so on.
In this jet-space, Eq. (\ref{schroedinger7}) becomes an algebraic
equation in the independent variables
$x,\,t,\,\psi,\,\psi^\ast,\,\psi_{_{\bfm x}},\,\psi_{_{\bfm
x}}^\ast,\,\psi_{_{\bfm{xx}}},\,\psi^\ast_{_{\bfm {xx}}}$ and
$\psi_t$ and defines an hyper-surface on $\widetilde{F}$. We
consider now an element $g\in{\cal G}$ of a transformation Lie
group which acts on the element of $\widetilde{F}$. We say that
$g$ is a symmetry transformation for Eq. (\ref{schroedinger7}) if
a point $P$ is mapped on the hyper-surface in a point on itself.
In particular, if we consider infinitesimal transformations, we
must require that those generated by the element $g$ lie on the
tangent plane on the hyper-surface in $P$.\\ Because $\cal G$ is a
Lie group, its elements around the identity can be written as:
\begin{eqnarray}
g=e^{i\,\epsilon\,{\bfm v}} \ ,
\end{eqnarray}
where $\epsilon$ is a parameter and $\bfm v$ is a vector in the tangent plane on the hyper-surface.
Its general expression will be:
\begin{eqnarray}
\nonumber
{\bfm v}&=&\tau\,\partial_t+\xi^{^{\bfm x}}\cdot\partial_{_{\bfm
x}}+\phi\,\partial_\psi+\widetilde{\phi}\,\partial_{\psi^\ast}
+\phi^t\,\partial_{\psi_t}\\
&+&\phi^{^{\bfm x}}\cdot\partial_{\psi_{\bfm x}}+\widetilde{\phi}^{\,\,^{\bfm x}}\cdot\partial_{\psi_{\bfm
x}}+\phi^{\,\,^{\bfm{xx}}}\cdot\partial_{\psi_{\bfm{xx}}}+
\widetilde{\phi}^{\,\,^{\bfm{xx}}}\cdot\partial_{\psi_{\bfm{xx}}}
 \ ,\label{generatore}
\end{eqnarray}
where the coefficients:
\begin{eqnarray}
\nonumber
\xi^{^{\bfm x}}=\left(
\begin{array}{c}
\xi^x\\
\xi^y\\
\xi^z
\end{array}
\right) \ ,\hspace{10mm}
\phi^{^{\bfm x}}=\left(
\begin{array}{c}
\phi^x\\
\phi^y\\
\phi^z
\end{array}
\right) \ ,\hspace{10mm}
\widetilde{\phi}^{^{\,\,{\bfm x}}}=\left(
\begin{array}{c}
\widetilde{\phi}^{\,\,x}\\
\widetilde{\phi}^{\,\,y}\\
\widetilde{\phi}^{\,\,z}
\end{array}
\right) \ ,
\end{eqnarray}
\begin{eqnarray}
\phi^{^{\bfm{xx}}}=\left(
\begin{array}{c}
\phi^{xx}\\
\phi^{yy}\\
\phi^{zz}
\end{array}
\right) \ ,\hspace{10mm}
\widetilde{\phi}^{^{\,\,{\bfm{xx}}}}=\left(
\begin{array}{c}
\widetilde{\phi}^{\,\,xx}\\
\widetilde{\phi}^{\,\,yy}\\
\widetilde{\phi}^{\,\,zz}
\end{array}
\right) \ ,
\end{eqnarray}
are three-vectors. All this quantities together with $\tau,\,\phi$
and $\widetilde{\phi}$ are function of ${\bfm x},\,t,\,\psi$ and
$\psi^\ast$. The quantities $\phi^{^{\bfm
x}},\,\widetilde{\phi}^{^{\,\,{\bfm
x}}},\,\phi^{^{\bfm{xx}}},\,\widetilde{\phi}^{^{\,\,{\bfm{xx}}}}$
and $\phi^t$ are related to the coefficients $\xi^{^{\bfm x}}$ and
$\tau$ by means of:
\begin{eqnarray}
\phi^i={\rm D}_i\,\phi-\left({\rm D}_i\,\xi^x\right)\,\psi_x
-\left({\rm D}_i\,\xi^y\right)\,\psi_y-\left({\rm
D}_i\,\xi^y\right)\,\psi_y -\left({\rm D}_i\,\tau\right)\,\psi_t \
,\label{dx}
\end{eqnarray}
with $i\equiv x,\,y,\,z,\,t$ where
\begin{eqnarray}
{\rm
D}_i\equiv\partial_i+\psi_i\,\partial_\psi+\psi_i^\ast\,\partial_{\psi^\ast}
 \ ,
\end{eqnarray}
is the total derivative, and:
\begin{eqnarray}
\phi^{ii}={\rm D}_{ii}\,\phi-2\,\left({\rm D}_i\,\xi^x\right)\,\psi_{xx}
-\left({\rm D}_{ii}\,\xi^x\right)\,\psi_x+(x\rightarrow y)+(x\rightarrow
z)+(x\rightarrow t)\label{dxx}
\ .
\end{eqnarray}
In Eqs. (\ref{dx}) and (\ref{dxx}) the time derivative of the
fields $\psi$ and $\psi^\ast$ can be eliminated by using the
evolution equation (\ref{schroedinger7}). Of course, analogous
expressions hold for $\widetilde{\phi}^{\,\,i}$ and
$\widetilde{\phi}^{\,\,ii}$. At this point we must determine the
coefficients $\tau,\,\xi^{\bfm x},\,\phi$ and $\widetilde{\phi}$
so that the vector (\ref{generatore}) lies on the tangent plane to
the hyper-surface generated by Eq.
(\ref{schroedinger7})\footnote{In Ref. \cite{Olver} the generator
of the transformation ${\bfm v}$ acting on the jet-space
$\widetilde{F}$ was called prolongation of the vector $\bfm u$ and
denoted by ${\bfm v}=pr^{(2)}{\bfm u}$ where $\bfm u$ is a vector
on the tangent plane on $M$. Here we prefer to avoid this
terminology for purposes of simplicity.}. Let
$L(\psi,\,\psi^\ast)$ denote the PDE:
\begin{eqnarray}
L(\psi,\,\psi^\ast)\equiv i\,\psi_t+\psi_{_{\bfm{xx}}}+\kappa\,(\psi^\ast\,\psi_{_{\bfm
x}}-\psi\,\psi^\ast_{_{\bfm x}})\,\psi_{_{\bfm x}}+
{\kappa\over2}\,(\psi^\ast\,\psi_{_{\bfm{xx}}}-\psi\,\psi_{_{\bfm{xx}}}^\ast)\,\psi
 \ ,
\end{eqnarray}
$g$ is a symmetry transformation if:
\begin{eqnarray}
L(g\,\psi,\,g\,\psi^\ast)=0 \ ,
\end{eqnarray}
whenever $\psi$ and $\psi^\ast$ are solutions of $L(\psi,\,\psi^\ast)=0$.
The condition for the vector $\bfm v$ to lie on the tangent plane
is given by \cite{Olver}:
\begin{eqnarray}
{\bfm v}\left[i\,\psi_t+\psi_{_{\bfm{xx}}}+\kappa\,(\psi^\ast\,\psi_{_{\bfm
x}}-\psi\,\psi^\ast_{_{\bfm x}})\,\psi_{_{\bfm x}}+
{\kappa\over2}\,(\psi^\ast\,\psi_{_{\bfm{xx}}}-\psi\,\psi_{_{\bfm{xx}}}^\ast)\,\psi\right]=0 \ .
\end{eqnarray}
Taking into account the expression of the vector $\bfm v$ (\ref{generatore}) we obtain:
\begin{eqnarray}
\nonumber
&&i\,\phi^t+\left(1+\frac{\kappa}{2}\,\psi\,\psi^\ast\right)\,\phi^{\bfm{xx}}-\frac{\kappa}{2}\,\psi^2\,
\widetilde{\phi}^{\,\,{\bfm{xx}}}+\kappa\,\left(2\,\psi^\ast\,\psi_{\bfm
x}-\psi\,\psi_{\bfm x}^\ast\right)\,\phi^{\bfm
x}-\kappa\,\psi\,\psi_{\bfm x}\,\widetilde{\phi}^{\,\,\bfm x}\\
&&+\frac{\kappa}{2}\left(\psi^\ast\,\psi_{\bfm{xx}}-2\,\psi\,\psi^\ast_{\bfm
{xx}}-2\psi_{\bfm x}\,\psi_{\bfm x}^\ast\right)\,\phi+\frac{\kappa}{2}\left(
\psi\,\psi_{\bfm{xx}}+2\psi_{\bfm x}^2\right)\,\widetilde{\phi}=0 \ ,
\end{eqnarray}
which must be solved using the expression of the quantities
$\phi,\,\phi^i,\,\phi^{ii},\,\tilde{\phi}^{\,\,i}$ and
$\tilde{\phi}^{\,\,ii}$. By requiring that the coefficient of each
monomial in $\psi,\,\psi^\ast$ and their derivative vanishes
separately, we obtain a sequence of derivative equations for the
coefficients $\tau,\,\xi^{^{\bfm x}},\,\phi,\,\widetilde{\phi}$
that when satisfied, give us the expression of the generators of
the full Lie symmetries. The computation was developed with
MATHEMATICA 2.0 package. We report in table the results:
\begin{eqnarray}
\begin{array}{cccc}
\partial_{xx}\xi^x+\partial_{yy}\xi^x+\partial_{zz}\xi^x=0 \
,&\partial_y\xi^x=-\partial_x\xi^y \ ,&
\partial_t\xi^t=2\,\partial_x\xi^x \ ,&\phi=\psi \ ,\\
\partial_{xx}\xi^y+\partial_{yy}\xi^y+\partial_{zz}\xi^y=0 \
,&\partial_z\xi^x=-\partial_x\xi^z \ ,&
\partial_x\xi^x=\partial_y\xi^y \ ,&\widetilde{\phi}=-\psi^\ast \ ,\\
\partial_{xx}\xi^z+\partial_{yy}\xi^z+\partial_{zz}\xi^z=0 \
,&\partial_z\xi^y=-\partial_y\xi^z \ ,&
\partial_y\xi^y=\partial_z\xi^z \ ,&
\end{array}
\end{eqnarray}
and all the other derivatives are equal to zero. The general solution of this
system can be easily obtained. Thus, we conclude that the most general
infinitesimal symmetry of the EIP-Schr\"odinger equation has coefficient
functions of the form:
\begin{eqnarray}
&&\xi^x=\lambda\,(y-z)+\mu\,x+\alpha^x \ ,\\
&&\xi^y=\lambda\,(z-x)+\mu\,y+\alpha^y \ ,\\
&&\xi^z=\lambda\,(x-y)+\mu\,z+\alpha^z \ ,\\
&&\tau=2\,\mu\,t+\alpha^t \ ,\\
&&\phi=\beta\,\psi \ ,\\
&&\widetilde{\phi}=-\beta\,\psi^\ast \ ,
\end{eqnarray}
where $\lambda,\,\mu,\,\beta,\,\alpha^i$ are real coefficients.
Thus the Lie algebra of infinitesimal symmetries is spanned by the nine
vector fields:
\begin{eqnarray}
&&\left.
\begin{array}{c}
{\bfm v}_1=\partial_x\\
{\bfm v}_2=\partial_y\\
{\bfm v}_3=\partial_z\\
{\bfm v}_4=\partial_t\\
\end{array}
\right\}\hspace{50mm}\mbox{translation,}\label{trasla}\\ \nonumber
&&\\ &&\left.
\begin{array}{c}
{\bfm v}_5=y\,\partial_x-x\,\partial_x\\
{\bfm v}_6=x\,\partial_z-z\,\partial_x\\
{\bfm v}_7=z\,\partial_y-y\,\partial_z\\
\end{array}
\right\}\hspace{35mm}\mbox{rotational,}\label{rota}\\ \nonumber
&&\\ &&\hspace{2mm}{\bfm
v}_8=t\,\partial_t+2\,(x\,\partial_x+y\,\partial_y+z\,\partial_z)
\hspace{12mm}\mbox{dilation,}\label{scala}\\ \nonumber &&\\
&&\hspace{2mm}{\bfm
v}_9=\psi\,\partial_\psi-\psi^\ast\,\partial_{\psi^\ast}\hspace{35mm}\mbox{$U(1)$
global transformation.}\hspace{15mm}\label{gau}
\end{eqnarray}
The groups ${\cal G}_i$ generated by the ${\bfm v}_i$ are
given in the following table. The entries give the transformed point
$\exp(\epsilon\,{\bfm v}_i)\,(t,\,{\bfm
x},\,\psi,\,\psi^\ast)=(\tilde{t},\,\tilde{\bfm
x},\,\tilde{\psi},\,\tilde{\psi^\ast})$:
\begin{eqnarray}
{\cal G}_{1,2,3}:&&\hspace{5mm}(t,\,{\bfm x}+{\bfm\epsilon},\,\psi,\,\psi^\ast) \ ,\\
{\cal G}_{4}:&&\hspace{5mm}(t+\epsilon,\,{\bfm x},\,\psi,\,\psi^\ast) \ ,\\
{\cal G}_{5,6,7}:&&\hspace{5mm}(t,\,R^{ij}\,x_j,\,\psi,\,\psi^\ast) \ ,\\
{\cal G}_{8}:&&\hspace{5mm}(e^{2\,\epsilon}\,t,\,e^\epsilon\,{\bfm x},\,\psi,\,\psi^\ast) \ ,\\
{\cal G}_{9}:&&\hspace{5mm}(t,\,{\bfm x},\,e^{i\,\epsilon}\,\psi,e^{-i\,\epsilon}\,\psi^\ast) \ ,
\end{eqnarray}
where $R^{ij}$ is the $3\times3$ relational matrix. Since each group
${\cal G}_i$ is a symmetry group we have that if $\psi(t,\,{\bfm x})$ is a
solution of Eq. (\ref{schroedinger7}), so are the functions:
\begin{eqnarray}
&&\psi^{(1)}=\psi(t,\,{\bfm x}+{\bfm\epsilon}) \ ,\\
&&\psi^{(2)}=\psi(t+\epsilon,\,{\bfm x}) \ ,\\
&&\psi^{(3)}=\psi(t,\,R^{ij}\,x_j) \ ,\\
&&\psi^{(4)}=\psi(e^{2\,\epsilon}\,t,\,e^{\epsilon}\,{\bfm x}) \ ,\\
&&\psi^{(5)}=e^{i\,\epsilon}\,\psi(t,\,{\bfm x}) \ ,
\end{eqnarray}
where $\epsilon$ and $\bfm \epsilon$ are real quantities.

\setcounter{equation}{0}
\section{Conserved quantities}
Having obtained the generators of the Lie symmetries and their
algebrae, here we study the conserved quantities
related to these symmetries, starting from the action of the system and by
means of the N\"other theorem. In this section it is convenient to restore the standard unity
of $\hbar$ and $m$. As it was said previously, not all the
symmetries of the PDE (\ref{schroedinger7}) obtained in the last section are
also symmetries for the action:
\begin{equation}
{\cal A}=\int{\cal L}\,d^3x\,dt \ .\label{action1}
\end{equation}
As a consequence do not all those symmetries are related to conserved
quantities.\\
To start with, we remember that "{\sl Lagrangian}" symmetries are
those that do not change the formal expression of the action. Therefore,
they are coordinates and fields
transformations satisfying the relationship:
\begin{equation}
\delta\,{\cal A}=\int\left[ \delta\,{\cal L}\,d^3x\,dt+{\cal
L}\,\delta(\,d^3x\,dt)\right]=0 \ .\label{cond3}
\end{equation}
The N\"other theorem states that a current ${\cal
J}^\nu$ exists, given by:
\begin{equation}
{\cal J}^\nu=\frac{\partial\,{\cal
L}}{\partial\,(\partial_\nu\psi)}
\,\delta\,\psi+\frac{\partial\,{\cal
L}}{\partial\,(\partial_\nu\psi^\ast)} \,\delta\,\psi^\ast-f^\nu \
,  \label{ncurrent}
\end{equation}
satisfying the continuity equation:
\begin{equation}
\partial_\nu{\cal J}^\nu=0 \ . \label{cn}
\end{equation}
Therefore the quantity:
\begin{equation}
Q=\int\limits_{\cal D}{\cal J}^0\,d^2x \ ,\label{ncarica}
\end{equation}
is time conserved if the spatial component ${\cal J}^i$ vanishes steeply
on the boundary $\partial\,{\cal D}$.\\
Taking into account the result of the last section we begin considering
the invariance of (\ref{schroedinger7}) under $U(1)$ transformations (\ref{gau}):
\begin{equation}
\psi^\prime(t,\,{\bfm x})=e^{-i\alpha/\hbar}\,\psi(t,\,{\bfm x}) \
.\label{u1}
\end{equation}
The current is:
\begin{equation}
j^\mu\equiv\left(c\,\rho,-\frac{i\,\hbar}{2\,m}\,(1+\kappa\,\rho)\,(\psi^\ast\,{\bf\nabla}\psi-\psi\,{\bf\nabla}\psi)\right)
 \ ,\label{current}
\end{equation}
and satisfies the continuity equation:
\begin{equation}
\partial_\mu\,j^\mu=0 \ .
\end{equation}
The time component of Eq. (\ref{current}) is the matter density
$\rho$, the spatial part is the quantum current $\bfm j$ with
EIP.\\ Consider now the space-time translations with generator
(\ref{trasla}):
\begin{equation}
t\rightarrow t-t_0 \ ,\hspace{10mm}{\bfm x}\rightarrow{\bfm
x}-{\bfm x}_0 \ ,\label{traslation1}
\end{equation}
the field $\psi(t,\,{\bfm x})$ transforms as:
\begin{equation}
\psi(t,\,{\bfm x})\rightarrow\psi(t+t_0,\,{\bfm x}+{\bfm x}_0) \
.\label{traslation2}
\end{equation}
These symmetries imply the conservation law:
\begin{equation}
\partial_\mu\,T^{\mu\,\nu}=0 \ ,\label{cont1}
\end{equation}
where the energy-momentum tensor $T^{\mu\nu}$ obeys to EIP and is
defined as:
\begin{eqnarray}
T^{00}&=&\frac{\hbar^2}{2\,m}|{\bfm\nabla}\,\psi|^2
-\kappa\,\frac{\hbar^2}{8m}\left(\psi^\ast\,{\bfm\nabla}\psi-
\psi\,{\bfm\nabla}\psi^\ast\right)^2 \ ,\label{prima}\\
T^{0i}&=&-i\,c\,{\hbar\over2}\left(\psi^\ast\,\partial_i\psi-
\psi\,\partial_i\psi^\ast\right) \ ,\label{seconda}\\ \nonumber
T^{i0}&=&-\frac{\hbar^2}{2\,m}\left(\partial_i\psi^\ast\,\partial_0\psi+\partial_i\psi\,
\partial_0\psi^\ast\right)\\
&&+\kappa\,\frac{\hbar^2}{4\,m}\,\left(\psi^\ast\,\partial_i\psi-
\psi\,\partial_i\psi^\ast\right)\left(\psi^\ast\,\partial_0\psi-\psi\,\partial_0\psi^\ast\right) \ ,\label{terza}\\
\nonumber
T^{ij}&=&\frac{\hbar^2}{2\,m}\,\left(\partial_i\psi^\ast\,\partial_j\psi+\partial_i\psi\,
\partial_j\psi^\ast\right)\\
\nonumber &&-\kappa\,\frac{\hbar^2}{4\,m}\,
\left(\psi^\ast\,\partial_i\psi-\psi\,\partial_i\psi^\ast\right)
\left(\psi^\ast\,\partial_j\psi-\psi\,\partial_j\psi^\ast\right)\\
&&-\delta_{ij}\left[\frac{\hbar^2}{4\,m}\,\Delta\rho+
\kappa\,\frac{\hbar^2}{8m}\left(\psi^\ast\,{\bfm\nabla}\psi-
\psi\,{\bfm\nabla}\,\psi^\ast\right)^2 \right] \ .\label{quarta}
\end{eqnarray}
Equations (\ref{prima}) and (\ref{seconda}) define the generators
of the transformations (\ref{traslation1}) and are the energy
density and the momentum density of the system, respectively. From
Eq. (\ref{prima}) we may define the quantity:
\begin{equation}
U_{_{\rm EIP}}=-\kappa\,\frac{\hbar^2}{8\,m}
\left(\psi^\ast\,{\bfm\nabla}\psi-\psi\,{\bfm\nabla}\psi^\ast\right)^2
\ ,\label{ueip1}
\end{equation}
representing the interaction energy density introduced by EIP. The
quantity $T^{00}$ is the sum of the kinetic energy density
$(\hbar^2/2\,m)\,|{\bfm\nabla}\psi|^2$ and the interaction energy
density:
\begin{equation}
T^{00}=\frac{\hbar^2}{2\,m}\,|{\bfm\nabla}\,\psi|^2+U_{_{\rm EIP}}
\ .
\end{equation}
From (\ref{seconda}) we can see that the momentum density is not
influenced by EIP. The energy flux $T^{i0}$ and the momentum flux
$T^{ij}$ are modified by EIP, as we can see from Eqs.
(\ref{terza}) and (\ref{quarta}). Using Eq. (\ref{cont1}) we
remark that the quantities:
\begin{equation}
E=\int T^{00}\,d^3x \ ,\label{energy}
\end{equation}
\begin{equation}
P^i={1\over c}\,\int T^{0i}\,d^3x \ ,
\end{equation}
are constants of motion. By comparing Eqs. (\ref{seconda}) and
(\ref{terza}) we can see that the momentum density $T^{0i}$ does
not coincide with the energy flux density $T^{i0}$ because of the
non Lorentz invariance of the system. On the contrary we have
$T^{ij}=T^{ji}$, suggesting spatial rotations invariance of the
system.

In a nonrelativistic theory, like the one we are considering, the energy
of the system must be a semidefinite positive quantity. This is a true
statement for our model. In fact, taking into account the expression of
the spatial component of (\ref{current}), which represents the quantum
current density, Eq. (\ref{prima}) of the energy density $T^{00}$
becomes:
\begin{eqnarray}
T^{00}=\frac{\hbar^2}{2\,m}|{\bfm\nabla}\,\psi|^2
+\kappa\,\frac{m}{2}\left(\frac{\bfm j}{1+\kappa\,\rho}\right)^2 \
.
\end{eqnarray}
This expression appears to be a semidefinite quantity when the
parameter $\kappa$ is positive. We can transform this
expression in the form:
\begin{eqnarray}
T^{00}=\frac{\hbar^2}{2\,m}|{\bfm\nabla}\,\psi|^2\,(1+\kappa\,\rho)
-\kappa\,\frac{\hbar^2}{8\,m}\left({\bfm\nabla}\,\rho\right)^2 \ ,
\end{eqnarray}
which now results a semidefinite positive quantity when $\kappa$
is negative, if we remember that the quantities $1+\kappa\,\rho$
is always positive.

Let us consider the transformations on the coordinate $\bfm x$,
produced by the orthogonal matrix $R$:
\begin{eqnarray}
&&x^i\rightarrow R^i_{~j}\,x^j \
,\hspace{10mm}R^i_{~j}\,R^j_{~k}=\delta^i_{~k} \ ,\\
&&\psi(t,\,{\bfm x})\rightarrow\psi(t,\,R^{-1}\,{\bfm x}) \
.\label{rotation}
\end{eqnarray}
The action invariance allows us to define the angular momentum density:
\begin{eqnarray}
M^{\mu ij}=x^j\,T^{\mu i}-x^i\,T^{\mu j} \ ,\label{quinta}
\end{eqnarray}
with $\mu=0,\dots,3;~~i,\,j=1,\,2,\,3$, obeying the continuity equation:
\begin{equation}
\partial_\mu M^{\mu ij}=0 \ .\label{cont2}
\end{equation}
We may note from  Eqs. (\ref{seconda}) and (\ref{quarta}) that the
generators of the transformation (\ref{rotation})
\begin{equation}
<L_i>=\varepsilon_{ijk}\,\int M^{0jk}\,d^3x \ ,\label{angular}
\end{equation}
are constants of motion, not affected by $U_{_{\rm EIP}}$ which,
on the contrary, modifies the flux densities.\\ We are left with
the scaling symmetry generated by Eq. (\ref{scala}). It is easy to
see that it is not a symmetry for the Lagrangian, independently of
the number of the dimension $D$ in which the system is immersed.
Therefore we are not able to apply the N\"other theorem to obtain
the correspondent conserved quantities.

We ask now what happens to the other generators of the Schr\"odinger group.
In particular, from the list of symmetries obtained in section 3.1, the Galilei and the
conformal symmetries are missed.

Let us consider the conformal group transformations:
\begin{equation}
t=\frac{\alpha\,t+\beta}{\gamma\,t+\delta} \
,\hspace{10mm}\alpha\,\delta-\beta\,\gamma=1 \ ,\label{conform}
\end{equation}
with $\alpha,\,\beta,\,\gamma,\,\delta$ arbitrary constants, that
can be decomposed in three independent transformations:
$t\rightarrow1/t,\,t\rightarrow\alpha\,t$ and
$1/t\rightarrow\/t+\delta$. The first is a discrete transformation
and does not produce constants of motion, the other two, the
dilation and the special conformal transformation, makes the
action of the system not invariant.\\ The reason why we loose the
conformal symmetry lies in the fact that the parameter $\kappa$
has a proper dimension. In order to analyze how the potential
$U_{_{\rm EIP}}$ breaks the symmetry, consider the case of
dilation:
\begin{equation}
\left[{\bfm x},\,t,\,\kappa,\,\psi(t,{\bfm
x})\right]\longrightarrow \left[\lambda\,{\bfm
x},\,\lambda^2\,t,\,\lambda^3\,\kappa,\,\lambda^{-3/2}\,\psi(t,\,{\bfm
x})\right] \ .\label{dil1}
\end{equation}
In this case, Eq. (\ref{cond3}) becomes:
\begin{equation}
\delta{\cal A}=\int\left(\partial_\mu\,D^\mu+U_{_{\rm
EIP}}\right)\,d^3x\,dt=0 \ ,
\end{equation}
where:
\begin{equation}
D^\mu=2\,t\,T^{\mu 0}-x_i\,T^{\mu i} \ ,
\end{equation}
is the well-known dilation current of N\"other. Therefore the
potential $U_{_{\rm EIP}}$ prevents to write the quantity
$\partial_\mu D^\mu+U_{_{\rm EIP}}$ in the (\ref{dil1}) as a
tetradivergence. Note also that the conformal invariance is
generally broken when derivative potentials \cite{Hagen2}, like
$U_{_{\rm EIP}}$, are present. Nevertheless, if we consider the
transformation:
\begin{eqnarray}
\left[{\bfm x},\,t,\,\kappa,\,\psi(t,{\bfm
x})\right]\longrightarrow \left[\lambda\,{\bfm
x},\,\lambda^2\,t,\,|\mu|^{-2}\,\kappa,\,\mu\,\psi(t,\,{\bfm
x})\right] \ ,\label{dil}
\end{eqnarray}
with $\lambda$ and $\mu$ arbitrary constants, Eq.
(\ref{schroedinger6}) remains invariant. Remark that in Eq.
(\ref{dil}) the transformation with parameter $\lambda$ is the
same of the scale transformation found in the last section while,
the transformation with parameter $\mu$ is not a Lie symmetry. It
permits us to reduce the study of the system described by Eq.
(\ref{schroedinger6}) to consider only the two relevant cases with
$|\kappa|=1$.\\ We consider now the Galileo transformation on the
coordinates:
\begin{equation}
t\rightarrow t \ ,\hspace{10mm}{\bfm x}\rightarrow{\bfm x}+{\bfm
v}\,t \ ,\label{galileo1}
\end{equation}
and we set the field transformation ansatz:
\begin{equation}
\psi(t,\,{\bfm x})\rightarrow R(t,\,{\bfm
x})\,e^{i\,\alpha(t,\,{\bfm x})}\,\psi(t,\,{\bfm x}-{\bfm v}\,t) \
,\label{galileo2}
\end{equation}
with $\alpha(t,\,{\bfm x})$ and $R(t,\,{\bfm x})$ arbitrary real
functions. The requirement that the action (\ref{action1}) be
invariant under the transformations (\ref{galileo1}) and
(\ref{galileo2}) gives the following derivative on the functions
$\alpha(t,\,{\bfm x})$ and $R(t,\,{\bfm x})$:
\begin{eqnarray}
&&{\bfm\nabla}\alpha(t,\,{\bfm x})=-\frac{m\,{\bfm
v}}{\hbar}\frac{1}{1+\kappa\,\rho} \ ,\label{cond1}\\
&&\frac{\partial\alpha(t,\,{\bfm x})}{\partial t}=\frac{m\,{\bfm
v}^2}{2\,\hbar}\frac{1}{1+\kappa\,\rho} \ ,\label{cond2}\\
&&R(t,\,{\bfm x})=const \ ,\label{cond4}
\end{eqnarray}
that, in the limit $\kappa\rightarrow0$, reproduce the factor
$\alpha(t,\,{\bfm x})=-m\,{\bfm v}\cdot{\bfm x}+m\,{\bfm
v}^2\,t/2$ of the linear theory. We can easily realize that in the
general case $\kappa\not=0$ Eqs. (\ref{cond1})-(\ref{cond2}) are
not satisfied by an arbitrary function of $\rho$.\\ The Galileo
broken symmetry due to the potential (\ref{ueip1}) is not
surprising. In fact we know that the generator of the Galilei
transformation is given by:
\begin{equation}
{\bfm G}=<{\bfm P}>\,t-m\,N\,<{\bfm x}_c> \ ,\label{gal}
\end{equation}
which represent the velocity center of mass of the system.
If we derive with respect to time Eq. (\ref{gal}) and remembering that  $<{\bfm
P}>$ is time independent because is a constant of motion, we obtain:
\begin{equation}
\frac{d{\bfm G}}{dt}=<{\bfm P}>-m\,N\,\frac{d\,<{\bfm x}_c>}{dt} \ .
\end{equation}
Taking into account the relations (\ref{xxx}) and $<{\bfm
P}>=m\,\int{\bfm j}_0\,d^2x$ we have:
\begin{equation}
\frac{d{\bfm G}}{dt}=-\kappa\,m\,\int\rho\,{\bfm j}_0\,d^2x \ ,  \label{galilei}
\end{equation}
where ${\bfm j}_0$ is the current without EIP.
We see immediately that the presence of EIP breaks the Galilei symmetry
which is restored in the limit of $\kappa\rightarrow0$.

We conclude with a brief discussion of the discrete transformations $P$ and $T$.
These are not broken by the presence of the potential (\ref{ueip1}), as it
can be easily shown considering the following relations:
\begin{eqnarray}
\nonumber P:~~~&&(t,\,{\bfm x})\rightarrow(t,\,-{\bfm x}) \
,\hspace{5mm}\psi(t,\,{\bfm x})\rightarrow\psi(t,\,-{\bfm x}) \
,\\ &&{\bfm j}(t,\,{\bfm x})\rightarrow-{\bfm j}(t,\,-{\bfm x}) \
,\\ \nonumber T:~~~&&(t,\,{\bfm x})\rightarrow(-t,\,{\bfm x}) \
,\hspace{5mm} \psi(t,\,{\bfm x})\rightarrow\psi^\ast(-t,\,{\bfm
x}) \ ,\\ &&{\bfm j}(t,\,{\bfm x})\rightarrow-{\bfm j}(-t,\,{\bfm
x}) \ .
\end{eqnarray}
Therefore the expression of the potential (\ref{ueip1}) is
invariant under the above transformations.


\setcounter{equation}{0}
\section{Nonlinear gauge transformations}
In this section we describe a particular class of nonlinear gauge
transformations allowing us to linearize the continuity equation
making real the complex potential that appears in the evolution
equation (\ref{schroedinger6}). Since the method that we are
describing can be applied to a wide class of nonlinear
Schr\"odinger equations, let us describe it in a general fashion
and only at the end of the section we apply it to the EIP system.
We introduce the method on the more simple 1-dimensional system,
the generalization to the 3-dimensional case is in progress.\\ We
introduce the density of Lagrangian
\begin{equation}
{\cal L}=i\,\frac{\hbar}{2}\left(\psi^\ast\,\frac{\partial\,\psi}
{\partial\,t}-\psi\,\frac{\partial\,\psi^\ast}{\partial\,t}\right)-
\frac{\hbar^2}{2\,m}\,\Big|\frac{\partial\,\psi}{\partial\,x}\Big|^2-U([\psi^\ast],\,[\psi])
 \ ,\label{lagra}
\end{equation}
which describes a class of one dimensional nonrelativistic and
canonical quantum systems. In Eq. (\ref{lagra}) the nonlinear real
potential $U([\psi^\ast],\,[\psi])$ is a functional of the fields
$\psi$ and $\psi^\ast$. We will use the hydrodynamic fields
$\rho(x,\,t)$ and $S(x,\,t)$ [cfr. Eqs. (\ref{ro})-(\ref{s})]. The
evolution equations of the fields
$a\equiv\psi,\,\psi^\ast,\,\rho,\,S$ can be obtained from the
action of the system ${\cal A}=\int{\cal L}\,dx\,dt$ by using the
least action principle $E_a({\cal A})=0$. We make use of the
following notation for the spatial derivatives:
\begin{eqnarray}
a_n=\frac{\partial^n\,a}{\partial\,x^n} \ ,\hspace{20mm}a_0\equiv a \ .
\end{eqnarray}
It is immediate to obtain the evolution equation of the field
$\psi$ that is given by the following Schr\"odinger equation which
contains a complex nonlinearity:
\begin{equation}
i\,\hbar\,\frac{\partial\,\psi}{\partial\,t}=-\frac{\hbar^2}{2\,m}
\,\frac{\partial^2\,\psi}{\partial\,x^2}+W([\rho],\,[S])\,\psi+i\,{\cal
W}([\rho],\,[S])\,\psi \ .\label{sch}
\end{equation}
The real $W([\rho],\,[S])$ and the imaginary ${\cal
W}([\rho],\,[S])$ part are given by the following expressions:
\begin{eqnarray}
&&W([\rho],\,[S])=\frac{\delta}{\delta\,\rho}\,\int
U([\rho],\,[S])\,dx\,dt \ ,\\ &&{\cal
W}([\rho],\,[S])=\frac{\hbar}{2\,\rho}\frac{\delta}{\delta\,S}\,\int
U([\rho],\,[S])\,dx\,dt \ .
\end{eqnarray}
The evolution equation for the field $\rho$ is obtained directly from
E$_a({\cal A})=0$, posing $a\equiv S$ (now $S$ is the field canonically
conjugated to the field $-\rho$). We obtain the equation:
\begin{equation}
\frac{\partial\,\rho}{\partial\,t}+\frac{\partial\,j_{_\psi}}{\partial\,x}=
\frac{\partial}{\partial\,S}\,U([\rho],\,[S]) \ ,\label{ec}
\end{equation}
where the quantum current $j_{_\psi}$ takes the expression:
\begin{equation}
j_{_{\psi}}=\frac{S_1}{m}\,\rho+\sum_{n=0}\,(-1)^n\,\frac{\partial^n}
{\partial\,x^n}\left[\frac{\partial}{\partial\,S_{n+1}}\,U([\rho],\,[S])\right]
 \ .\label{cur1}
\end{equation}
We remark that Eq. (\ref{ec}) is the continuity equation and the
term in the right hand side represents a source for the field
$\rho$. When the conservation of the number of particles
$N=\int\rho\,dx$ is required, the hypothesis that the potential
$U([\rho],\,[S])$ does not depend on $S$ but only on its
derivative must be introduced, therefore the Eq. (\ref{ec}) takes
the form:
\begin{equation}
\frac{\partial\,\rho}{\partial\,t}+\frac{\partial\,j_{_\psi}}{\partial\,x}=0
 \ .\label{ec1}
\end{equation}
We note that Eq. (\ref{ec1}) can be obtained directly from
(\ref{sch}) and from its complex conjugate, performing the
standard procedure. We can see that the imaginary part ${\cal
W}([\rho],\,[S])$ is responsible for the nonlinearity of the
expression of the current $j_{_\psi}$ (\ref{cur1}).\\ Let us
introduce the following transformation for the field $\psi$:
\begin{equation}
\psi(x,\,t)\rightarrow \phi(x,\,t)={\cal
U}([\psi^\ast],\,[\psi])\,\psi(x,\,t) \ ,\label{transf}
\end{equation}
which allows us to eliminate the imaginary part of the evolution
equation of the field $\psi$, which corresponds also to
linearize the expression of the current $j_{_\psi}$.\\
The operator $\cal U$, generating this transformation, is unitary ${\cal U}^\dag={\cal
U}^{-1}$ and is defined by:
\begin{eqnarray}
{\cal U}([\psi^\ast],\,[\psi])=
\exp\left\{i\,\frac{m}{\hbar}\,\sum_{n=0}\,(-1)^n\,\int\frac{1}
{\rho}\frac{\partial^n}{\partial\,x^n}\left[\frac{\partial}{\partial\,S_{n+1}}\,
U([\rho],\,[S])\right]\,dx\right\} \ . \label{gauge}
\end{eqnarray}
If we write the field $\phi$ in terms of the hydrodynamic fields $\rho,\,\sigma$:
\begin{eqnarray}
\phi(x,\,t)=\rho^{1/2}(x,\,t)\,\exp\,\left[\frac{i}{\hbar}\,\sigma(x,\,t)\right] \ ,
\end{eqnarray}
and, due to the unitarity of the transformation, the modulo
of $\phi$ is equal to the modulo of the field $\psi$, while the
phase $\sigma$ is given by:
\begin{equation}
\sigma=S+m\,\sum_{n=0}\,(-1)^n\,\int\frac{1}
{\rho}\frac{\partial^n}{\partial\,x^n}\left[\frac{\partial}{\partial\,S_{n+1}}\,
U([\rho],\,[S])\right]\,dx \ .
\end{equation}
By accepting the statement made by Feynman and Hibbs (\cite{Feynman},
p.96): "{\sl Indeed all measurements of quantum-mechanical systems could be
made to reduce eventually to position and time measurements}", (see also
\cite{Doebner2}) the two wave functions $\psi$ and $\phi$ represent the same physical
system and, as a consequence, we can interpret the Eq. (\ref{transf}) as a
nonlinear gauge transformation of the function described by Eq.
(\ref{sch}).\\
From Eq. (\ref{sch}) and taking into
account the transformation (\ref{transf}), it is easy to obtain the following
evolution equation for the field $\phi$:
\begin{eqnarray}
\nonumber
i\,\hbar\,\frac{\partial\,\phi}{\partial\,t}&=&-\frac{\hbar^2}{2\,m}\,\frac{\partial^2\,\phi}
{\partial\,x^2}+W([\rho],\,[S])\,\phi\\ \nonumber
&-&\frac{1}{2}\,m\,\left\{\sum_{n=0}\,(-1)^n\,\frac{1}{\rho}\,
\frac{\partial^n}{\partial\,x^n}\left[\frac{\partial}{\partial\,S_{n+1}}\,U([\rho],\,[S])
\right]\right\}^2\,\phi\\ \nonumber
&-&m\,\sum_{n=0}\,(-1)^n\,\frac{\partial}{\partial\,t}
\left\{\,\int\frac{1}{\rho}\,\frac{\partial^n}{\partial\,x^n}
\,\left[\frac{\partial}{\partial\,S_{n+1}}\,U([\rho],\,[S])\right]\,dx\right\}\,\phi\\
&-&\sum_{n=0}\,(-1)^n\,\frac{S_1}{\rho}\,\frac{\partial^n}{\partial\,x^n}\left[\frac{\partial}
{\partial\,S_{n+1}}\,U([\rho],\,[S])\right]\,\phi \ .\label{schp}
\end{eqnarray}
Note that the nonlinearity appearing in Eq. (\ref{schp}) is now real.
The continuity equation of the system takes
the form:
\begin{equation}
\frac{\partial\,\rho}{\partial\,t}+\frac{\partial\,j_{_\phi}}{\partial\,x}=0
\end{equation}
where the current $j_{_{\phi}}$ has the standard expression of the
linear quantum mechanics:
\begin{equation}
j_{_{\phi}}=\frac{\sigma_1}{m}\,\rho \ .\label{cur2}
\end{equation}
The gauge transformation (\ref{transf}) and (\ref{gauge}) makes
real the complex nonlinearity in the evolution equation, and makes
non canonical the new dynamical system. However, this
transformation may be useful to describe the evolution of system
by means of an equation containing a real nonlinearity. We remark
that nonlinear transformations have been introduced and used
systematically for the first time in order to study nonlinear
Schr\"odinger equations as the Doebner-Goldin one in Ref.
\cite{Doebner2}.\\ The method here proposed can be found in
literature applied to equations describing systems of collectively
interacting particles.\\ We can quote for instance the canonical
Doebner-Goldin equation \cite{Doebner2,Doebner3} that can be
obtained from (\ref{lagra}) when the potential $U([\rho],\,[S])$
has the following form:
\begin{equation}
U([\rho],\,[S])=\frac{D}{2}\,(\rho_1\,S_1-\rho\,S_2) \
.\label{fokker1}
\end{equation}
A complex nonlinearity is generated in the evolution equation of the
field $\psi$, with real and imaginary part given respectively by:
\begin{eqnarray}
&&W([\rho],\,[S])=-m\,D\,\frac{\partial}{\partial\,x}\,\left(\frac{j_{_{\psi}}}{\rho}\right)
 \ ,\\
&&{\cal
W}([\rho],\,[S])=\frac{\hbar\,D}{2\,\rho}\,\frac{\partial^2\,\rho}{\partial\,x^2}
\ .
\end{eqnarray}
The quantum current $j_{_\psi}$ takes the form of a Fokker-Planck current:
\begin{equation}
j_{_{\psi}}=\frac{S_1}{m}\,\rho+D\,\rho_1 \ ,
\end{equation}
resulting to be the sum of two terms, the former is a drift current while the latter
is a Fick current. The generator of the transformation
$\cal U$ (\ref{gauge}) takes the form:
\begin{equation}
{\cal
U}([\psi^\ast],\,[\psi])=\exp\left(\frac{i}{\hbar}\,m\,D\,\log\rho\right)
\ .\label{tra1}
\end{equation}
This is a particular case of a class of transformations
introduced by Doebner and Goldin and permits to write the evolution
equation:
\begin{equation}
i\,\hbar\,\frac{\partial\,\phi}{\partial\,t}=-\frac{\hbar^2}{2\,m}\,\frac{\partial^2\,\phi}
{\partial\,x^2}+2\,m\,D^2\,\frac{1}{\rho^{1/2}}\,\frac{\partial^2\,\rho^{1/2}}{\partial\,x^2}\,\phi
 \ .\label{27}
\end{equation}
Equation (\ref{27}) was studied in Ref. \cite{Guerra}; after rescaling:
$\sigma\rightarrow\sqrt{1-(2\,m\,D/\hbar)^2}\,\sigma$,
it reduces to the linear Schr\"odinger equation.\\

Now we describe the nonlinear gauge transformation applied to the
Schr\"odinger equation with EIP. The system that we want transform
is gives by Eq. (\ref{schroedinger6}) that we rewrite here in
1-dimensional space:
\begin{equation}
i\,\hbar\,\frac{\partial\,\psi}{\partial\,t}=-\frac{\hbar^2}{2\,m}\,\frac{\partial^2\,\psi}{\partial\,x^2}
+\Lambda(\rho,\,j_{_\psi})\,\psi+F(\rho)\,\psi+V\,\psi \ ,
\end{equation}
with:
\begin{eqnarray}
\Lambda(\rho,\,j_{_\psi})=W(\rho,\,j_{_\psi})+i\,{\cal
W}(\rho,\,j_{_\psi}) \ ,\label{lambda1}
\end{eqnarray}
and
\begin{eqnarray}
W(\rho,\,j_{_\psi})&=&k\,\frac{m}{\rho}\,\left(\frac{j_{_\psi}}{1+\kappa\,\rho}\right)^2
\ ,\\ {\cal
W}(\rho,\,j_{_\psi})&=&-\kappa\,\frac{\hbar}{2\,\rho}\,\nabla
\left(\frac{j_{_\psi}\,\rho}{1+\kappa\,\rho}\right) \ .
\end{eqnarray}
Therefore, we introduce the unitary gauge transformation ${\cal
U}$ for the field $\psi$:
\begin{equation}
\psi(x,\,t)\rightarrow\phi(x,\,t)={\cal
U}([\rho],\,[S])\,\psi(x,\,t) \ ,\label{un}
\end{equation}
which acts on the phase of $\psi$ as:
\begin{equation}
\frac{\partial\,S}{\partial\,x}\rightarrow\frac{\partial\,\sigma}{\partial\,x}
=\frac{\partial\,S}{\partial\,x}\,(1+\kappa\,\rho)
 \ .
\end{equation}
It is easy to see that $\cal U$ is given by:
\begin{equation}
{\cal U}([\rho],\,[S])=\exp\left(i\,\frac{\kappa}{\hbar}\,\int
\rho\,\frac{\partial\,S}{\partial\,x}\,dx\right) \ .\label{unit}
\end{equation}
The current $j_{_\phi}$, associated to the new field $\psi$, takes
now the standard form of the linear quantum mechanics:
\begin{equation}
j_{_\phi}={1\over m}\frac{\partial\,\sigma}{\partial\,x}\,\rho \ ,
\end{equation}
while the continuity equation is written as:
\begin{equation}
\frac{\partial\,\rho}{\partial\,t}+\frac{\partial\,j_{_{\phi}}}{\partial\,x}=0
\ .
\end{equation}
The evolution equation for the field $\phi$ is again nonlinear:
\begin{eqnarray}
\nonumber
i\,\hbar\,\frac{\partial\,\phi}{\partial\,t}=-\frac{\hbar^2}{2\,m}\,\frac{\partial^2\,\phi}
{\partial\,x^2}+\widetilde{\Lambda}(\rho,\,j_{_\phi})\,\phi+F(\rho)\,\phi+V\,\phi
\ ,
\end{eqnarray}
with
\begin{eqnarray}
\widetilde{\Lambda}(\rho,\,j_{_\phi})= \kappa\,m\,\frac{
j_{_\phi}^2}{\rho\,(1+\kappa\,\rho)}\,\phi
-\kappa\,\frac{\hbar^2}{4\,m}\,\left[\frac{\partial^2\,\rho}{\partial\,x^2}-
{1\over\rho}\,\left(\frac{\partial\,\rho}{\partial\,x}\right)^2
\right]\,\phi \ ,
\end{eqnarray}
but now it is a real quantity. Note also that the transformation
(\ref{un}) does not affect the extra nonlinearity $F(\rho)$
because $|\psi|^2=|\phi|^2=\rho$. Of course the price that we pay
to make real the quantity $\Lambda(\rho,\,j)$ given by
(\ref{lambda1}) is that the new system, described by $\phi$, is
noncanonical because of the nonlinearity of the transformation.\\


\setcounter{chapter}{4}
\setcounter{section}{0}
\setcounter{equation}{0}
\chapter*{Chapter IV\\
\vspace{10mm}EIP-Gauged Schr\"odinger Model}
\markright{Chap. IV - EIP-Gauged Schr\"odinger Model}

In this chapter we introduce the gauged EIP-Schr\"odinger model
describing collective effects in interacting particles systems
with the Lagrangian (\ref{lagrangiana1}) coupled in a minimal
fashion with gauge fields $A_\mu$ that takes values in the abelian
group $U(1)$.\\ Many are the applications of NLSEs coupled with a
gauge field and found in literature. One of the more important
example is given by the Ginzburg-Landau theory of the
superconductivity \cite{Ginzburg}, where the same NLSE of Refs.
\cite {Gross,Pitaevskii} with nonlinearity $\propto\,\rho$ is
coupled with an Abelian gauge field, the interaction of this one
being described by the Maxwell Lagrangian
\cite{Papanicolaou,Stratopoulos,Kaper}.

In the model studied by us, the interaction of gauge fields are
described by the more complicated Maxwell-Chern-Simons Lagrangian (MCS).
Many are the motivation to consider the MCS interaction respect to
the more easy case when only the Maxwell term is present.
It is well known that the Maxwell theory could be defined in any space-time dimension; the
field strength tensor is still the antisymmetric one $F_{\mu\nu}$, the
Maxwell Lagrangian ${\cal L}\propto F_{\mu\nu}\,F^{\mu\nu}$ and the equations of
motion do not change their form. The only difference is in the number of
independent fields contained in the theory. Differently, when the
system is imbedded in a even space-time dimensions, there are two
possible expressions for a first order derivative gauge Lagrangian which are both gauge and Lorentz
invariant. The first one is the
standard Maxwell Lagrangian, while the other, ${\cal
L}\propto\epsilon^{\mu\nu\sigma\tau}\,F_{\mu\nu}\,F_{\sigma\tau}$, can
be shown to be a pure divergence and therefore does not give contribution to the
motion equations. The situation changes drastically when
the dynamics of a physical system is developed in odd space-time
dimensions. In this case the interaction of
gauge fields can be described either by the Maxwell that also the
Chern-Simons (CS) terms.\\
As a consequence, because of the presence of the CS term, the model is
useful to describe physical systems with planar dynamics. At this time
the reader might wonder if our discussion is merely of academic
interest. The answer to the question is negative. In fact, two dimensional
physics can occur in our three-dimensional world. This is
because of the third law of thermodynamics, which states that all the
degrees of freedom are frozen out in the limit of zero temperature; it is
possible to strictly confine the electrons to surfaces. Therefore it may
happen that in a strongly confining potentials, or at sufficiently low
temperatures, the excitation energy in one direction may be much higher
than the average thermal energy of the particles, so that these dimensions
are effectively frozen out.

How it was suggested by Wilczek \cite{Wilczek} the presence
of CS term confers to the system an anyonic behavior \cite{Lerda}, that now obeys
to a non conventional statistics \cite{Wu}. Therefore field
theory in presence of CS coupling can describe phenomenologies in which
particles or elementary excitations could be anyons. This hope has in
fact been realized in the case of the fractional Hall effect where the
quasi-particles are believed to be charged vortices obeying to anyonic
statistics \cite{Laughlin}. Recent experiments \cite{Khurana} seem to confirm the
existence of fractionally charged excitations and hence indirectly of anyons.\\
Another topic in condensed matter physics where CS term is believed to
be correct, is the recently discovered high-$T_c$ superconductors,
characterized by their two-dimensional nature
\cite{Wilczek1,Weinberg1}. This hypothesis on the usefulness of the CS term is confirmed
by the $P$ and $T$ violating symmetries which are observed  in this materials
\cite{Frohlich,Lyons,Weber}. \\
We remember also that CS term is an alternative method respect to the
Proca Lagrangian for giving mass to the gauge field, without breaking the gauge
invariance \cite{Siegel,Schonfeld,Templeton,Hagen,Deser}. Moreover, it provides an example of topological field theory
(for a review see \cite{Birmingham}) since even in a curve space-time,
the action of the CS term has the same form without any additional metric insertions.
This fact has important consequences because, how we will show, the CS term
does not give contribution to the energy-momentum
tensor of the system.

\setcounter{equation}{0}
\section{MCS model with EIP}
Let us introduce the density of Lagrangian of the Schr\"odinger equation obeying
to an exclusion-inclusion principle with Maxwell-Chern-Simons interaction:
\begin{equation}
{\cal L}={\cal L}_{\rm mat}+{\cal L}_{\rm gauge} \ ,  \label{EMCS}
\end{equation}
where ${\cal L}_{\rm mat}$ is obtained from Eq. (\ref{lagrangiana1}) after the substitution:
\begin{eqnarray}
\partial_\mu\rightarrow D_\mu\equiv\partial_\mu+\frac{i\,e}{\hbar\,c}A_\mu
\ ,\label{dcov}
\end{eqnarray}
and takes the final form:
\begin{eqnarray}
\nonumber &&{\cal L}_{\rm
mat}=i\,c\,\frac{\hbar}{2}\left[\psi^\ast\,D_0\psi-\psi\,(D_0\psi)^\ast\right]
-\frac{\hbar^2}{2\,m}\,|{\bfm D }\psi|^2\\
&&+\kappa\,\frac{\hbar^2}{8\,m}[\psi^\ast\,{\bfm
D}\psi-\psi\,({\bfm D} \psi)^\ast]^2-\widetilde
U(\psi^\ast\,\psi)-V\,\psi^\ast\,\psi \ ,\label{glagrangian}
\end{eqnarray}
where we have denoted the spatial component of the covariant
derivative with ${\bfm D}\equiv\bfm\nabla-i\,(e/\hbar\,c)\,{\bfm
A}$, the indices are lower and upper depending on metric tensor in
the Minkowski space $\eta_{\mu\nu}\equiv {\rm diag}(1,\,-1,\,-1)$:
$A^\mu=\eta^{\mu\nu}A_\nu$; $\widetilde U(\rho)$ is an analytic
real potential function of the field $\rho=|\psi|^2$, $V$ is an
external potential, $m$ is the mass parameter and $\kappa$ the
coupling constant for the EIP potential. The quantity:
\begin{equation}
U_{_{\rm EIP}}=-\kappa\,\frac{\hbar^2}{8\,m}\left[\psi^\ast\,
{\bfm D}\psi-\psi\,({\bfm D}\psi)^\ast\right]^2 \ ,
\end{equation}
is the EIP potential with minimal coupling [cfr. Eq. (\ref{eip})].
\\ Of course the model might be study in any spatial dimension $D$
(if $D$ is even the CS term is absent), but for the purposes of
application that will be developed in the following chapters, we
focus our attention in the planar case $D=2$. Because the system
is in (2+1) dimensions, the greek indices take the value
$0,\,1,\,2$ while the latin indices, that assume the value
$1,\,2,$ are the spatial ones. We have, $x_\mu\equiv
(c\,t,\,-{\bfm
x}),~~\partial_\mu\equiv(c^{-1}\,\partial/\partial\,t,\,
{\bfm\nabla})$. In (\ref{dcov}) $c$ is the speed of light and $e$
is the coupling constant of the Abelian gauge field described by
the scalar potential $A^0$ and the vector potential ${\bfm A}$,
with $A_\mu\equiv(A_0,\,-{\bfm A})$. Moreover, we assume the sum
convention when the indices are repeated.

The Lagrangian ${\cal L}_{\rm gauge}$ is given by:
\begin{equation}
{\cal L}_{\rm gauge}=-\frac{\gamma}{4}\,F_{\mu\nu}\,F^{\mu\nu}+\frac{g}{4}
\,\varepsilon^{\tau\mu\nu}\,A_\tau\,F_{\mu\nu} \ ,  \label{MCS}
\end{equation}
with $F_{\mu\nu}=\partial_\mu\,A_\nu-\partial_\nu\,A_\mu$.
The Levi-Civita tensor $\varepsilon^{\tau\mu\nu}$, fully
antisymmetric, is defined as $\varepsilon^{012}=1$.
The parameters $\gamma$ and $g$ in Eq. (\ref{MCS}) give the relative
weight between the Maxwell interaction and the CS one. The
result obtained in this chapter, posing $g=0$ or $\gamma=0$ respectively, hold
also when the Maxwell term or the Chern-Simons one are present alone.
Because of the planarity of the system, the electric field is a vector with component:
\begin{eqnarray}
E^i=-\partial_0\,A^i-
\partial_i\,A^0 \ ,
\end{eqnarray}
while the magnetic field becomes a scalar:
\begin{eqnarray}
B=-\epsilon^{ij}\,\partial_i\,A_j \ .
\end{eqnarray}
In terms of the component of the tensor $F^{\mu\nu}$ we have:
$E^i=F_{0i},\,\,B=-F_{12}$.\\
Now we introduce the action of the system:
\begin{equation}
{\cal A}=\int{\cal L}\,d^2x\,dt \ .
\end{equation}
The motion equations for the matter field $\psi$ and for the gauge fields
$A_\mu$ can be written as:
\begin{eqnarray}
{\rm E}_{\psi^\ast}{\cal A}=0 \ ,\hspace{30mm} {\rm
E}_{A_\mu}{\cal A}=0 \ ,  \label{motion1}
\end{eqnarray}
where E$_{A_\mu}$ is obtained from (\ref{euler}) with
the substitution $\psi\rightarrow A_\mu$.
Explicitly, Eq. (\ref{motion1}) becomes:
\begin{eqnarray}
&&i\,\hbar\,c\,D_0\,\psi =-\frac{\hbar^2}{2\,m}\,{\bfm
D}^2\psi-\kappa\,\frac{\hbar^2}{2\,m}\left[\psi^\ast\,{\bfm D}\psi-\psi\,({\bfm D}
\psi)^\ast\right]\,{\bfm D}\psi\nonumber\\
&&-\kappa\,\frac{\hbar^2}{4\,m}\,{\bfm D}\left[\psi^\ast\,{\bfm D}\psi-\psi\,(
{\bfm D}\psi)^\ast\right]\,\psi+F(\rho)\,\psi+V\,\psi \ .  \label{sch1}
\end{eqnarray}
If we introduce the spatial current:
\begin{equation}
{\bfm J}=-\frac{i\,\hbar}{2\,m}\,(1+\kappa\,\rho)\,\left[\psi^\ast\,{\bfm D}
\psi-\psi\,({\bfm D}\psi)^\ast\right] \ ,\label{covariantcurrent}
\end{equation}
given by $\bfm J=(1+\kappa\,\rho)\,\bfm J_0$, being
\begin{eqnarray}
{\bfm J}_0=-\frac{i\,\hbar}{2\,m}\,[\psi^\ast\,{\bfm D}\psi-\psi\,({\bfm
D}\psi^\ast)]={\bfm j}_0-\frac{e}{m\,c}\,{\bfm A}\,\rho \ ,
\end{eqnarray}
the spatial current of the MCS theory without the EIP, it is
easy to see that Eq. (\ref{sch1}) obeys to the following continuity equation:
\begin{equation}
\frac{\partial\,\rho}{\partial\,t}+\bfm\nabla\cdot{\bfm J}=0 \ .\label{eqco}
\end{equation}
Eq. (\ref{sch1}) can be rewritten in terms of the fields $\rho$ and $\bfm J$
in the form:
\begin{eqnarray}
i\,\hbar\,c\,D_0\,\psi=-\frac{\hbar^2}{2\,m}{\bfm D}^2\psi
+\Lambda(\rho,\,{\bfm J})\,\psi+F(\rho)\,\psi+V\,\psi \ ,\label{materfield}
\end{eqnarray}
where the complex nonlinearity $\Lambda(\rho,\,{\bfm J})$ is given by:
\begin{equation}
\Lambda(\rho,\,{\bfm J})=\kappa\,\frac{m
}{\rho}\left(\frac{{\bfm J}}{1+\kappa\,\rho}\right)^2
-i\,\kappa\,\frac{\hbar}{2\,\rho}\,{\bfm D} \left(\frac{{\bfm J}\,\rho}{
1+\kappa\,\rho}\right) \ .\label{lambda}
\end{equation}
The motion equations for the fields $A_\mu$, given from the second of
Eqs. (\ref{motion1}), are:
\begin{equation}
\gamma\,\partial_\mu\,F^{\mu\nu}+\frac{g}{2}\,\varepsilon^{\nu\tau\mu}
\,F_{\tau\mu}=\frac{e}{c}\,J^\nu \ ,  \label{gaugefield}
\end{equation}
where the covariant current $J^\nu$ is given by:
$J^\nu\equiv(c\,\rho,\,{\bfm J})$.\\ Eq. (\ref{gaugefield})
requires same comments. For $g=0$ we recognize the standard
Maxwell equations with sources. It is well known that this
equation does not admit any trivial solution also in absence of
the matter field ${\bfm J}$. Differently, when we set $\gamma=0$
the Chern-Simons equations for the gauge fields are obtained. This
equations does not admit trivial solutions only in presence of
matter. In fact how we will show in section 4.2, in absence of the
Maxwell term, the gauge fields can be expressed as nonlinear
functions of the field $\psi$, and vanish in absence of the matter
field. Therefore, when the Maxwell term is present, the gauge
fields describe dynamical fields with proper degrees of freedom,
while in absence of this one, the CS term can be see as a
constrain which the matter field must obey. Is this constrain that
confer to the system an anyonic behavior.\\ Now,we perform the
gauge transformations:
\begin{eqnarray}
A_{\mu}&\rightarrow& A_{\mu}+\partial_{\mu}\omega \ ,  \label{gauge1} \\
\psi&\rightarrow& e^{-i(e/\hbar\,c)\,\omega}\psi \ ,  \label{gauge2}
\end{eqnarray}
where $\omega$ is a well-behaved function so that $\epsilon^{\mu\nu}
\partial_\mu\,\partial_\nu\,\omega=0$ (with $\epsilon^{\mu\nu}=-\epsilon^{
\nu\mu}$). The Lagrangian (\ref{EMCS}) changes as:
\begin{equation}
{\cal L}\rightarrow{\cal
L}+\frac{g}{4}\,\epsilon^{\mu\nu\tau}\,\partial_\mu\left(\omega\,F_{\nu\tau}\right)
 \ ,
\end{equation}
with an extra surface term that does not change the motion equations of
the fields $\psi$ and $A_\mu$.
This property of the system is typical in the presence of CS term and
continues to be valid also when the EIP interaction is introduced.
From Eq. (\ref{gaugefield}), contracting with the differential operator $\partial_\nu$,
we may obtain again the conservation law of the current (\ref{eqco}) in the
covariant form:
\begin{equation}
\partial_\nu\,J^\nu=0 \ .  \label{continuity}
\end{equation}
The time component of Eq. (\ref{gaugefield}) is the Gauss law:
\begin{equation}
\gamma\,\bfm\nabla\cdot{\bfm E}-g\,B=e\,\rho \ ,  \label{Gauss}
\end{equation}
which is the expression usually reported in literature, because the EIP
does not modify the time component of the current.\\
After integration on the whole plane of Eq. (\ref{Gauss}), and taking into
account that the CS term is dominant over the Maxwell term at long
distance \cite{Jackiw,Pi1}, we obtain the
important property that every configuration with charge $Q=e\,\int\rho(t,\,{\bfm x}
)\,d^2x$ transports also a magnetic flux $\Phi=\int B(t,\,{\bfm x})\,d^2x$:
\begin{equation}
-g\,\Phi=Q \ .\label{Gauss1}
\end{equation}
This relation suggests the following interpretation:
The system described by the Lagrangian (\ref{EMCS}) can be interpreted
as analogous to a system of magnetic monopoles in (2+1) dimension
obeying to a generalized exclusion-inclusion principle in the
configuration space.\\
Finally, the spatial component of (\ref{gaugefield}) is the Amp\`ere law
in $(2+1)$ dimensions with CS contribution:
\begin{equation}
\gamma\,\bfm\nabla\wedge B-\frac{\gamma}{c}\,\frac{\partial\,{\bfm E}}{\partial\,t
}-g\,{\bfm E}^\ast =\frac{e}{c}\,{\bfm J} \ ,  \label{ampere}
\end{equation}
where ${\bfm E}^\ast$ is the dual vector of the electric field with components
$E_i^\ast=\epsilon_{ij}\,E_j$.\\

\setcounter{equation}{0}
\section{Hamiltonian formulation}
The Hamiltonian formulation is crucial for the construction of
self-dual solution which we have analyzed in the last chapter and
for a variety of other purposes. In presence of gauge fields the
Hamiltonian formulation reserves a special treatment. In fact,
because of the non vanishing Maxwell term ($\gamma\not=0$), the
Lagrangian (\ref{glagrangian}), (\ref{MCS}) is degenerate in the
velocity and the system can be described only as a constrained
Hamiltonian, the constraint being given by the Gauss law. In its
treatment we follow the Dirac-Bergmann approach \cite{Dirac}.

The field $\pi_{_\phi}$ canonically conjugated of the field $\phi$, is
given by Eq. (\ref{defmomenti}): $
\pi_{_\phi}=\partial{\cal L}/\partial\dot\phi$
where the dot indicates the time derivative. Taking into account the
expression of the Lagrangian given by Eqs. (\ref{glagrangian}), (\ref{MCS})
and setting for $\phi$ the fields
$\psi,\,\psi^\ast,\,A_\mu$, we obtain the following expressions:
\begin{eqnarray}
\pi_{_\psi}&=&i\,\frac{\hbar}{2}\,\psi^\ast \ ,  \label{conjugate1} \\
\pi_{_{\psi^\ast}}&=&-i\,\frac{\hbar}{2}\,\psi \ ,  \label{conjugate2} \\
\Pi_{\mu}&=&{\frac{\gamma}{c}}F_{\mu0}+\frac{g}{2\,c}\epsilon_{0\mu\nu}\,A^\nu \
. \label{conjugate3}
\end{eqnarray}
Eqs. (\ref{conjugate1}) and (\ref{conjugate2}) give rise to
primary derivative:
\begin{eqnarray}
&&\xi_1=\pi_{_\psi}-i\,\frac{\hbar}{2}\,\psi^\ast \ ,\label{cost1}\\
&&\xi_2=\pi_{_{\psi^\ast}}+i\,\frac{\hbar}{2}\,\psi \ ,\label{cost2}
\end{eqnarray}
while Eq. (\ref{conjugate3}) show that $\Pi_{0}$ vanishes and therefore
it is a third primary constraint.
Performing the Legendre transformation:
\begin{equation}
{\cal
H}=\pi_{_\psi}\,\dot\psi+\pi_{_{\psi^\ast}}\,\dot\psi^\ast+\Pi_{i}\,\dot A^i-{\cal L} \ ,
\end{equation}
we obtain the Hamiltonian density of the system:
\begin{equation}
{\cal H}={\cal H}_{\rm mat}+{\cal H}_{\rm gauge} \ ,\label{htot}
\end{equation}
where the Hamiltonian of the matter field ${\cal H}_{\rm mat}$ is:
\begin{eqnarray}
{\cal H}_{\rm mat}=\frac{\hbar^2}{2\,m}\,|{\bfm
D}\psi|^2-\kappa\,\frac{\hbar^2}{ 8\,m}\left[\psi^\ast\,{\bfm
D}\psi-\psi\,({\bfm D}\psi)^\ast\right]^2 +\widetilde
U(\psi^\ast\,\psi)+V\,\psi^\ast\,\psi \ , \label{ghamilton}
\end{eqnarray}
while the one of the gauge field ${\cal H}_{\rm gauge}$ is:
\begin{eqnarray}
{\cal H}_{\rm gauge}={\frac{\gamma}{2}}\,({\bfm E}^2+B^2)+A_0\,(g\,B-\gamma\,\bfm\nabla
\cdot{\bfm E}+e\,\rho)
+\partial_i\,(A_0\,\Pi^{i}) \ .\label{mcsham}
\end{eqnarray}
Let us introduce now the Poisson brackets (cfr. section 1.3)
between two functionals:
\begin{eqnarray}
\nonumber \{f({\bfm x}),\,g({\bfm y})\}&=&\int
\left[\frac{\delta\,f( {\bfm x})}{\delta\,\psi({\bfm {z})}}
\frac{\delta\,g({\bfm y})}{ \delta\,\pi_{_{\psi}}({\bfm {z})}}-
\frac{\delta\,f({\bfm x})}{\delta\,\pi_{_{\psi}}({\bfm z)}}
\frac{\delta\,g({\bfm y})}{\delta\,\psi({\bfm {z})}}\right]\,d^2z
\\ \nonumber &+&\int \left[\frac{\delta\,f( {\bfm
x})}{\delta\,\psi^\ast({\bfm {z})}} \frac{\delta\,g({\bfm y})}{
\delta\,\pi_{_{\psi^\ast}}({\bfm {z})}}- \frac{\delta\,f({\bfm
x})}{\delta\,\pi_{_{\psi^\ast}}({\bfm z)}} \frac{\delta\,g({\bfm
y})}{\delta\,\psi^\ast({\bfm {z})}}\right]\,d^2z \\
&+&\int\left[\frac{\delta\,f({\bfm x})}{\delta\,A^\mu({\bfm {z})}}
\frac{\delta\,g({\bfm y})}{\delta\,\Pi_{\mu}({\bfm {z})}}-
\frac{\delta\,f( {\bfm x})}{\delta\,\Pi_{\mu}({\bfm {z})}}
\frac{\delta\,g({\bfm y})}{ \delta\,A^\mu({\bfm
{z})}}\right]\,d^2z \ . \label{poi}
\end{eqnarray}
The second class primary derivative (\ref{cost1}) and
(\ref{cost2}) satisfy the relation:
\begin{eqnarray}
\{\xi_1(t,\,{\bfm x}),\,\xi_2(t,\,{\bfm y})\}=-i\,\hbar\,\delta^{(2)}({\bfm x}-
{\bfm y})
\end{eqnarray}
and can be accommodated by the introduction of the Dirac brackets:
\small
\begin{eqnarray}
\nonumber \{f({\bfm x}),\,g({\bfm y})\}_D=\{f({\bfm x}),\,g({\bfm
y})\}&+&\frac{i}{\hbar}\,\int\{f({\bfm x}),\,\xi_1({\bfm z})\}\,\{
\xi_2({\bfm z}),\,g({\bfm y})\}\,d^2z\\
&-&\frac{i}{\hbar}\,\int\{f({\bfm x}),\,\xi_2({\bfm z})\}\,\{
\xi_1({\bfm z}),\,g({\bfm y})\}\,d^2z \ .\label{bp}
\end{eqnarray}
\normalsize
Expression (\ref{bp}) can be simplified if one solves
the relation (\ref{cost1}), (\ref{cost2}) for $\pi_{_{\psi}}$ and
$\pi_{_{\psi^\ast}}$ and treats $f$ and $g$ as functionals of
$\psi$ and $\psi^\ast$ only. A straightforward calculations give:
\begin{eqnarray}
\nonumber \{f({\bfm x}),\,g({\bfm y})\}_D=&-&\frac{i}{\hbar}\,\int
\left[\frac{\delta\,f( {\bfm x})}{\delta\,\psi({\bfm {z})}}
\frac{\delta\,g({\bfm y})}{ \delta\,\psi^\ast({\bfm {z})}}-
\frac{\delta\,f({\bfm x})}{\delta\,\psi^\ast({\bfm z)}}
\frac{\delta\,g({\bfm y})}{\delta\,\psi({\bfm {z})}}\right]\,d^2z
\\
&+&\int\left[\frac{\delta\,f({\bfm x})}{\delta\,A^\mu({\bfm {z})}}
\frac{\delta\,g({\bfm y})}{\delta\,\Pi_{\mu}({\bfm {z})}}-
\frac{\delta\,f( {\bfm x})}{\delta\,\Pi_{\mu}({\bfm {z})}}
\frac{\delta\,g({\bfm y})}{ \delta\,A^\mu({\bfm
{z})}}\right]\,d^2z \ . \label{bp1}
\end{eqnarray}
The constraint $\Pi_{0}=0$ is first class and the requirement
of its conservation leads to a secondary constraint:
\begin{eqnarray}
\eta=\{\Pi_{0},\,H\}=\gamma\,{\bfm\nabla}\cdot{\bfm E}-g\,B-e\,\rho=0 \ ,\label{eta}
\end{eqnarray}
where $H=\int{\cal H}\,d^2x$ is the Hamiltonian of the system given by (\ref{htot}). The secondary
constraint $\eta=0$ does not involve any further constraint since, how is easy to
verify, $\{\eta,\,H\}_D=0$.\\
The total Hamiltonian is now $H_T=H+\int\lambda\,\Pi_{_0}\,d^2x$, where $\lambda$ is a Lagrange multiplier.
How it is well known, $A_0$ and $\Pi_{_0}$ have no physical meaning, and $\Pi_{_0}=0$ for all the time,
while $A_0$ can take arbitrary values.
Accordingly we may drop them out from the set of dynamical variables of the models.
This can be accomplished by discarding the term $\lambda\,\Pi_{_0}$ in $H_T$, the only role of which is to let
$A_0$ vary arbitrarily, and by treating $A_0$ as an arbitrary multiplier.
As a result, the Hamiltonian of the gauge fields
${\cal H}_{\rm gauge}$ becomes:
\begin{equation}
{\cal H}_{\rm gauge}={\frac{\gamma}{2}}\,({\bfm E}^2+B^2) \ ,\label{mcshamilton}
\end{equation}
and the total Hamiltonian density ${\cal H}_{\rm mat}+{\cal
H}_{\rm gauge}$ is given in terms of canonical fields
$\psi,\,\psi^\ast,\,A_i,\,\Pi_{i}$ with $i=1,\,2$. We note that in
${\cal H}_{\rm gauge}$ the coupling constant $g$, introduced by
the CS interaction, does not appear.\\ So far we have assumed
$\gamma\not=0$. In the pure CS model ($\gamma=0$) the situation is
somewhat different. The theory has two more second class primary
constraint:
\begin{eqnarray}
&&\xi_3=\Pi_{1}+\frac{g}{2\,c}\,A_2=0 \ ,\label{cost3}\\
&&\xi_4=\Pi_{2}-\frac{g}{2\,c}\,A_1=0 \ ,\label{cost4}
\end{eqnarray}
resulting from the definition of momenta $\Pi_{i}$. As
\begin{eqnarray}
\{\xi_3(t,\,{\bfm x}),\,\xi_4(t,\,{\bfm y})\}=\frac{g}{c}\,\delta^{(2)}
({\bfm x}-{\bfm y}) \ ,
\end{eqnarray}
these can be accommodated by modifying the brackets: \small
\begin{eqnarray}
\nonumber
&&\{f({\bfm x}),\,g({\bfm y})\}_D=\{f({\bfm x}),\,g({\bfm y})\}\\
\nonumber
&&+\frac{i}{\hbar}\,\int\left[\{f({\bfm x}),\,\xi_1({\bfm z})\}\,\{
\xi_2({\bfm z}),\,g({\bfm y})\}-\{f({\bfm x}),\,\xi_2({\bfm z})\}\,\{
\xi_1({\bfm z}),\,g({\bfm y})\}\right]\,d^2z\\
&&-\frac{c}{g}\,\int\left[\{f({\bfm x}),\,\xi_3({\bfm z})\}\,\{
\xi_4({\bfm z}),\,g({\bfm y})\}-\{f({\bfm x}),\,\xi_4({\bfm z})\}\,\{
\xi_3({\bfm z}),\,g({\bfm y})\}\right]\,d^2z \ .\label{bd2}
\end{eqnarray}
\normalsize By solving Eqs. (\ref{cost3}) and (\ref{cost4}) for
$\Pi_{1}$ and $\Pi_{2}$, this can be trough to the form:
\begin{eqnarray}
\nonumber \{f({\bfm x}),\,g({\bfm y})\}=&-&\frac{i}{\hbar}\,\int
\left[\frac{\delta\,f( {\bfm x})}{\delta\,\psi({\bfm {z})}}
\frac{\delta\,g({\bfm y})}{ \delta\,\psi^\ast({\bfm
{z})}}-\frac{\delta\,f({\bfm x})}{\delta\,\psi^\ast({\bfm z)}}
\frac{\delta\,g({\bfm y})}{\delta\,\psi({\bfm {z})}}\right]\,d^2z
\\ &+&\int\left[\frac{\delta\,f({\bfm x})}{\delta\,A_1({\bfm
{z})}} \frac{\delta\,g({\bfm y})}{\delta\,A_2({\bfm {z})}}-
\frac{\delta\,f( {\bfm x})}{\delta\,A_2({\bfm {z})}}
\frac{\delta\,g({\bfm y})}{ \delta\,A_1({\bfm {z})}}\right]\,d^2z
\ , \label{bp3}
\end{eqnarray}
where $f$ and $g$ are considered as functionals of only
$\psi,\,\psi^\ast,\, A_1$ and $A_2$.\\ In the pure CS case a
further reduction is possible. In fact as we will show in the next
section, one can explicitly solve the derivative (\ref{cost3}) and
(\ref{cost4}) and end with $\psi$ and $\psi^\ast$ as the only
canonical variables. In the general case, however, the fact that
the constraint involves both $\Pi_{i}$ and $A_i$ makes such
reduction impossible. Using the Poisson brackets "{\sl naked}"
from the constraint we can evaluate the evolution equations of the
fields $\psi$ and $\psi^\ast$:
\begin{eqnarray}
i\,\hbar\,\frac{\partial\psi}{\partial\, t}&=&\frac{\delta\,H}{
\delta\,\psi^\ast} \ ,\\
i\,\hbar\,\frac{\partial\psi^\ast}{\partial\, t}&=&-\frac{\delta\,H}{
\delta\,\psi} \ ,
\end{eqnarray}
that are equal to Eq. (\ref{materfield}) and its conjugate.
The analogous equations for the fields $A_i$ and $\Pi_{i}$ become:
\begin{eqnarray}
\frac{\partial\,A^i}{\partial\, t}&=&\frac{\delta\,H}{\delta\,\Pi_{i}} \ ,
\\
\frac{\partial\,\Pi_{i}}{\partial\, t}&=&-\frac{\delta\,H}{\delta\,A^i} \ ,
\end{eqnarray}
and correspond respectively  to Eq. (\ref{conjugate3}) and Eq.
(\ref{gaugefield}).

Let us now define the following quantities:
\begin{eqnarray}
&&N=\int\psi^\ast\,\psi\,d^2x \ ,\label{co1}\\ &&<{\bfm
x}_c>={1\over N}\,\int\psi^\ast\,{\bfm x}\,\psi\,d^2x \
,\label{co2}\\ &&<{\bfm
P}>=-i\,{\frac{\hbar}{2}}\,\int\left[\psi^\ast\,{\bfm D}\psi
-\psi\,( {\bfm D}\psi)^\ast\right]\,d^2x \ ,  \label{co3} \\
&&<M>=-i\,{\frac{\hbar}{2}}\,\int{\bfm
x}\wedge\left[\psi^\ast\,{\bfm D} \psi-\psi\,({\bfm
D}\psi)^\ast\right]\,d^2x \ ,  \label{co4} \\ &&E=\int{\cal
H}\,d^2x \ ,\label{co5}
\end{eqnarray}
that represent, respectively, the particle number, the position of
mass center, the linear momentum, the angular momentum and the energy of
the system.
The time evolution of these quantities is given by:
\begin{equation}
\frac{d\,f}{d\,t}=\{f,\,H\}_D+\frac{\partial\,f}{\partial\,t} \ ,
\label{poisson}
\end{equation}
for a generic functional $f(\psi,\,\psi^\ast,\,A_i,\,\Pi_{_i})$,
where the last term in the right hand side takes into account the
explicit time dependence of the functional $f$.
After the computation we are left with the following Ehrenfest relations:
\begin{eqnarray}
&&\frac{d}{d\,t}\,N=0 \ ,  \label{eh1} \\ &&\frac{d}{d\,t}<{\bfm
x}_c>={1\over N}\,\int{\bfm J}\,d^2x \ ,  \label{eh2} \\
&&\frac{d}{d\,t}<{\bfm P}>=\int {\bfm
f}\,d^2x-\int\psi^\ast\,{\bfm\nabla} V\,\psi\,d^2x \ , \label{eh3}
\\ \nonumber &&\frac{d}{d\,t}<M>=\int {\bfm x}\wedge{\bfm
f}\,d^2x-\int\psi^\ast\,\left({\bfm x} \wedge{\bfm\nabla}
V\right)\,\psi\,d^2x \ ,\label{eh4} \\
&&\frac{d}{d\,t}\,E=\int\psi^\ast\,\frac{d\,V}{d\,t}\,\psi\,d^2x \
,  \label{eh5}
\end{eqnarray}
where the quantity:
\begin{equation}
{\bfm f}=e\,\rho+\frac{e}{c}\,{\bfm J}\wedge\,B
\end{equation}
is the density of the Lorentz force with EIP. We note that Eqs.
(\ref{eh2})-(\ref{eh4}) are formally identical to the analogous
relations of the theory without the EIP ($\kappa=0$). The
difference is in the expression of the current: ${\bfm J}= {\bfm
J}_0\,(1+\kappa\,\rho)$. From Eq. (\ref{eh1}) we see that the
number of particles of the system is a constant of motion and from
(\ref{eh5}) we see that for external time-independent potential
also the energy is preserved. Note also that EIP potential does
not appear explicitly in the relations (\ref{eh3}) and (\ref{eh4})
as shown in section 2.3. The same statement is true for the
nonlinear potential $\widetilde U(\rho)$ when it has a good
behavior at infinity. Without the external force the linear and
angular momentum are constants of motion. In the presence of a
central force, while the linear momentum is not conserved, we have
the conservation of the angular momentum. Finally, it is trivial
to note that from Eqs. (\ref{Gauss1}) and (\ref{eh1}) we have the
conservation of charge and magnetic flux attached to the system.

\subsection{Derivation of Ehrenfest relations}
The Ehrenfest relations (\ref{eh1})-(\ref{eh5}) for the observable
(\ref{co1})-(\ref{co5}) can be obtained following the computation reported
in section 2.3.2 for the analogue quantity without the gauge coupling.
The only quantities that involve more complicate steps are Eqs.
(\ref{eh3}) and (\ref{eh4}). Therefore we derive in a detailed way only
these two relations.

In the following we make the hypothesis that the fields vanish steeply at
infinity and that we can neglect surface terms.\\
The first relation, Eq. (\ref{eh1}), is a trivial consequence of the
continuity equation (\ref{eqco}), while the (\ref{eh2}) is easily obtained
from Eq. (\ref{eqco}) as:
\begin{eqnarray}
\frac{d}{d\,t}<{\bfm x}_c>={1\over N}\int{\bfm
x}\,\frac{d\,\rho}{d\,t}\,d^2x =-{1\over N}\int{\bfm
x}\,{\bfm\nabla}\cdot{\bfm J}\,d^2x={1\over N}\int{\bfm J}\,d^2x \
,
\end{eqnarray}
where we have performed an integration by parts in the last
step.\\ Relation (\ref{eh5}) is immediately obtained posing $f=E$
being $E$ the Hamiltonian (\ref{co5}) and taking into account the
property of the Poisson brackets $\{f,\,f\}=0$ valid for every
functional $f$. Eq. (\ref{poisson}) becomes:
\begin{equation}
\frac{d\,f}{d\,t}=\frac{\partial\,f}{\partial\,t} \ ,
\end{equation}
that is the relation (\ref{eh5}), if we take into account that the
explicit time dependence in the Hamiltonian can occur only in the
external potential $V(t,\,{\bfm x})$.

Let us consider now Eq. (\ref{eh3}) for the component $i$ of the
momentum. Using Eqs. (\ref{poisson}), (\ref{bp1}) for $f=<P^i>$
given by (\ref{co3}) and taking into account the following
functional derivative:
\begin{eqnarray}
\frac{\delta f}{\delta\,\psi}&=&-i\hbar\,(D^i\psi)^\ast \ ,\\
\nonumber &&\\ \frac{\delta
H}{\delta\,\psi^\ast}&=&\frac{\hbar^2}{2\,m}\,D_j\,D^j \psi
+\kappa\,\frac{\hbar^2}{2\,m}\left[\psi^\ast\,D^j\psi-\psi\,(D^j
\psi)^\ast\right]\,D_j\psi\nonumber\\
&+&\kappa\,\frac{\hbar^2}{4\,m}\,D_j\left[\psi^\ast\,D^j\psi-\psi\,(
D^j\psi)^\ast\right]\,\psi +F(\rho)\,\psi+V\,\psi \ ,\\ \nonumber
&&\\ \frac{\delta f}{\delta\,A^i}&=&-\frac{e}{c}\,\rho \ ,\\
\nonumber &&\\ \frac{\delta f}{\delta\,\Pi_{i}}&=&0 \ , \\
\nonumber &&\\ \frac{\delta H}{\delta\,\Pi_{i}}&=&-c\,F^{i0} \ ,
\end{eqnarray}
we obtain:
\begin{eqnarray}
\frac{d}{d\,t}<P^i>&=&\int\Biggl\{-\frac{\hbar^2}{2\,m}
\left[(D^i\psi)^\ast\,D_jD^j\psi+D^i\psi\,(D_jD^j\psi)^\ast\right]\nonumber\\
&+&\kappa\,\frac{\hbar^2}{2\,m}\left[\psi^\ast\,D^j\psi-\psi\,(D^j\psi)^\ast\right]
\left[(D_j\psi)^\ast\,D^i\psi-D_j\psi\,(D^i\psi)^\ast\right]\nonumber\\
&+&\kappa\,\frac{\hbar^2}{4\,m}\,D_j\,\left[\psi^\ast\,D^j\psi-\psi\,(D^j\psi)^\ast\right]
\left[\psi^\ast\,D^i\psi-\psi\,(D^i\psi)^\ast\right]\nonumber\\
&-&\left[F(\rho)+V\right]\,\partial^i\,\rho+e\,\rho\,F^{i0}\Biggr\}\,d^2x
\ .\label{a71}
\end{eqnarray}
\normalsize
Integrating by parts two times the first term and one time the third in
the right hand side of (\ref{a71}) we obtain:
\small
\begin{eqnarray}
\nonumber &&\frac{d}{d\,t}<P^i>=\int\Biggl\{-\frac{\hbar^2}{2\,m}
\left[(D_jD^jD^i\psi)^\ast\,\psi+D^i\psi\,(D_jD^j\psi)^\ast\right]\\
\nonumber
&&+\kappa\,\frac{\hbar^2}{4\,m}\left[\psi^\ast\,D^j\psi-\psi\,(D^j\psi)^\ast\right]\,
\left[(D_j\psi)^\ast\,D^i\psi-D_j\psi\,(D^i\psi)^\ast -\psi^\ast
D_j^{~i}\psi+\psi\,(D_j^{~i}\psi)^\ast\right]\\
&&-\left[F(\rho)+V\right]\,\partial^i\,\rho+e\,\rho
\,F^{i0}\Biggr\}\,d^2x \ .\label{tre}
\end{eqnarray}
\normalsize Starting from the identity:
\small
\begin{eqnarray}
(D_jD^jD^i\psi)^\ast\,\psi+D^i\psi\,(D_jD^j\psi)^\ast
&=&(D^iD_jD^j\psi)^\ast\,\psi+\left([D_j,\,D^i]\,D^j\psi\right)^\ast\,\psi\nonumber\\
&+&\left(D^j[D_j,\,D^i]\psi\right)^\ast\,\psi+D^i\psi\,(D_jD^j\psi)^\ast \ ,\label{a9}
\end{eqnarray}
\normalsize integrating by parts the third term in the right hand
side of (\ref{a9}) and noting that:
\begin{equation}
[D_j,\,D^i]=-\frac{i\,e}{\hbar\,c}\left(\partial_jA^i-\partial^iA_j
\right) \ ,
\end{equation}
we have:
\small
\begin{eqnarray}
(D_jD^jD^i\psi)^\ast\,\psi+D^i\psi\,(D_jD^j\psi)^\ast
=\frac{i\,e}{\hbar\,c}\left(\partial_jA^i-\partial^iA_j\right)\,
\left[\psi^\ast\,D^j\psi-\psi\,(D^j\psi)^\ast\right] \ ,\label{uno}
\end{eqnarray}
\normalsize and moreover:
\small
\begin{eqnarray}
\left[(D_j\psi)^\ast\,D^i\psi-D_j\psi\,(D^i\psi)^\ast-\psi^\ast
D_j^{~i}\psi+\psi\,(D_j^{~i}\psi)^\ast\right]
=-i\,\frac{2\,e}{\hbar\,c}\left(\partial_jA^i-\partial^iA_j\right)\,\rho
 \ .\label{due}
\end{eqnarray}
\normalsize Inserting relations (\ref{uno}), (\ref{due}) in
(\ref{tre}) we can obtain the relation:
\small
\begin{eqnarray}
\frac{d}{d\,t}<P^i>&=&\int\Biggl\{-\frac{i\,e\,\hbar}{2\,m\,c}
\left(\partial_jA^i-\partial^iA_j\right)
\left[\psi^\ast\,D^j\psi-\psi\,(D^j\psi)^\ast\right]
\,(1+\kappa\,\rho)+e\,\rho\,F^{i0}\Biggr\}\,d^2x\nonumber\\
&+&\int\rho\,\partial^i\,F(\rho)\,d^2x+\int\rho\,\partial^i\,V\,d^2x
\ .\label{a131}
\end{eqnarray}
\normalsize By taking into account the relation $F=d\widetilde
U/d\rho$, the last two integrals in (\ref{a131}) can be written
as:
\begin{equation}
\int\partial^i\,(\rho\,F-\widetilde U)\,d^2x \ ,
\end{equation}
and therefore, if the potential and the field $\rho$ have a good behavior at
infinity, it can be ignored. Then Eq. (\ref{a131}) is
equals Eq. (\ref{eh3}) if we take into account the
definition of electric and magnetic fields and of current (\ref{covariantcurrent}).\\
Eq. (\ref{eh4}) can be obtained following the same iter used to deduce the relation (\ref{eh3}),
we leave the easy but tedious computation to the reader.\\

\setcounter{equation}{0}
\section{Conservation laws}
Let us set, in the Ehrenfest relations (\ref{eh1})-(\ref{eh5}),
the external potential $V=0$, we can deduce the following
conserved quantities:
\begin{eqnarray}
&&\frac{d}{d\,t}\,N=0 \ ,\hspace{10mm}N=\int\rho\,d^2x \ ,\label{c1}\\
&&\frac{d}{d\,t}{\bfm P}_{\rm tot}=0 \ ,\hspace{6mm}{\bfm P}_{\rm tot}=<{\bfm P}>+\int
{\bfm E}\wedge B\,d^2x \ ,\\
&&\frac{d}{d\,t}M_{\rm tot}=0 \ ,\hspace{6mm}M_{\rm tot}=<M>+\int{\bfm x}\wedge({\bfm E}\wedge B)\,d^2x  \
,\\
&&\frac{d}{d\,t}\,E=0 \ ,\hspace{10mm}E=\int{\cal H}\,d^2x \ ,\label{c4}
\end{eqnarray}
$N$ is the particle number, ${\bfm P}_{\rm
tot}$ is the total linear momentum, $M_{\rm tot}$ the total
angular momentum and finally $E$ is the total energy of the system.
\\
In this section we want to obtain the above constants of motion
(\ref{c1})-(\ref{c4}) by means of an approach different from that
used in the previous section, i.e. by means of fundamental
principles and of the N\"other theorem. The deduction of these
quantities is close to the content of chapter III to obtain the
energy-momentum tensor for the EIP-Schr\"odinger equation, but
here, the presence of the CS term makes the iter a little more
complicate.

We consider the symmetry related to the invariance of the system over
space-time translations. If we consider an infinitesimal transformation:
\begin{equation}
t\rightarrow t-a \ ,\hspace{10mm}{\bfm x}\rightarrow{\bfm x}-{\bfm a} \ ,
\end{equation}
it is easy to see that the variations of the fields are given by:
\begin{equation}
\delta_a\psi=a^\mu\,\partial_\mu\psi,\hspace{10mm}
\delta_a\psi^\ast=a^\mu\,\partial_\mu\psi^\ast,\hspace{10mm}
\delta_a A_\nu=a^\mu\,\partial_\mu A_\nu \ ,\label{var}
\end{equation}
and the function ${\cal L}$ changes for a quantity $\delta\,{\cal L}=\delta^{\mu\nu}
\partial_\nu {\cal L}$, (here there is no sum over the repeated index $\nu$);
where the symbol $\delta^{\mu\nu}$ is the Kronecker tensor and the index $\mu$
selects the variation of ${\cal L}$ over space translations from time
translations.\\
Introducing the tensor $T^{\mu\nu}$, defined as:
\begin{equation}
T^{i\nu}={\cal J}^\nu_{\rm space} \ ,\label{sp}
\end{equation}
where ${\cal J}^\nu_{\rm space}$ is the N\"other current generated
from a space translation, and
\begin{equation}
T^{0\nu}={\cal J}^\nu_{\rm time} \ ,\label{ti}
\end{equation}
where now ${\cal J}^\nu_{\rm time}$ is the N\"other current
generated from a time translation. Then, from the expression of
the density of Lagrangian (\ref{EMCS}) and using the N\"other
formula:
\begin{equation}
{\cal J}^\nu=\frac{\partial\,{\cal
L}}{\partial\,(\partial_\nu\psi)}
\,\delta\,\psi+\frac{\partial\,{\cal
L}}{\partial\,(\partial_\nu\psi^\ast)} \,\delta\,\psi^\ast
+\frac{\partial\,{\cal L}}{\partial\,(\partial_\nu A_\mu)}
\,\delta\,A_\mu-\delta^{\mu\nu}\,\partial_\mu{\cal L} \ ,
\end{equation}
and the definitions (\ref{sp}) and (\ref{ti}) we obtain the
quantities:
\small
\begin{eqnarray}
{\widetilde T}^{00}&=&\frac{\hbar^2}{2\,m}|{\bfm D}\psi|^2
-\kappa\,\frac{\hbar^2}{8m} \left[\psi^\ast\,{\bfm D}\psi-
\psi\,({\bfm D}\psi)^\ast\right]^2+\widetilde
U(\rho)+e\,A^0\,\rho\nonumber\\ &+&\gamma\,\left(
\frac{1}{4}F_{\mu\nu}\,F^{\mu\nu}-F^{0\tau}\,\partial^0
A_\tau\right)
-\frac{g}{2}\left(\frac{1}{2}\epsilon^{\tau\mu\nu}A_\tau\,F_{\mu\nu}
-\epsilon^{0\lambda\tau}\,A_\lambda\,\partial^0A_\tau\right) \
,\label{uni}\\ \nonumber &&\\ {\widetilde
T}^{0i}&=&i\,c\,{\frac{\hbar}{2}}\left(\psi^\ast\,\partial^i\psi-
\psi\,\partial^i\psi^\ast\right)-\gamma\,F^{0\tau}\,\partial^iA_\tau
-\frac{g}{2}\,\epsilon^{0\lambda\tau}\,A_\lambda\,\partial^iA_\tau
\ ,\\ \nonumber &&\\ \nonumber {\widetilde
T}^{i0}&=&\frac{\hbar^2}{2\,m}\left[(D^i\psi)^\ast\,\partial^0\psi+D^i\psi\,
\partial^0\psi^\ast\right]
-\kappa\,\frac{\hbar^2}{4\,m}\,\left[\psi^\ast\,D^i\psi-
\psi\,(D^i\psi)^\ast\right]\left(\psi^\ast\,\partial^0\psi-\psi\,\partial^0\psi^\ast
\right)\\
&-&\gamma\,F^{i\tau}\,\partial^0A_\tau-\frac{g}{2}\,\epsilon^{i\lambda\tau}
\,A_\lambda\,\partial^0A_\tau \ , \\
\nonumber
&&\\
{\widetilde T}^{ij}&=&\frac{\hbar^2}{2\,m}\,\left[(D^i\psi)^\ast\,\partial^j\psi+D^i\psi\,
\partial^j\psi^\ast\right]-\kappa\,\frac{\hbar^2}{4\,m}\,
\left[\psi^\ast\,D^i\psi-\psi\,(D^i\psi)^\ast\right]
\left(\psi^\ast\,\partial^j\psi-\psi\,\partial^j\psi^\ast\right)\nonumber\\
&-&\gamma\,F^{i\tau}\,\partial^j\,A_\tau
-\frac{g}{2}\,\epsilon^{i\lambda\tau}\,A_\lambda\,\partial^j\,A_\tau
+\delta^{ij}\left\{i\,c\,\frac{\hbar}{2}\,\left[\psi^\ast\,D_0\psi-
\psi\,(D_0\psi)^\ast\right]-\frac{\hbar^2}{2\,m}\,|{\bfm
D}\psi|^2\right.\nonumber\\
&+&\left.\kappa\,\frac{\hbar^2}{4m}\left[\psi^\ast\,{\bfm D}\psi-
\psi\,({\bfm D}\psi)^\ast\right]^2-\widetilde U(\rho)
-\frac{\gamma}{4}\,F_{\mu\nu}\,F^{\mu\nu}+{\frac{g}{4}}
\epsilon^{\tau\mu\nu}\,A_\tau\,F_{\mu\nu}\right\} \
.\label{quattro}
\end{eqnarray}
\normalsize In (\ref{uni})-(\ref{quattro}) the gauge invariance of
the tensor ${\widetilde T}^{\mu\nu}$ does not appear explicitly.
We consider now the new tensor ${\widetilde{\widetilde
T}}\,^{\mu\nu}$ given by:
\begin{equation}
{\widetilde{\widetilde T}}\,^{\mu\nu}={\widetilde T}^{\mu\nu}+{\widetilde
f}^{\mu\nu} \ ,\label{so}
\end{equation}
where the quantity:
\begin{equation}
{\widetilde f}^{\mu\nu}=A^\nu\,\left(\gamma\,\partial_\tau F^{\tau\mu}+\frac{g}{2}
\,\epsilon^{\mu\lambda\tau}\,F_{\lambda\tau} -\frac{e}{c}\,J^\mu\right) \ ,
\end{equation}
vanishes because of the motion equations (\ref{gaugefield}). We
observe that the tensor ${\widetilde{\widetilde T}}\,^{\mu\nu}$ is
gauge invariant. Finally, if we remember that the expression of
the energy-momentum tensor is defined modulo a quantity
$\partial_\tau\,X^{\tau\mu\nu}$ where $X^{\tau\mu\nu}$ is a third
rank tensor which is antisimetric in the two indices $\tau$ and
$\mu$ given by:
\begin{equation}
X^{\tau\mu\nu}\equiv-\gamma\,F^{\tau\mu}\,A^\nu+
\frac{g}{2}\,\epsilon^{\tau\mu\lambda}\,A_\lambda\,A^\nu \ ,
\end{equation}
it is easy to verify that the tensor $T^{\mu\nu}$ defined as:
\begin{equation}
T^{\mu\nu}={\widetilde{\widetilde T}}\,^{\mu\nu}-\gamma\,\partial_\tau\,(F^{\tau\mu}\,A^\nu)+
\frac{g}{2}\,\epsilon^{\tau\mu\lambda} \,\partial_\tau\,(A^\nu\,A_\lambda) \
, \label{sostit}
\end{equation}
is gauge invariant and obeys to the continuity equation:
\begin{equation}
\partial_\mu\,T^{\mu\nu}\equiv\partial_\mu\,{\widetilde{\widetilde T}}\,^{\mu\nu}=0 \ .\end{equation}
As a consequence of (\ref{sostit}) the motion constants of the system:
\begin{equation}
\int T^{\mu 0}\,d^2x=\int{\widetilde{\widetilde T}}\, ^{\mu 0}\,d^2x \ ,
\end{equation}
are not modified by the substitutions (\ref{so}) and (\ref{sostit}).
The final expression of the components of the energy-momentum tensor
$T^{\mu\nu}$ are:
\begin{eqnarray}
\nonumber T^{00}&=&\frac{\hbar^2}{2\,m}|{\bfm D}\psi|^2
-\kappa\,\frac{\hbar^2}{8m} \left[\psi^\ast\,{\bfm D}\,\psi-
\psi\,({\bfm D}\,\psi)^\ast\right]^2 +\widetilde U(\rho)\\
&+&{\frac{\gamma}{4}}\,F_{ij}\,F^{ij}-{\frac{\gamma}{2}}
\,F_{0i}\,F^{0i} \ ,  \label{prima1} \\ \nonumber &&\\
T^{0i}&=&i\,c\,{\frac{\hbar}{2}}\left[\psi^\ast\,D^i\psi-
\psi\,(D^{~i}\psi)^\ast\right]-\gamma\,F^{0j}\,F^i_{~j} \ ,
\\
\nonumber &&\\
T^{i0}&=&\frac{\hbar^2}{2\,m}\left[(D^i\psi)^\ast\,D^0\psi+D^i\psi\,
(D^0\psi)^\ast\right]  \nonumber \\
&-&\kappa\,\frac{\hbar^2}{4\,m}\,\left[\psi^\ast\,D^i\psi-
\psi\,(D^i\psi)^\ast\right]\left[\psi^\ast\,D^0\psi-\psi\,(D^0\psi)^\ast
\right]-\gamma\,F^{0j}\,F^i_{~j} \\ \nonumber &&\\ \nonumber
T^{ij}&=&\frac{\hbar^2}{2\,m}\,\left[(D^i\psi)^\ast\,D^j\psi+D^i\psi\,
(D^j\psi)^\ast\right]\\ &-&\kappa\,\frac{\hbar^2}{4\,m}\,
\left[\psi^\ast\,D^i\psi-\psi\,(D^i\psi)^\ast\right]
\left[\psi^\ast\,D^j\psi-\psi\,(D^j\psi)^\ast\right]  \nonumber \\
&-&\delta^{ij}\left\{\frac{\hbar^2}{4\,m}\,\Delta\rho+
\kappa\,\frac{\hbar^2}{ 8m}\left[\psi^\ast\,{\bfm D}\,\psi-
\psi\,({\bfm D}\,\psi)^\ast\right]^2\right.\nonumber\\
&+&\left.\widetilde U(\rho)-\rho\,\frac{d\,\widetilde
U(\rho)}{d\,\rho}\right\}
-\gamma\,F^{i\lambda}\,F^j_{~\lambda}-{\frac{\gamma}{4}}
\delta^{ij}\,F^{\mu\nu}\,F_{\mu\nu} \ .
\end{eqnarray}
We note that in the above expression of $T^{\mu\nu}$ the CS
contribution does not appear explicitly. This is due to the
topological nature of the CS interaction. How we have noted in the
introduction of this chapter, independently of the geometry of the
variety in which the system is imbedded, the expression of the CS
Lagrangian appears the same without the necessity to introduce the
metric tensor $g_{\mu\nu}$. In fact the quantity
$\epsilon_{\mu\nu\lambda}$ is a good tensor. As a consequence, if
we take into account that the $T^{\mu\nu}$ tensor can be also
obtained by a functional variation of the Lagrangian density with
respect to the metric we obtain immediately the statement that the
contributes of CS to the energy-momentum tensor are null due to
the independence of its from $g_{\mu\nu}$. Notwithstanding the
presence of the CS term modifies the form of the gauge fields
$A_\mu$ because they must satisfies Eq. (\ref{gaugefield}).\\ The
tensor $T^{\mu\nu}$ is symmetric only in the spatial indices
$T^{ij}=T^{ji}$, because the theory is not Lorentz invariant but
only rotation invariant and satisfies the continuity equation:
\begin{equation}
\partial_\mu\,T^{\mu\,\nu}=0 \ .
\end{equation}
Of course, the two following quantities are conserved:
\begin{equation}
E=\int T^{00}\,d^2x \ ,\hspace{10mm}P^i_{\rm tot}={1\over c}\int T^{0i}\,d^2x \ .
\label{conserve1}
\end{equation}
In section 3.2 we have shown that the energy density $T^{00}$ is a semidefinite positive
quantity. This properties hold also in presence of the gauge field.
In fact, in the case $\kappa>0$ by using Eqs.
(\ref{covariantcurrent}), (\ref{prima1}) and the expression of the fields $\bfm
E$ and $B$ as functions of the potential $A_\mu$, the quantity $T^{00}$ assume the form
\begin{eqnarray}
T^{00}=\frac{\hbar^2}{2\,m}|{\bfm D}\psi|^2
+\kappa\,\frac{m}{2}\left(\frac{ {\bfm
J}}{1+\kappa\,\rho}\right)^2 +\widetilde U(\rho)
+{\frac{\gamma}{2}}\,({\bfm E}^2+B^2) \ . \label{energy1}
\end{eqnarray}
Therefore, if $\widetilde{U}(\rho)=0$, $T^{00}$ is a semidefinite
positive quantity. In the case $\kappa<0$, after rewriting the
quantity $T^{00}$ as:
\begin{eqnarray}
T^{00}=\frac{\hbar^2}{2\,m}|{\bfm D}\psi|^2\,(1+\kappa\rho)
-\kappa\,\frac{ \hbar^2}{8\,m}\,(\bfm\nabla\,\rho)^2 +\widetilde
U(\rho)+{\frac{\gamma}{2}}\,({\bfm E}^2+B^2) \ , \label{energy2}
\end{eqnarray}
we can obtain the same conclusion, if we take into account that
$1+\kappa\,\rho\geq0$.\\
We consider now the invariance of the
system under spatial rotations defined by the transformations:
\begin{equation}
x^i\rightarrow \Omega^i_{~j}\,x^j \ ,\hspace{10mm}\Omega^i_{~j}\,\Omega^j_{~k}=
\delta^i_{~k} \ .
\end{equation}
The angular momentum density $M^0$ and the relative flux $M^i$ form the vector
$M^\mu=(M^0,\,\bfm M)$, that is obtained from the tensor $T^{\mu\nu}$ by
the relation:
\begin{eqnarray}
M^\mu=\epsilon_{ij}\,x^i\,T^{\mu j} \ .
\end{eqnarray}
From the symmetry $T^{ij}=T^{ji}$ we infer the continuity equation for $M^\mu$:
\begin{equation}
\partial_\mu M^\mu=0 \ .
\end{equation}
The total angular momentum:
\begin{equation}
M_{\rm tot}={1\over c}\int M^0\,d^2x \ ,  \label{conserve2}
\end{equation}
is time conserved and, like ${\bfm P}_{\rm tot}$, is not modified by the presence of EIP
potential. \\
The quantity (\ref{conserve1}) and (\ref{conserve2}) are the generators of
the space-time translation and rotations supplemented by local gauge
transformations when performed on gauge not invariant fields. In fact if we consider for
example the quantity:
\begin{eqnarray}
g=e^{\frac{i}{\hbar}\,a_i\,P^i} \ ,
\end{eqnarray}
which generate a space translation with parameter $a_i$, the
linear part of the variations of the fields can be performed
trough the "{\sl naked}" Poisson brackets (\ref{bp1}):
\small
\begin{eqnarray}
\nonumber \psi(t,\,{\bfm x})\rightarrow\psi_a(t,\,{\bfm
x})&=&g\,\psi(t,\,{\bfm x})=\psi(t,\,{\bfm
x})+a_i\,\{\psi(t,\,{\bfm x}),\,P^i\}_D\\ &=&\psi(t,\,{\bfm
x})+a_i\,\partial^i\psi(t,\,{\bfm x})+i\,e\,a_i\,\omega(t,\,{\bfm
x})\,\psi(t,\,{\bfm x}) \ ,\\ \nonumber &&\\ \nonumber
\psi^\ast(t,\,{\bfm x})\rightarrow\psi_a^\ast(t,\,{\bfm
x})&=&g\,\psi^\ast(t,\,{\bfm x})=\psi^\ast(t,\,{\bfm
x})+a_i\,\{\psi^\ast(t,\,{\bfm x}),\,P^i\}_D\\
&=&\psi^\ast(t,\,{\bfm x})+a_i\,\partial^i\psi^\ast(t,\,{\bfm
x})+i\,e\,a_i\,\omega(t,\,{\bfm x})\,\psi^\ast(t,\,{\bfm x})
\\
\nonumber &&\\ \nonumber {\bfm A}(t,\,{\bfm x})\rightarrow{\bfm
A}(t,\,{\bfm x})&=&g\,{\bfm A}(t,\,{\bfm x})={\bfm A}(t,\,{\bfm
x})+a_i\,\{{\bfm A}(t,\,{\bfm x}),\,P^i\}_D\\ &=&{\bfm
A}(t,\,{\bfm x})+a_i\,\partial^i{\bfm A}(t,\,{\bfm
x})+i\,e\,a_i\,\omega(t,\,{\bfm x})\,{\bfm A}(t,\,{\bfm x}) \ ,\\
\nonumber &&\\ \nonumber {\bfm\Pi}(t,\,{\bfm
x})\rightarrow{\bfm\Pi}(t,\,{\bfm x})&=&g\,{\bfm\Pi}(t,\,{\bfm
x})={\bfm\Pi}(t,\,{\bfm x})+a_i\,\{{\bfm\Pi}(t,\,{\bfm
x}),\,P^i\}_D\\ &=&{\bfm\Pi}(t,\,{\bfm
x})+a_i\,\partial^i{\bfm\Pi}(t,\,{\bfm
x})+i\,e\,a_i\,\omega(t,\,{\bfm x})\,{\bfm\Pi}(t,\,{\bfm x})
\end{eqnarray}
\normalsize
with $\omega$ the gauge parameter. The fact that the
quantities (\ref{conserve1}) and (\ref{conserve2}) generate
translations supplied by gauge transformations is due to structure
of the Poisson brackets (\ref{bp1}) which are obtained after
taking into account the constraint (\ref{eta}) which is the
generator of the gauge transformation (see below). However, if we
restrict ourselves to gauge-invariant fields, e.g. $\rho(t,\,{\bfm
x}),\,{\bfm E}(t,\,{\bfm x}),$ and $B(t,\,{\bfm x})$, Eqs.
(\ref{conserve1}), (\ref{conserve2}) generates pure translations:
\begin{eqnarray}
\{{\bfm P}(t,\,{\bfm x}),\,\rho(t,\,{\bfm
x})\}&=&{\bfm\nabla}\,\rho(t,\,{\bfm x}) \ ,\\
\{{\bfm P}(t,\,{\bfm x}),\,F_{\mu\nu}(t,\,{\bfm
x})\}&=&{\bfm\nabla}\,F_{\mu\nu}(t,\,{\bfm x}) \ .
\end{eqnarray}

It is known that the introduction of the CS interaction confers to
the matter field a non conventional statistical behavior \cite{Jackiw,Hagen}.
This statistical behavior is the result of the Aharonov-Bohm effect \cite{Aharonov},
because, in presence of the CS term, each charge field carries forward,
intrinsically, also point magnetic vortices. This statement is true
also in the presence of the Maxwell term, because only the asymptotic
behavior is invoked, which, because of the high-derivative terms
of the Maxwell coupling, is dominated by the CS term.
This anomalous situation in not modified by the presence of EIP because
its does not modify the expression of $<{\bfm P}>$ and $<M>$. In fact, posing
for simplicity $\gamma=0$,  the second relation of (\ref
{conserve1}) and (\ref{conserve2}) becomes:
\begin{eqnarray}
&&<{\bfm
P}>=-\frac{i\,\hbar}{2}\,\int(\psi^\ast\bfm\nabla\psi-\psi\,\bfm\nabla\psi^
\ast)\,d^2x-\frac{e}{c}\,\int{\bfm A}\,\rho\,d^2x \ , \label{mom1}
\\
&&<M>=-\frac{i\,\hbar}{2}\,\int {\bfm
x}\wedge(\psi^\ast\bfm\nabla\psi-\psi\,
\bfm\nabla\psi^\ast)\,d^2x-\frac{e}{c}\,\int{\bfm x}\wedge{\bfm
A}\,\rho\,d^2x \ . \label{mom2}
\end{eqnarray}
Eq. (\ref{Gauss}) can be solved without ambiguity in the Coulomb gauge
($\bfm\nabla\cdot{\bfm A}=0$):
\begin{equation}
{\bfm A}({\bfm x})=-\frac{e}{2\,\pi\,g}\int\bfm\nabla\cdot\arctan\left(\frac{
x_2-y_2}{x_1-y_1}\right)\,\rho({\bfm y})\,d^2y \ ,
\end{equation}
and inserting it into Eqs. (\ref{mom1}) and (\ref{mom2}), after integration,
we obtain:
\begin{eqnarray}
&&<{\bfm
P}>=-\frac{i\,\hbar}{2}\,\int(\psi^\ast\bfm\nabla\psi-\psi\,\bfm\nabla\psi^
\ast)\,d^2x \ , \\ &&<M>=-\frac{i\,\hbar}{2}\,\int {\bfm
x}\wedge(\psi^\ast\bfm\nabla\psi-\psi\,
\bfm\nabla\psi^\ast)\,d^2x-\frac{Q^2}{4\,\pi\,c\,g} \ .
\label{mom3}
\end{eqnarray}
From Eq. (\ref{mom3}) we can see that the angular momentum of the field
$\psi$ is the sum of two terms: the former represents the orbital angular
momentum, while the last, the spin, is responsible of the anomalous behavior of the
system. This result, already known, is not too surprising: its origin is in the Gauss law
(\ref{Gauss}) that is not modified by the presence of EIP. As a
consequence of Eq. (\ref{mom3}) it follows that EIP does not change the anyonic behavior
of the system and so that the spin-statistics
relation of the anyonic systems holds. Therefore the Lagrangian density
(\ref{glagrangian}), (\ref{MCS}) describe a model of anyonic particle
with kinetic obeying to the EIP. This one generate an
exclusion-inclusion effect in the configuration space wilts the CS term
is responsible of the not conventional quantum statistics.

Finally, we discuss briefly the gauge transformations. In section
4.1 we have seen that the Lagrangian (\ref{glagrangian}),
(\ref{MCS}) is invariant up to a divergence over global gauge
$U(1)$ transformations:
\begin{eqnarray}
\psi(t,\,{\bfm x})&\rightarrow&
e^{-i(e/\hbar\,c)\,\omega}\,\psi(t,\,{\bfm x}) \ , \\
A_\mu(t,\,{\bfm x})&\rightarrow& A_\mu(t,\,{\bfm x}) \ ,
\end{eqnarray}
with $\delta{\cal L}=0$. The N\"other current is now:
\begin{equation}
{\cal J}^\mu=\left\{c\,\rho,\,-\frac{i\,\hbar}{2\,m}\,(1+\kappa\,\rho)\,[\psi^\ast\,{\bfm D}
\psi-\psi\,({\bfm D}\psi^\ast)]\right\} \ ,
\end{equation}
which satisfies the continuity equation (\ref {continuity}).\\ We
consider now the local gauge $U(1)$ transformation:
\begin{eqnarray}
\psi(t,\,{\bfm x})&\rightarrow& e^{-i\,(e/\hbar\,c)\,\omega(t,\,{\bfm x})}\,\psi(t,\,{\bfm x}
) \ ,  \label{gaugec1} \\
A_\mu(t,\,{\bfm x})&\rightarrow& A_\mu(t,{\bfm x})+\partial_\mu\omega(t,\,{\bfm
x}) \ .  \label{gaugec2}
\end{eqnarray}
It is well known, that the N\"other theorem, in the case of symmetries related to
continuous transformations, does not give conserved quantities but from it we
obtain the same identities.
In fact, the gauge charge associated with this continuous symmetry vanishes
identically.\\
In the case of the transformations (\ref{gauge1}) and
(\ref{gauge2}) we have (using the notations of section 3.2):
\begin{eqnarray}
&&\delta\,\psi=-\frac{i\,e}{\hbar\,c}\,\psi\,\omega \ ,\\
&&\delta\,A_\mu=\partial_\mu\omega \
,\\
&&f^\mu=\frac{g}{4}\,\epsilon^{\mu\nu\tau}\,F_{\nu\tau}\,\omega
 \ .
\end{eqnarray}
The N\"other vector ${\cal J}^\mu\equiv({\cal J}^0,\,{\cal J}^i)$
can be calculated by using (\ref{ncurrent}), we obtain:
\begin{eqnarray}
&&{\cal J}^0=\left(e\,\rho-\gamma\,{\bfm\nabla}\cdot{\bfm
E}+g\,B\right)\omega \ ,\\
&&{\cal J}^i=\left(\frac{e}{c}\,J^i-\gamma\,\partial_\mu F^{\mu
i}-\frac{g}{2}\,\epsilon^{i\mu\nu}F_{\mu\nu}\right)\,\omega \ .
\end{eqnarray}
By means of Eq. (\ref{cn}) the charge $Q_\omega$, given by
$Q_\omega=\int{\cal J}_0\,d^2x$, takes the expression:
\begin{equation}
Q_\omega=\int\left(\gamma\,\bfm\nabla\cdot{\bfm E}-g\,B-e\,\rho\right)\,\omega
\,d^2x \ .
\end{equation}
We can easily see, after using the Gauss theorem (\ref{Gauss}), that $Q_\omega$ is identically zero .
This means that the Gauss condition (\ref{Gauss}) is satisfied at
each time, during the motion.


\setcounter{chapter}{5}
\setcounter{section}{0}
\setcounter{equation}{0}
\chapter*{Chapter V\\
\vspace{10mm}Canonical Systems\\ Obeying to EIP: Solitons}
\markright{Chap. V - Canonical Systems Obeying to EIP: Solitons}

In the previous chapters we have considered the definition of
collectively interacting particles obeying to EIP and we have
studied the main properties of a particles system.\\ In this
chapter and in the following one, we study a special class of
states of these systems called {\sl solitons}. Solitons are
special solutions of nonlinear evolution equations, not obtainable
within perturbative methods, that preserve their shape during the
propagation. Soliton solutions appear in many topics in physics.
For instance, the cubic Schr\"odinger equation solutions describe
propagation of deep water waves and of modulated ion-acoustic
waves in plasma, three-dimensional diffractive patterns of a laser
beam \cite{Scott} and more importantly describe the recently
observed Bose-Einstein condensation in rarefied vapors of metal
$^7$Li, $^{23}$Na $^{87}$Rb
\cite{Holland,Sinha,Reinhardt,Yukalov}.\\ Due to their not
spreading property and to their particle-like behavior, solitons
can be used in field theory in order to describe elementary
particles. Rigorously speaking, solitons are wave packets which
preserve their shape during their propagation and after collision
with other solitons. Usually it is required that solitons preserve
asymptotically their speed after collisions, but this is hard to
met with the interpretation of solitons as particles.\\ Typically,
in literature, soliton term is misused to denote solitary waves.
These are also wave packets preserving their shapes but no
assumptions are made about their behavior after collisions. Thus,
a single soliton is a solitary wave but solitary waves are not
necessarily solitons. In order to see if a solution is a "{\sl
genuine}" soliton it is necessary to follow it after collision
with another soliton requiring the knowledge of multi-solitons
states. This solutions generally are very hard to be obtained.\\
In the following, we use the word soliton but all our solutions
are simply solitary waves.

\section{Solitons}
We study a particular class of solutions of Eq.
(\ref{schroedinger6}) in the free case, i.e. when $V=0$. In this
situation, we can consider the motion of the mass center on a
straight line with uniform velocity (for a discussion of solitons
in an external potential see for example Ref. \cite{Morgan}). In
addition to the condition $V=0$, we limit our attention to the one
dimensional case when the EIP holds. The extension to highest
dimensions is not straightforward. In particular the easy case of
waves with amplitude modulated only in one space-dimension is non
physical situation because they carry infinite energy. \\ In
chapter III we have studied the symmetry Lie group of the system
obeying to the EIP. The result was that Eq. (\ref{schroedinger6})
is invariant over roto-translation transformations, scaling
transformations and the global unitary $U(1)$ group
transformations. We have a nine parameter full symmetry group.  We
can use this information to search special solutions obeying to
EIP. In fact, we can require that the solution should be invariant
under a proper selected subgroup of the full symmetry group so
that the solution itself can be obtained solving an ordinary
differential equation rather than a more complicate partial
differential equation.\\ In this section we focus our attention on
solutions that are constants in between the $y$ and $z$ spatial
direction and moreover are invariant over the $x$ and $t$
translation:
\begin{eqnarray}
x\rightarrow x\pm u\,\varepsilon \ ,\hspace{20mm}t\rightarrow t+\varepsilon \ , \label{tra}
\end{eqnarray}
with $\varepsilon$ and $u$ constants. The global invariant of this
transformation is the quantity $x\mp u\,t$. Therefore we require
that the field $\psi$ depends only on the time $t$ and on the
coordinate of the soliton mass center $\xi=x\mp u\,t$ where $u$
now has the meaning of velocity of the soliton (as usual, the sign
minus stands for a soliton moving from the left to the right side
of the $x$ axis, while the sign plus stands for an antisoliton
moving in the opposite versus). Thus the wave function becomes
$\psi(x,\,t)\equiv\psi(\xi,\,t)$ where we include a possible
explicit time dependence in the phase. (Rigorously speaking, this
solution is not invariant over the transformation (\ref{tra})
because the time dependence on $S$. Notwithstanding, this
assumption does not modify the general procedure to find soliton
solutions, allowing to obtain a more general expression). In the
following we use the method described in ref. \cite{Hasse2}, valid
for the NLSEs that are most frequently encountered in physical
problems. We assume that, for $\xi\rightarrow\pm\infty$, the
particle density $\rho(\xi)\rightarrow0$ so that
$\int_{-\infty}^{+\infty}\rho(\xi)\,d\xi=N$, where $N$ represents
the collective particle number contained in the soliton. Moreover,
the phase $S$ is written as:
\begin{equation}
S=s(\xi)-\epsilon\,t \ ,\label{ph}
\end{equation}
and the field $\psi$ assumes the form:
\begin{equation}
\psi=\rho(\xi)^{1/2}\,\exp\left\{{i\over\hbar}[s(\xi)-\epsilon\,t]
\right\} \ .\label{on}
\end{equation}
It is now easy to verify that the {\it Hamilton-Jacobi} equations (\ref{hj}) and the
continuity equation (\ref{cont}) describing the solitonic state
can be reduced to the following system of coupling equations:
\begin{eqnarray}
&&\pm u\,\frac{\partial s}{\partial\xi}=\frac{1+2\,\kappa\,\rho}{2\,m}\left(
\frac{\partial s}{\partial\xi}\right)^2+U_q(\rho)+F(\rho)-\epsilon \ ,\label{hj1}\\
&&\pm u\,\frac{\partial\rho}{\partial\xi}={1\over
m}\frac{\partial}{\partial\xi}\left[
\frac{\partial s}{\partial\xi}\,\rho\,(1+\kappa\,\rho)\right]  \ ,\label{cont3}
\end{eqnarray}
where we have indicated with $U_q(\rho)$ the one dimensional quantum
potential in the $\xi$ coordinate:
\begin{equation}
U_q(\rho)=-\frac{\hbar^2}{2\,m}\,{1\over\sqrt{\rho}}\frac{\partial^2\sqrt{\rho}}
{\partial\xi^2} \ .
\end{equation}
The quantum velocity $v_q(\xi)=m^{-1}\partial s(\xi)/\partial\xi$
must be finite when $\xi\rightarrow\pm\infty$; with this condition
Eq. (\ref{cont3}) can be integrated a first time, obtaining:
\begin{eqnarray}
&&\frac{\partial\,s}{\partial\,\xi}=
\pm\frac{m\,u}{1+\kappa\,\rho} \ ,\label{difase}
\end{eqnarray}
and after second integration with the condition $\xi(0)=0$,
we have:
\begin{equation}
s(\xi)=\pm m\,u\,\int\limits^\xi\limits_0\frac{d\,\xi^\prime}
{1+\kappa\,\rho(\xi^\prime)} \ .\label{fases}
\end{equation}
Eq. (\ref{ph}) and (\ref{fases}) allow us to calculate the phase
$S(\xi)$ provided that the quantity $\rho(\xi)$ is known.

To evaluate the density
$\rho(\xi)$ we note that, if we take into account Eq. (\ref{difase}),
Eq. (\ref{hj1}) reduces to the following second order differential equation:
\begin{eqnarray}
{2\over\rho}\frac{d^2\,\rho}{d\,\xi^2}
-\left({1\over\rho}\frac{d\,\rho}{d\,\xi}\right)^2
+{(2\,m\,u/\hbar)^2\over(1+\kappa\,\rho)^2}
-\frac{8\,m}{\hbar^2}\,F(\rho)+\frac{8\,m\,\epsilon}{\hbar^2}=0 \
.\label{diff}
\end{eqnarray}
Before solving this equation, we observe that $\epsilon$ can be
written in the form:
\begin{equation}
\epsilon=U_q+F(\rho)-{1\over2}\,m\,u^2{1\over(1+\kappa\,\rho)^2} \ .\label{energy11}
\end{equation}
Equation (\ref{energy11}) has an immediate physical interpretation: the
quantum potential causes the spreading of the ordinary Schr\"odinger wave
packet; this spreading is compensated by the nonlinearity $F(\rho)$ and by
the EIP contribution $-(m\,u^2/2)\,(1+\kappa\,\rho)^{-2}$.
Therefore it is possible to build up a non-spreading
solitary wave.\\
After the introduction of the function
\begin{equation}
y(\rho)=\left({1\over\rho}\frac{d\,\rho}{d\,\xi}\right)^2 \ ,\label{sub}
\end{equation}
so that
\begin{eqnarray}
\frac{d\,y}{d\,\rho}=\frac{2}{\rho}\,\frac{d}{d\,\xi}\left(\frac{1}{\rho}\,\frac{d\,\rho}{d\,\xi}\right)
 \ ,
\end{eqnarray}
Eq. (\ref{diff}) reduces to a first order linear
differential equation:
\begin{eqnarray}
&&\frac{d\,y}{d\,\rho}+{y\over\rho}+{(2\,m\,u/\hbar)^2
\over\rho\,(1+\kappa\,\rho)^2}-
\frac{8\,m}{\hbar^2}{F(\rho)\over\rho}
+\frac{8\,m\,\epsilon}{\hbar^2}{1\over\rho}=0 \ ,\label{eq3}
\end{eqnarray}
that can be easily integrated, giving:
\begin{eqnarray}
&&y(\rho)={A\over\rho}-\frac{8\,m\,\epsilon}{\hbar^2}
-{(2\,m\,u/\hbar)^2 \over1+\kappa\,\rho}
+\frac{8\,m}{\hbar^2}\,{\widetilde U(\rho)\over\rho} \
,\label{integro}
\end{eqnarray}
where the integration constant is $A-1/\kappa$ in order to obtain the
right limit for $\kappa\rightarrow0$.\\
By comparing Eq. (\ref{sub}) to Eq. (\ref{integro}), we obtain:
\begin{eqnarray}
\left(\frac{d\,\rho}{d\,\xi}\right)^2=A\,\rho
-\frac{8\,m\,\epsilon}{\hbar^2}\rho^2
-\left(\frac{2\,m\,u}{\hbar}\right)^2\,
\frac{\rho^2}{1+\kappa\,\rho}
+\frac{8\,m}{\hbar^2}\,\rho\,\widetilde U(\rho) \ .\label{eq2}
\end{eqnarray}
The evaluation of the soliton shape is thus reduced to the
solution of the first order ordinary differential equation (\ref{eq2}).
By introducing the dimensionless variables:
\begin{equation}
n=|\kappa|\,\rho \ ,
\end{equation}
and
\begin{equation}
\chi=\frac{2\,m\,u}{\hbar}\,(x\mp u\,t) \ ,
\end{equation}
Eq. (\ref{eq2}) takes the form:
\begin{eqnarray}
\left(\frac{d\,n}{d\,\chi}\right)^2=\alpha\,n+\beta\,n^2+\gamma\,n\,\widehat{U}(n)
-\frac{n^2}{1+\sigma\,n} \ ,\label{eq}
\end{eqnarray}
where $\alpha=A\,|\kappa|\,\left(\hbar/2\,m\,u\right)^2$ and
$\beta=-2\,\varepsilon/m\,u^2$ are the new integration constants,
$\gamma=2\,|\kappa|/m\,u^2$ and $\widehat {U}(n)\equiv \widetilde
U(\rho)$. The parameter $\sigma=\kappa/|\kappa|$ assumes the value
$+1$ when the inclusion principle holds $(\kappa>0,\,n\geq0)$.
Accordingly, for the exclusion principle $(\kappa<0,\,0\leq
n\leq1)$ we have $\sigma=-1$. Here we have made use of the scaling
properties (\ref{dil1}) which permit us to take into account only
the two special case $\kappa=\pm1$. We note, finally, that while
solving the Eq. (\ref{eq}), we have to take into account the two
arbitrary constants $\alpha$ and $\beta$ that define a family of
solutions.

To determine the soliton
solutions we must search the solutions
of the first order differential equation (\ref{eq}),
varying the arbitrary constants $\alpha$ and $\beta$, while the sign
$\sigma=\pm1$ is kept fixed. Eq. (\ref{eq}) after integration gives:
\begin{equation}
\pm\,\chi=\int\limits^n\,\left(\frac{\alpha\,n+
(\sigma\,\alpha+\beta-1)\,n^2+\sigma\,\beta\,n^3+\gamma\,n\,
(1+\sigma\,n)\,\widehat{U}}{1+\sigma\,n}\right)^{-{1\over2}}\,dn \
.\label{integral}
\end{equation}
From Eq. (\ref{en1}), we have for the soliton case:
\begin{equation}
<\widehat{P}>=\pm M\,u \ ,
\end{equation}
where $M$ is defined by:
\begin{equation}
M=m\,\int\limits_{-\infty}\limits^{+\infty}\frac{\rho}{1+\kappa\,\rho}\,dx \ .\label{mm}
\end{equation}
By using Eq. (\ref{p1}) we can write the soliton energy as:
\begin{eqnarray}
E=\frac{<\widehat{P}^2>}{2\,m}+ \frac{\kappa}{2}\,m\,u^2\,
\int\limits_{-\infty}\limits^{+\infty}{\left[\left(\frac{\rho}{1+\kappa\,\rho}\right)^2+\widetilde
U(\rho)\right]\,dx} \ ,\label{en}
\end{eqnarray}
with
\begin{eqnarray}
<\widehat{P}^2>\equiv\hbar^2\int\Big|\frac{d\,\psi}{d\,x}\Big|^2\,dx \ .
\end{eqnarray}
To evaluate $<\widehat{P}^2>/2\,m$ we take into account (\ref{on}):
\begin{eqnarray}
\nonumber
\frac{<\widehat{P}^2>}{2\,m}&=&\frac{\hbar^2}{4}\int\frac{1}{\rho}\left(\frac{d\,\rho}{d
 x}\right)^2\,dx+\int\left(\frac{d\,s}{d\,x}\right)^2\,\rho\,dx\\
&=&\frac{\hbar^2}{4}\int\frac{1}{\rho}\left(\frac{d\,\rho}{d
 x}\right)^2\,dx+(m\,u)^2\int\frac{\rho}{(1+\kappa\,\rho)^2}\,dx \ ,
\end{eqnarray}
using Eq. (\ref{diff}) we obtain:
\begin{eqnarray}
\frac{<\widehat{P}^2>}{2\,m}=
m\,u^2\,\int\limits_{-\infty}\limits^{+\infty}
\frac{\rho}{(1+\kappa\,\rho)^2}\,dx-
\int\limits_{-\infty}\limits^{+\infty}\rho\,\frac{d\,\widetilde
U(\rho)}{d\rho}\,dx+\epsilon\,N
 \ .\label{p2}
\end{eqnarray}
Considering that $u=<\widehat{P}>/M$ and Eq. (\ref{p2}), the energy
(\ref{en}) satisfies the following soliton energy-momentum dispersion relation:
\begin{eqnarray}
\nonumber E&=&\frac{<\widehat{P}>^2}{2\,M}\,\left[1+\int
\limits_{-\infty}\limits^{+\infty}\frac{\rho}{(1+\kappa\,\rho)^2}\,dx\right]\,\left[
\int\limits_{-\infty}\limits^{+\infty}\frac{\rho}{1+\kappa\,\rho}\,dx\right]^{-1}\\
&+&\int\limits_{-\infty}\limits^{+\infty}\left[\widetilde
U(\rho)-\rho\,\frac{d\,\widetilde
U(\rho)}{d\rho}\right]\,dx+\epsilon\,N
 \ .
\end{eqnarray}
If we chose the constant $\epsilon$, appearing in the phase of $\psi$, as:
\begin{eqnarray}
\epsilon\,N=\int\limits_{-\infty}\limits^{+\infty}\left[\rho\,
\frac{d\,\widetilde U(\rho)}{d\rho}-\widetilde U(\rho)\right]\,dx
-\frac{1}{2}\,m\,u^2\int\limits_{-\infty}\limits^{+\infty}\frac{\rho}{(1+\kappa\,\rho)^2}\,dx
\ ,\label{eps}
\end{eqnarray}
the energy-momentum dispersion relation assumes the expression:
\begin{equation}
E=\frac{<\widehat{P}>^2}{2\,M} \ ,
\end{equation}
which is related to a free particle of mass $M$, traveling with
momentum $<\widehat{P}>=\pm M\,u$.

Let us study Eq. (\ref{eq}) defining the shape of the solitons. We
discuss the case $\widehat {U}(n)=0$ and $\beta=1$. As an example,
we consider the case where the EIP is reduced to the inclusion
principle (boson case) and the particle interaction is attractive
($\sigma=1$). When $\alpha=0$ we can obtain the expression of the
soliton explicitly. In fact, Eq. (\ref{integral}) now becomes:
\begin{eqnarray}
\pm\chi=\int\limits^n\sqrt{\frac{1+n}{n^3}}\,dn \ ,
\end{eqnarray}
and after integration we obtain the following
implicit solution:
\begin{equation}
\mbox{arcoth}\sqrt{\frac{1+n}
{n}}-\sqrt{\frac{1+n}{n}}=\pm{1\over2}\chi \ .
\end{equation}
The value $n(\chi)$, at the origin $n(0)=n_\circ$, is the solution of the
transcendent equation:
\begin{equation}
\tanh\sqrt{\frac{1+n_\circ}
{n_\circ}}=\sqrt{\frac{n_\circ}{1+n_\circ}} \ ,
\end{equation}
with $n_\circ=2.27671$. We note that for
$\chi\rightarrow\pm\infty$, the asymptotic form is
$n\simeq4/\chi^2$. A soliton with this behavior for
$\chi\rightarrow\pm\infty$ was taken into account recently in Ref.
\cite{Polychronakos}. We note also that at the origin the soliton
has an angular point because its first kinetic is discontinuous:
\begin{eqnarray}
\nonumber
\left(\frac{d\,n}{d\,\chi}\right)_{\chi=0}=
\pm\sqrt{\frac{n_\circ^3}{n_\circ +1}} \ .
\end{eqnarray}
The wave function of the soliton is:
\begin{eqnarray}
\psi(\chi,\,t)=\sqrt{n(\chi)\over \kappa}\,\exp\left\{
-i\left[\sqrt{\frac{1+n(\chi)}{n(\chi)}}-\sqrt{\frac{1+n_\circ}
{n_\circ}}+\frac{\epsilon}{\hbar}\,t\right]\right\} \ ,\label{esplicit}
\end{eqnarray}
where
\begin{eqnarray}
\epsilon=-m\,u^2\,\frac{n^2_\circ}{(1+n_\circ)^2} \ ,\label{eps1}
\end{eqnarray}
obtained using Eq. (\ref{eps}), so that the soliton has a particle-like behavior.
The expression (\ref{esplicit}) of the soliton can be used to calculate
explicitly the quantities $N, \ <P>$ and $E$.
Because of the Eqs. (\ref{eps}) and (\ref{eps1}) the number $N$ is:
\begin{equation}
N={\hbar\over \kappa\,m\,u}\,(1+n_\circ)\sqrt{1+n_\circ\over n_\circ} \ .
\end{equation}
On the other hand if we define the mass $M$ as in Eq. (\ref{mm}):
\begin{equation}
M={2\over 1+n_\circ}\,m\,N \ ,
\end{equation}
the momentum is given by $<P>=M\,u$ and the energy is $E=<P>^2/2\,M$.

\section{Effective potential}
In this section we prove that the unitary nonlinear transformation
(\ref{transf}) ($\psi(\xi,\,t)\rightarrow\phi(\xi,\,t)$) studied
in section 3.3, when applied on the solitonic states
$\psi(\xi,\,t)$ exists and that the new states $\phi(\xi,\,t)$ are
solutions of a Schr\"odinger equation with an algebraic real
nonlinearity.

Let us consider the unitary transformation:
\begin{equation}
\psi(\xi,\,t)\rightarrow\phi(\xi,\,t)={\cal U}(\xi)\,\psi(\xi,\,t) \ ,\label{trasf}
\end{equation}
where ${\cal U}(\xi)$ is given by:
\begin{equation}
{\cal U}(\xi)=\exp\left\{{i\over\hbar}\,
[\pm m\,u\,\xi-s(\xi)]\right\} \ .\label{unity}
\end{equation}
The new wave function $\phi(\xi,\,t)$ has the same amplitude of
the wave function $\psi(\xi,\,t)$ but a different phase:
\begin{equation}
\phi(\xi,\,t)=\rho(\xi)^{1/2}\,\exp\left\{{i\over\hbar}\,
(\pm m\,u\,\xi-\epsilon\,t)\right\} \ .\label{ph1}
\end{equation}
The unitary transformation can be rewritten as:
\begin{equation}
\psi(\xi,\,t)=\exp\left[-\frac{i}{\hbar}\,\Gamma(\xi)\right]\,\phi(\xi,\,t) \ ,\label{b1}
\end{equation}
where
\begin{equation}
\Gamma(\xi)=\pm
m\,u\,\left(\xi-\int\limits_0\limits^\xi\frac{d\,\xi^\prime}{1+\kappa\,\rho(\xi^\prime)}\right)
\ .\label{b2}
\end{equation}
After deriving (\ref{b1}) we obtain the following relations:
\begin{eqnarray}
\frac{\partial\,\psi}{\partial\,t}&=&\left[\frac{\partial\,\phi}{\partial\,t}-\frac{i}{\hbar}
\frac{\partial\,\Gamma}{\partial\,t}\,\phi\right]\,e^{-i\,\Gamma/\hbar}
\ ,\label{b4}\\
\frac{\partial^2\,\psi}{\partial\,\xi^2}&=&\left[-\,\frac{i}{\hbar}\,\frac{\partial^2\,\Gamma}{\partial\,\xi^2}\,\phi
-\frac{1}{\hbar^2}\,\left(\frac{\partial\,\Gamma}
{\partial\,\xi}\right)^2\,\phi-i\,\frac{2}{\hbar}
\frac{\partial\,\Gamma}{\partial\,\xi}\frac{\partial\,\phi}{\partial\,\xi}+\frac{\partial^2\,\phi}{\partial\,\xi^2}
\right]\,\,e^{-i\,\Gamma/\hbar} \ .\label{b6}
\end{eqnarray}
By considering (\ref{ph}), (\ref{difase}) and (\ref{b2}) we obtain:
\begin{eqnarray}
\frac{\partial\,S}{\partial\,\xi}=\frac{1}{\kappa\,\rho}\,\frac{\partial\,\Gamma}{\partial\,\xi}
 \ ,\label{b7}
\end{eqnarray}
and then the current $j$, given by (\ref{corrente2}) becomes:
\begin{equation}
j=\frac{1}{\kappa\,m}\,(1+\kappa\,\rho)\,\frac{\partial\,\Gamma}{\partial\,\xi} \ .\label{b8}
\end{equation}
In the case of soliton states the Schr\"odinger equation
(\ref{schroedinger6}) is:
\begin{eqnarray}
\nonumber
i\,\hbar\,\frac{\partial\,\psi}{\partial\,t}
&=&-\frac{\hbar^2}{2\,m}\,\frac{\partial^2\,\psi}{\partial\,\xi^2}+F(\rho)\,\psi\\
&+&\kappa\,\frac{m}{\rho}\left(\frac{j}{1+\kappa\,\rho}\right)^2\,\psi
-i\,\kappa\,\frac{\hbar}{2\,\rho}\,\frac{\partial}{\partial\,\xi}\,
\left(\frac{j\,\rho}{1+\kappa\,\rho}\right)\,\psi \ .\label{b71}
\end{eqnarray}
By using (\ref{b4}), (\ref{b6}) and (\ref{b8}) we may write Eq
(\ref{b71}) in the form:
\begin{eqnarray}
\nonumber
\frac{\partial\,\Gamma}{\partial\,t}\,\phi+i\,\hbar\,\frac{\partial\,\phi}{\partial\,t}&=
&-\frac{\hbar^2}{2\,m}\,\frac{\partial^2\,\phi}{\partial\,\xi^2}+\frac{i\,\hbar}{2\,m}
\,\frac{\partial^2\,\Gamma}{\partial\,\xi^2}\,\phi+\frac{2+\kappa\,\rho}{2\,m\,\kappa\,\rho}
\,\left(\frac{\partial\,\Gamma}{\partial\,\xi}\right)^2\,\phi\\ &&
+\frac{i\,\hbar}{m}\,\frac{\partial\,\Gamma}{\partial\,\xi}\,\frac{\partial\,\phi}{\partial\,\xi}
-\frac{i\,\hbar}{2\,m\,\rho}\,\frac{\partial}{\partial\,\xi}
\left(\frac{\partial\,\Gamma}{\partial\,\xi}\,\rho\right)\,\phi+F(\rho)\,\phi
\ .\label{b10}
\end{eqnarray}
We use now the relation $\partial\,\Gamma/\partial\,t=\mp
u\,\partial\,\Gamma/\partial\,\xi$ with:
\begin{eqnarray}
\frac{\partial\,\Gamma}{\partial\,\xi}=\pm
m\,u\,\frac{\kappa\,\rho}{1+\kappa\,\rho}
\end{eqnarray}
(easily derivable from (\ref{difase}) and
(\ref{b8})), Eq. (\ref{b10}) and:
\begin{eqnarray}
-\frac{i\,\hbar}{2\,m\,\rho}\,\frac{\partial\,\Gamma}{\partial\,\xi}\,\frac{\partial\,\rho}{\partial\,\xi}\,\phi
+\frac{i\,\hbar}{m}\,\frac{\partial\,\Gamma}{\partial\,\xi}\,\frac{\partial\,\phi}{\partial\,\xi}=
\frac{i\,\hbar}{2\,m}\,\left(\phi^\ast\,\frac{\partial\,\phi}{\partial\,\xi}-
\phi\,\frac{\partial\,\phi^\ast}{\partial\,\xi}\right)\,\frac{\partial\,\Gamma}{\partial\,\xi}\,\frac{\phi}{\rho}
 \ .
\end{eqnarray}
If we take into account that $j_\phi=\pm u\,\rho$ we arrive to the
following NLSE:
\begin{eqnarray}
i\,\hbar\,\frac{\partial\,\phi}{\partial\,t}=
-\frac{\hbar^2}{2\,m}\,\frac{\partial^2\,\phi}{\partial\,\xi^2}
+{1\over2}\,m\,u^2\,\kappa\,\rho\,\frac{2+\kappa\,\rho}{(1+\kappa\,\rho)^2}\,\phi+
F(\rho)\,\phi \ .\label{b12}
\end{eqnarray}
Let us now introduce the variable $x$ and define $F_{\rm eff}(\rho)$:
\begin{eqnarray}
F_{\rm eff}(\rho)=F(\rho)+{1\over2}\,m\,u^2\,\kappa\,\rho\,\frac{2+\kappa\,\rho}{(1+\kappa\,\rho)^2}
 \ ,\label{b13}
\end{eqnarray}
then Eq. (\ref{b12}) can be rewritten as:
\begin{eqnarray}
i\,\hbar\,\frac{\partial\,\phi}{\partial\,t}=-\frac{\hbar^2}{2\,m}\,\frac{\partial^2\,\phi}{\partial\,x^2}
+F_{\rm eff}(\rho)\,\phi \ .\label{b14}
\end{eqnarray}
By means of $F_{\rm eff}(\rho)=d\,U_{\rm eff}(\rho)/d\,\rho$ we can
introduce the potential $U_{\rm eff}(\rho)$ which is given by:
\begin{equation}
U_{\rm eff}(\rho)=\widetilde
U(\rho)+{1\over2}\,m\,u^2\,\kappa\,\frac{\rho^2}{1+\kappa\,\rho}
 \ .\label{veff}
\end{equation}
We remark that the term $(\kappa\,m\,u^2/2)\,\rho^2/(1+\kappa\,\rho)$
originates from the $U_{\rm EIP}(\rho,\,j)$ and represent the EIP
effect on the shape of the soliton. We remember also that the
transformation (\ref{trasf}) is noncanonical. In the case of soliton
solution (and only for this special solution) we are able to write a
new evolution equation derivable by a variational principle, starting
from the density Lagrangian:
\begin{eqnarray}
{\cal L}_{\rm eff}={\cal L}_0-U_{\rm eff} \ ,
\end{eqnarray}
where ${\cal L}_0$ is the density of Lagrangian of the linear
Schr\"odinger equation. Because of the noncanonicity of (\ref{trasf})
$U_{\rm eff}$ is not obtainable directly from $U_{\rm EIP}$.

Let us consider the transformation recently introduced by Doebner and
Goldin \cite{Doebner2,Doebner3}:
\small
\begin{eqnarray}
\psi(t,\,x)\rightarrow\phi(t,\,x)=\sqrt{\rho(t,\,x)}\,
\exp\left[i\,\left({\gamma(t)\over2}\,\log\,\rho(t,\,
x)+{\lambda(t)\over\hbar}\,S(t,\,x)+\theta(t,\,x)\right)\right] \label{goldin}
\end{eqnarray}
\normalsize Transformation (\ref{goldin}) defines a class of
non-linear gauge transformations, varying the parameters
$\gamma(t)$, $\lambda(t)$ and $\theta(t,\,x)$ and has the
important property of making linear a particular sub-family of
equations belonging to the Doebner-Goldin equation family. By
comparing Eq. (\ref{unity}) to Eq. (\ref{goldin}) we can note that
the transformation introduced in this work is a particular case of
the more general transformation introduced by Doebner and Goldin.
Transformation (\ref{unity}) is limited to the soliton states
without linearizing the Schr\"odinger equation describing these
states. It reduces the complex nonlinearity of the Schr\"odinger
equation to another real one.\\ By following the same procedure of
the previous section, we can determine the shape of the solitonic
solutions of Eq. (\ref{b14}), that are the same states of Eq.
(\ref{schroedinger6}) modulo the phase $S(\xi,\,t)$. Of course,
standing on the unitarity of Eq. (\ref{unity}), the equation for
the shape of the soliton is the same of Eq. (\ref{eq}). The
problem of searching the solitonic solutions of Eq.
(\ref{schroedinger6}) with derivative complex nonlinearities due
to the EIP, is reduced now to the search of the solitonic
solutions of a Schr\"odinger equation with analytic real
nonlinearity.

We consider now the nonlinear potential:
\begin{equation}
\widetilde U(\rho)=\widetilde
U_0(\rho)-{1\over2}\,m\,u^2\,\kappa\,\frac{\rho^2}{1+\kappa\,\rho}
\ ,\label{pr}
\end{equation}
where $\widetilde U_0(\rho)$ is again an analytic real arbitrary
potential in $\rho$ and the second term is selected with the scope
of eliminating the effect of the EIP. Eq. (\ref{eq}) now takes the
form:
\begin{equation}
\left(\frac{d\,n}{d\,\chi}\right)^2=\alpha\,n-\frac{2\,\epsilon}{m\,u^2}\,n^2
+\frac{2\,|\kappa|}{m\,u^2}\,n\,\widehat {U}_0(n) \
.\label{secondas}
\end{equation}
Equation (\ref{secondas}) is identical to the equation of the
solitonic shape that we may find in literature if we take a NLSE
with the analytic non-linear potential $\widetilde U_0(\rho)$
\cite{Hasse2}. Then Eq. (\ref{secondas}) allows us to use the
soliton solutions of NLSEs available in literature.

\section{Applications}

As a first application of the results obtained in the previous
section, we derive the nonlinear potential $\widetilde U(\rho)$
which, when is present simultaneously with the EIP potential
$U_{_{\rm EIP}}(\rho,\,{\bfm j})$, permits the formulation of a
soliton with shape given by:
$\rho(\xi)\propto[\cosh(b\,\xi)]^{-2}$. We start considering the
non-linearity \cite{Gross,Ginzburg1}:
\begin{equation}
U_0(\rho)=-{\mu\over2}\,\rho^2 \ .\label{scelta}
\end{equation}
The Schr\"odinger equation with this non-linearity has been recently
used to study the Bose-Einstein condensation (BEC)
\cite{Stringari,Holland,Edwards,Ruprecht,Smerzi}.
In the Gross-Pitayevski equation the parameter
$\mu$ is given by:
\begin{equation}
\mu=\frac{4\,\pi\,\hbar^2\,N\,a}{m} \ ,
\end{equation}
where $N$ is the number of atoms in the condensate, $m$ their mass and
$a$ is the $s$-wave triplet scattering length. Its value is assumed to
range in
the interval $85\,a_0\,<\,a\,<\,140\,a_0$, $a_0$ being the Bohr radius \cite{Gardner}.\\
Set $\alpha^\prime=0$ and
$\mu>0$, Eq. (\ref{secondas}) with the potential (\ref{scelta}) is easily integrable
obtaining:
\begin{equation}
\rho(\xi)={N\,b\over2}\,\left[\cosh(b\,\xi)\right]^{-2} \ ,\label{solit}
\end{equation}
where $b$ is a dimensionless constant defined as:
\begin{equation}
b=\frac{\mu\,m\,N}{2\,\hbar^2} \ ,
\end{equation}
and the normalization $N=\int|\psi|^2\,d\xi$, that fixes the
parameter $\epsilon=-\mu^2\,m\,N^2/8\,\hbar^2$, has been taken
into account.\\ The phase $S(\xi,\,t)$ of the soliton takes the
form:
\begin{eqnarray}
S(\xi,\,t)=-\epsilon\,t\pm m\,u\,\xi\mp m\,u\,{c\over b}\,
\tanh^{-1}\,\left[c\,\tanh(b\,\xi)\right] \ ,\label{phase1}
\end{eqnarray}
with:
\begin{equation}
c=\left(1+{2\over\kappa\,b\,N}\right)^{-\frac{1}{2}} \ ,
\end{equation}
a dimensionless constant.\\
The EIP effect modifies the phase of the soliton.
In the limit $\kappa\rightarrow0$, i.e. when the EIP is switched off, the
phase of the soliton becomes equal to the phase of the soliton of the cubic
Schr\"odinger equation.

Finally, we remark that in the case of a pure exclusion principle $(\kappa<0)$,
the soliton exists, as we can see from (\ref{phase1}), only if:
\begin{equation}
4\,\hbar^2>|\kappa|\,\mu\,m\,N^2 \ .\label{condition}
\end{equation}
If we take into account the maximum value of the quantity $\rho(\xi)$, that is
$\rho(0)=\mu\,m\,N^2/4\,\hbar^2$ and the maximum number of particles that
can be put in a site:
\begin{equation}
\rho_{\rm max}={1\over|\kappa|} \ ,
\end{equation}
Eq. (\ref{condition}) can be written in the form:
\begin{equation}
\rho(0)<\rho_{\rm max} \ .
\end{equation}
This imposes no violation of the exclusion principle in the
central site, where the maximal occupation exists and,
consequently, no violation of the exclusion principle on all the
other points of the space.\\ Taking into account Eqs. (\ref{pr})
and (\ref{scelta}), we can write the potential $\widetilde
U(\rho)$, which generates the soliton given by Eqs. (\ref{solit})
and (\ref{phase1}), as:
\begin{equation}
\widetilde
U(\rho)=-\frac{\mu}{2}\,\rho^2-\kappa\,{1\over2}\,m\,u^2\frac{\rho^2}{1+\kappa\,\rho}
 \ .
\end{equation}


\setcounter{chapter}{6}
\setcounter{section}{0}
\setcounter{equation}{0}
\chapter*{Chapter VI\\
\vspace{10mm}EIP-Gauged Schr\"odinger Model: Chern-Simon Vortices}
\markright{Chap. VI - EIP-Gauged Schr\"odinger Model: Chern-Simon Vortices}

In this chapter we describe a possible application of the model
introduced in the chapter IV, describing systems of interacting
particles obeying to EIP. Of course, the model can be used
whenever the physical circumstances are such that collective
excitations of charge particles occur like, for instance, in the
study of degenerate plasmas. Here we study in some detail the
properties of static, self-dual, Chern-Simons (CS) vortices. It
was emphasized by several authors that CS theories could describe
effects observed in the recently discovered high-$T_c$
superconductors. \\ Before discussing the EIP-vortex solutions, it
might be worthwhile to mention how such solutions were
historically discovered.\\ One of the most discussed topics in
condensed matter is undoubtedly the superconductivity phenomenon.
In the original Ginzburg-Landau model \cite{Ginzburg}, the order
parameter described by the field $\psi$ interacts with a Maxwell
like gauge field. Although the equations depend on three free
parameters, two of them can be eliminated by an appropriated
scaling, leaving only one relevant physical parameter $\lambda$.
Superconductors of type I and II correspond respectively to the
value of the parameter $\lambda$ less or greater than one.
Subsequently it was shown that the Ginzburg-Landau model admits
vortex like solutions \cite{Abrikosov}: localized flux tube
surrounded by a circulating supercurrent. These vortices were
experimentally observed in superconductors of type II. In the
Bogomol'nyi limit, $\lambda=1$ vortex like solutions acquire
interesting properties, in particular static solutions are
admitted because of the absence of forces exchanged among vortices
\cite{Bogomolnyi}.\\ Nielsen and Olesen \cite{Nielsen}
rediscovered these solutions in the context of the relativistic
generalization of the Ginzburg-Laudau model, known as the Abelian
Higgs model \cite{Hong,Vega,Weinberg}. These authors were looking
for string-like objects in relativistic field theory. It turns out
that these vortices have finite energy per unit length in 3+1
dimensions (i.e. finite energy in 2+1 dimensions as the vortex
dynamics is essentially confined to the $x-y$ plane) quantized
flux, but are electrically neutral and have zero angular momentum.
In particle physics theories these solutions may be interpreted as
strings joining confined quarks, while in cosmology theories may
be interpreted as cosmic strings produced at a phase transition in
the early history of our universe \cite{Vilenkin}.\\ Subsequently,
Julia and Zee \cite{Julia} showed that the $SO(3)$ Georgi-Glashow
model, which admits t'Hooft-Polyakov monopole solutions, also
admits its charged generalization named dyon solutions with finite
energy and finite, non zero, electric charge. It was then natural
for them to enquire whether the Abelian Higgs model, which admits
neutral vortex solutions with finite energy (in 2+1 dimensions),
also admits its charged generalization or not. In the same paper,
Julia and Zee discussed this question and showed that the answer
is negative, i.e. unlike the monopole case, the Abelian Higgs
model does not admit charged vortices with finite energy and
finite and non zero electric charge. More than ten years later,
Paul and Khare \cite{Paul} showed that the Julia-Zee negative
result can be overcome if one adds the CS term to the Abelian
Higgs model. In particular, was showed that the Abelian Higgs
model with CS term in 2+1 dimensions admits charged vortex
solutions of finite energy and quantized, finite, N\"other charge
as well as flux. As an extra bonus, it was found that these
vortices also have non zero, finite angular momentum that is in
general fractional. This strongly suggested that these charged
vortices could in fact be charged anyons, as it was rigorously
shown by Fr\"ohlich and Machetti \cite{Marchetti}.\\ In the last
years, Jackiw and Pi \cite{Pi1}, studying the nonrelativistic
reduction of the Abelian Higgs-CS model to the Schr\"odinger-CS
model, where a nonlinear potential $U(\rho)\propto\rho^2$ appears,
were able to resolve it in the self-dual limit. In particular,
this model admits nontopological vortex solutions in which the
electric charge, the magnetic flux and the angular momentum are
altogether quantized quantities, while energy and momentum are
equal to zero, a feature of the self-dual solutions. The same
model was studied in presence of an external magnetic field
\cite{Ezawa}, the importance of it being evident if we take into
account the applications of the CS theory to the fractional
quantum Hall effect \cite{Laughlin}.\\ Also the CS-Abelian Higgs
model admits self-dual solutions when the nonlinear potential
takes the form: $U(\rho)\propto\rho\,(\rho-v^2)^2$. This potential
introduces a spontaneous symmetry-breaking mechanism and
consequently the topological vortex makes its appearance in the
symmetry-breaking phase. The solutions of the relativistic model
may be considered as the analogous in (2+1) dimensions of the
magnetic monopoles of 'tHooft-Polyakov \cite{Jackiw,Hong,Lee}.\\
Self-dual planar solitons can be also found in many theories of
fermions where the dynamics is described both in the frame of the
Dirac equation with gauge field interacting by means of CS
\cite{Bhaduri} or Maxwell-CS terms \cite{Hyun}, and in the
nonrelativistic model of the L\'evy-Leblond-CS equation
\cite{Duval}. In these theories, as in the scalar ones, after a
calibration of the nonlinear potential, it is possible to show
that each component of the spinorial field satisfies the
Bogomol'nyi equation reducible to the Liouville differential
equation whose solutions are known.\\ An interesting
generalization of the Abelian Higgs-CS model was formulated by
Torres \cite{Torres} and extended by other authors
\cite{Lee1,Ghosh,Antillon}. In this model a matter field is
coupled in a non minimal way with a gauge field through the
presence of a term that behaves as an anomalous magnetic moment.
In the Bogomol'nyi limit the model admits topological and
nontopological vortex solutions with non quantized electric
charge, magnetic flux and angular momentum while in the
topological sector the energy is quantized and infinitely
degenerate.\\ For an extensive review on CS theories see Ref.
\cite{Dunne}.\\

\setcounter{equation}{0}
\section{Static solutions}

Let us describe the model studied in this chapter. We consider the
particular situation in absence of the external potential
$V(t,\,{\bfm x})$ and when the interaction of the gauge field is
described exclusively by means of the CS term. This approximation
is specially relevant in the context of condensed matter systems
since in the long wave length limit, (low energy domain), the
high-derivative Maxwell terms are dominated by the first order
derivative CS one \cite{Jackiw,Pi1,Lee}. Thus we set $\gamma=0$
and $V=0$ so that the motion equations for the matter field
obtained from (\ref{glagrangian}) and (\ref{MCS}) become:
\begin{eqnarray}
i\,\hbar\,c\,D_0\,\psi =-\frac{\hbar^2}{2\,m}{\bfm D}^2\psi
+\Lambda(\rho,\,{\bfm J})\,\psi+F(\rho)\,\psi \ ,\label{sch2}
\end{eqnarray}
with $F(\rho)=d\,\widetilde U(\rho)/\rho$ and the nonlinear term
$\Lambda(\rho,\,{\bfm J})$ given by Eq. (\ref{lambda}). As we will
show, the arbitrary nonlinear potential $U(\rho)$ can be selected
in order to permit the existence of self-dual vortex solutions for
Eq. (\ref{sch2}). In the self-dual limit we are able to decouple
the gauge fields equations from the matter ones and reduce the
evolution equation of the $\psi$ field to an ordinary differential
equation which will be solved numerically by means of the
Runge-Kutta algorithm. \\ The motion equations for the gauge
fields, when only the CS term is present, become:
\begin{equation}
\frac{g}{2}\,\varepsilon^{\nu\rho\mu}
\,F_{\rho\mu}=\frac{e}{c}\,J^\nu \ . \label{gaugefield1}
\end{equation}
In section 4.3 we have seen that Eqs. (\ref{gaugefield1}) can be solved allowing us
to write the gauge fields as a function of the
sources $\rho$ and $\bfm J$. Therefore Eq. (\ref{sch2})
can be seen as an highly nonlinear
Schr\"odinger equation for the only field $\psi$.
The model that we come to study is  a continuous
deformation, in the parameter $\kappa$, of the model presented and studied
in Ref. \cite{Jackiw} by Jackiw and Pi.

To search the static solutions of Eqs. (\ref{sch2}) and (\ref{gaugefield1})
we make use of the property of invariance under gauge
transformation of the Lagrangian:
\begin{eqnarray}
\psi&\rightarrow& e^{-i(e/\hbar\,c)\,\omega}\psi \ , \label{gauge11}\\
A_{\mu}&\rightarrow& A_{\mu}+\partial_{\mu}\omega \ .\label{gauge21}
\end{eqnarray}
We choose to work in the London gauge where the matter
field $\psi$ becomes real,
because its phase $Arg(\psi)$ is absorbed by the field $A_\mu$.
By using Eqs. (\ref{gauge11}) and (\ref{gauge21}), and
setting the parameter $\omega=(\hbar\,c/e)\,Arg(\psi)$, we can see that
after the introduction of the new fields:
\begin{eqnarray}
&&\phi ({\bfm x})=|\psi ({\bfm x})| \ , \\
&&\chi _{\mu }({\bfm x})=A_{\mu }({\bfm x})+\frac{\hbar\,c}{e}\partial _{\mu
}\,Arg\,[\psi ({\bfm x})] \ ,
\end{eqnarray}
the expression of the Hamiltonian becomes:
\begin{equation}
H=\int \left[ \frac{\hbar ^{2}}{8\,m}\,\frac{(\bfm\nabla
\,\rho)^{2}}{\rho}+
\frac{e^{2}}{2\,m\,c^{2}}\,\rho\,(1+\kappa\,\rho)\,\mbox{\boldmath$\chi$}
^2+\widetilde U(\rho )\right] \,d^{2}x\ ,\label{ee}
\end{equation}
where $\rho=\phi^2$ is the particle density. We observe that for
$\kappa<0$ we have $0\leq\rho\leq1/|\kappa|$ while for $\kappa>0$
we have $\rho>0$ so that $1+\kappa\,\rho\geq0$. Therefore, from
(\ref{ee}) we see that, if $\widetilde U(\rho)=0$, $H$ is a
semidefinite positive quantity limited below by zero. If we take
into account the relation:
\begin{eqnarray}
\nonumber
&&\Big|\partial _{\pm }\log \rho +i\,\frac{2\,e}{\hbar \,c}\,(1+\kappa
\,\rho)^{1/2}\,\chi _{\pm }\Big|^2=\left( \frac{\bfm\nabla \rho }{\rho }\right)
^2+\left( \frac{2\,e}{\hbar \,c}\right)^2\,(1+\kappa \,\rho )\,\mbox{\boldmath$\chi$}
^2 \\
&&+i\frac{2\,e}{\hbar \,c}\,(1+\kappa\,\rho)^{1/2}\,(\chi_\pm
\,\partial_\mp
\log \rho-\chi _\mp\,\partial _\pm\log \rho )\ ,
\end{eqnarray}
with $\partial_\pm=\partial_1\pm i\,\partial_2$ and $\chi_\pm=\chi_1\pm i\,\chi_2$, the Hamiltonian becomes:
\begin{eqnarray}
H &=&\frac{\hbar ^{2}}{8\,m}\int \Big| \partial _{\pm }\log \rho
+i\, \frac{2\,e}{\hbar \,c}\,(1+\kappa \,\rho)^{1/2}\,\chi _{\pm
}\Big| ^2\,\rho\,d^{2}x  \nonumber \\ &\mp& \frac{e\,\hbar
}{3\,m\,c\,\kappa}\,\epsilon ^{ij}\int\chi _{i}\,\partial _j\left[
(1+\kappa \,\rho )^{3/2}-1\right] \ d^2x +\int\widetilde U(\rho
)\,d^2x \ . \label{en2}
\end{eqnarray}
The second term in Eq. (\ref{en2}) is obtained by taking into account the identity:
\begin{equation}
(1+\kappa \,\rho)^{1/2}\,\partial_i\rho =\frac{2}{3\,\kappa}\,\partial
_i\,\left[ (1+\kappa \,\rho )^{3/2}-1\right] \ .\label{exp}
\end{equation}
This term can be integrated by part. If we neglect the
surface terms and take into account the Gauss law given by the time component of Eq. (\ref{gaugefield1}):
\begin{equation}
-g\,B=e\,\rho \ ,\label{Gauss2}
\end{equation}
the Hamiltonian assumes the form:
\begin{eqnarray}
\nonumber H&=&\frac{\hbar^{2}}{8\,m}\int\Big|\partial_{\pm
}\log\rho+i\,
\frac{2\,e}{\hbar\,c}\,(1+\kappa\,\rho)^{1/2}\,\chi_{\pm}\Big|
^{2}\,\rho\,d^{2}x\\
&&+\int\left\{\widetilde{U}(\rho)\mp\frac{e^{2}\,\hbar}{3\,m\,c\,g\,\kappa}\left[(1+\kappa\,\rho
)^{3/2}-1\right]\,\rho\right\}\,d^{2}x \ .\nonumber\\ \label{ee1}
\end{eqnarray}
We introduce now the quantities:
\begin{eqnarray}
{\cal H}_0&=&\frac{\hbar^2}{8\,m}\,\Big|\partial_{\pm }\log\rho+i\,
\frac{2\,e}{\hbar\,c}\,(1+\kappa\,\rho)^{1/2}\,\chi_{\pm}\Big|
^{2}\,\rho \ ,\\
K(\rho)&=&{1\over\kappa}\,\left[(1+\kappa\,\rho)^{3/2}-1\right]\,\rho \ ,\\
\alpha&=&\mp\frac{e^2}{\hbar\,c\,g} \ ,
\end{eqnarray}
and observe that $K(\rho)\geq0$, $\forall\,\kappa\in I\!\!R$ and
$\rho\geq0$. The Hamiltonian assumes the form:
\begin{eqnarray}
H=\int{\cal H}_0\,d^2x+\int\left[\widetilde
U(\rho)+\frac{\hbar^2\,\alpha}{3\,m}\,K(\rho)\right]\,d^2x \ .
\end{eqnarray}
We observe that, independently on the sign of $\alpha$ (which will
be discuss below), the quantity ${\cal
H}_0+(\hbar^2\,\alpha/3\,m)\,K(\rho)\geq0$.\\ Due to the
canonicity of the system, the motion equation (\ref{sch2}) in the
Hamilton formalism and in the case of static solutions becomes:
\begin{equation}
\frac{\delta \,H}{\delta \,\psi^*}=0 \ ,
\end{equation}
which can be written as:
\begin{eqnarray}
\frac{\delta}{\delta\,\psi^\ast}\,\int{\cal
H}_0\,d^2x+\frac{\partial}{\partial\,\rho} \left[\widetilde
U(\rho)+\frac{\hbar^2\,\alpha}{3\,m}\,K(\rho)\right]\,\psi=0 \ .
\end{eqnarray}
Following Ref. \cite{Jackiw,Pi1} in which the Jackiw and Pi model
($\kappa=0$) is discussed, it is easy to realize that the
relations:
\begin{eqnarray}
&&\partial_{\pm}\log\rho+i\,\frac{2\,e}{\hbar\,c}\,(1+\kappa
\,\rho)^{1/2}\,\chi_{\pm }=0\ . \label{bogo}\\
&&U(\rho)+\frac{\alpha\,\hbar^2}{3\,m}\,K(\rho)=0 \ ,\label{potential}
\end{eqnarray}
ensure that the energy of the system vanishes. Eqs. (\ref{bogo})
are two differential equations of the first order that permit to
write the gauge field $\chi_+$ and $\chi_-$ as functions of the
field $\rho$, while the (\ref{potential}) fixes the analytical
expression of the nonlinear potential $\widetilde U(\rho)$. In the
limit $\kappa\rightarrow0$, Eq. (\ref{bogo}) reduces to the
Bogomol'nyi equation \cite{Bogomolnyi}, while in the same limit
the potential $\widetilde U(\rho)$ given by (\ref{potential})
becomes proportional to $\rho^2$, in agreement with the results of
Ref. \cite{Jackiw,Pi1}. Making use of Eqs. (\ref{bogo}) we can
determine the field $\bfm\chi$ if the expression of the field
$\rho$ is known:
\begin{equation}
\mbox{\boldmath$\chi$}=\pm\frac{\hbar\,c}{2\,e}\frac{\bfm\nabla\wedge\log\rho}{
(1+\kappa\,\rho)^{1/2}} \ .\label{chi}
\end{equation}
If we take into account the expression of the current (\ref{covariantcurrent}), in the London gauge:
\begin{equation}
{\bfm
J}=-\frac{e}{m\,c}\,\rho\,(1+\kappa\,\rho)\,\mbox{\boldmath$\chi$} \ ,
\end{equation}
and using Eq. (\ref{chi}), we can write $\bfm J$ as a function of the field $\rho$:
\begin{equation}
{\bfm J}=\mp\frac{\hbar}{2\,m}\,(1+\kappa\,\rho)^{1/2}
\,\bfm\nabla\wedge\rho \ .\label{cur}
\end{equation}
It is easy to verify that $\bfm J$ can be rewritten as a curl and therefore it is
a fully transverse current.
Finally, after integration of the relation:
\begin{equation}
\partial _{i}\,A^{0}=-E^i=-\frac{e}{c\,g}\,\epsilon_{ij}\,J^{j}\ ,
\end{equation}
we obtain:
\begin{equation}
A_{_{0}}\equiv \chi _{_{0}}=\frac{\hbar^2\,\alpha}
{3\,m\,e}\,{1\over\rho}\,K(\rho) \ .\label{chio}
\end{equation}
Equations (\ref{chi}) and (\ref{chio}) allow us to obtain the
gauge field $\chi_\mu$ when the shape of the matter field $\rho$
is known. In the limit $\kappa\rightarrow0$, the field $\chi_\mu$
assumes the expression given in Ref. \cite{Jackiw,Pi1}, so we
conclude that the model described by the Lagrangians
(\ref{glagrangian}) and (\ref{MCS}) generalizes the Jackiw and Pi
model when the system obeys to EIP.

Now we calculate the field $\rho$. Taking into account the relation
$B={\bfm\nabla}\wedge{\bfm\chi}$, from (\ref{chi}) it follows that:
\begin{equation}
B=\pm\frac{\hbar\,c}{2\,e}{\bfm\nabla}\wedge\left[\frac{{\bfm\nabla}\wedge\log\rho}
{(1+\kappa\,\rho)^{1/2}}\right] \ .
\end{equation}
By using the Gauss law (\ref{Gauss2}), we obtain the following
second order differential equation for the field $\rho$:
\begin{equation}
\Delta\,\log\left[\frac{4}{\kappa}\,\,\frac{(1+\kappa\,\rho)^{1/2}-1
}{(1+\kappa\,\rho)^{1/2}+1}\right]=-2\,\alpha\,\rho \ , \label{lio}
\end{equation}
which, in the limit of $\kappa\rightarrow0$, reduces to the
Liouville differential equation:
$\Delta\,\log\rho=-2\,\alpha\,\rho$. We are not able to find the
analytical solutions of Eq. (\ref{lio}). Numerical radially
symmetrical solutions for the fields $\rho$ will be discussed in
section 6.3. Here we only anticipate that, how in the limit case
$\kappa=0$, nonsingular, nonnegative solutions will be obtained
when the numerical constant $\alpha$ is positive. Hence the $\mp$
sign must be chosen according that of $g$.\\ We consider now Eq.
(\ref{sch2}) in the London gauge for the static configurations:
\begin{equation}
\frac{\hbar
^2}{2\,m}\frac{\Delta\rho^{1/2}}{\rho^{1/2}}=\frac{e^2}{
2\,m\,c^2}(1+2\,\kappa\,\rho)\,\mbox{\boldmath$\chi$}^2+e\,\chi_{0}+F(\rho)
\ .
\end{equation}
It is easy to verify that the fields $\bfm\chi$, $\chi_0$ and $\rho$ given by Eqs. (\ref{chi}), (\ref{chio})
and (\ref{lio}), are the required solutions.\\
We remark that the self-dual
character of Jackiw and Pi static solutions is not altered
by the presence of EIP. This self-duality can be easily
recognized if we define the following transformation:
\begin{equation}
\psi\rightarrow\eta=R(r)^{1/2}\,e^{-i\,(e/\hbar\,c)\,S} \ ,
\end{equation}
that changes the amplitude of the field $\psi$ in:
\begin{equation}
\rho(r)\rightarrow R(r)=\frac{4}{\kappa}\,\,\frac{(1+\kappa\,\rho)^{1/2}-1
}{(1+\kappa\,\rho)^{1/2}+1} \ .
\end{equation}
Now Eq. (\ref{bogo}) in the $\eta$ field becomes:
\begin{equation}
{\bfm D}_\pm\,\eta=0 \ .
\label{duale}
\end{equation}
As a consequence, the solutions of Eq. (\ref{bogo}) are the ground
states of the system. This can be easily demonstrated by computing
the component of the energy-momentum tensor. We have previously
noted that the energy vanishes for the solutions of Eq.
(\ref{bogo}) well-behaved at infinity. The density of momentum
$T^{0i}$ obtained in section 4.3, can be written by using
(\ref{bogo}) as:
\begin{equation}
T^{0i}=\mp\frac{c\,\hbar}{\kappa}\,\epsilon^{ij}\partial_j\left[(1+\kappa\,\rho)^{1/2}-1\right] \ ,
\end{equation}
that appears to be transverse. In the same way, we can show that the
flux density of energy and of linear momentum are:
\begin{eqnarray}
T^{i0}
&=&\frac{\epsilon^{ij}}{2\,g}\left(\frac{e\,\hbar}{3\,m\,c\,k}\right)^2\,
\partial_j\left[(1+\kappa\,\rho)^3-2\,(1+\kappa\,\rho)^{3/2}+1\right] \ , \label{tio}\\
T^{ij} &=&0 \ ,
\end{eqnarray}
and (\ref{tio}) appears to be a transverse quantity. Then, apart from total
derivative terms, the energy-momentum tensor vanishes for the solutions of Eq. (\ref{bogo}).\\
Finally, the expression of the density of angular momentum is given by:
\begin{eqnarray}
\epsilon_{ij}\,x^i\,T^{0j}=\pm\frac{\hbar\,c}{\kappa}\,{\bfm\nabla}\cdot
\left\{[(1+\kappa\,\rho)^{1/2}-1]\,{\bfm x}\right\}
\mp\frac{2\,\hbar\,c}{\kappa}
\left[(1+\kappa\,\rho)^{1/2}-1\right] \ ,
\end{eqnarray}
and, if the field $\rho$ vanishes at infinity more rapidly than $1/x^2$, the expression of the angular
momentum becomes:
\begin{equation}
<\,M\,>=\mp\frac{2\,\hbar}{\kappa}\int[(1+\kappa\,\rho)^{1/2}-1]\,d^2x \ .
\end{equation}
An alternative expression, useful in the next section, can be obtained starting from
Eq. (\ref{mom2}) that we rewrite for convenience:
\begin{eqnarray}
<\,M\,>=-\frac{i\,\hbar}{2}\int{\bfm x}\wedge(\psi^\ast\,{\bfm\nabla}
\psi-\psi\,{\bfm\nabla}\psi^\ast)\,d^2x-\frac{e}{c}\int{\bfm
x}\wedge{\bfm A}\,\rho\,d^2x \ ,
\end{eqnarray}
that, in the London gauge, assumes the form:
\begin{equation}
<\,M\,>=-\frac{e}{c}\int({\bfm x}\wedge{\bfm\chi})\,\rho\,d^2x \ .
\end{equation}
Using Eq. (\ref{Gauss2}) it becomes:
\begin{equation}
<\,M\,>=\frac{g}{c}\int({\bfm x}\wedge{\bfm\chi})\cdot({\bfm\nabla}\wedge{\bfm\chi})\,d^2x \ ,
\end{equation}
which, with easy algebraic computation and taking into account
the transversally of the field $\bfm\chi$, can be written as:
\begin{equation}
<\,M\,>=\frac{g}{c}\oint\left[\frac{1}{2}\,({\bfm\chi}^2)\,{\bfm x}-
({\bfm x}\cdot{\bfm\chi})\,{\bfm\chi}\right]\wedge
d{\bfm l} \ ,\label{mom}
\end{equation}
where the integral is to be taken both around a circle at infinity
and around infinitesimal contours surrounding the poles of $\bfm\chi$ (zeros of $\rho$).

\setcounter{equation}{0}
\section{Vortex like solutions}
We study the solutions with angular symmetry (vortices) of the system.
The wave function $\psi$ for a vortex takes the form:
\begin{equation}
\psi(r,\,\theta)=\rho(r)^{1/2}\,\exp\left[-\frac{i\,e}{\hbar\,c}\,S(r,\,\theta)\right]
 \ ,
\end{equation}
where $\rho(r)$ satisfies the equation (\ref{lio}) that in polar
coordinates becomes:
\begin{equation}
\frac{d}{d\,r}\left[r\,\frac{d}{d\,r}\log\frac{4}{\kappa}\,\,
\frac{(1+\kappa\,\rho)^{1/2}-1}{(1+\kappa\,\rho)^{1/2}
+1}\right]=
-2\alpha\,\rho\,r \ .\label{polar}
\end{equation}
Analytical solutions of (\ref{polar}) are not known, however in
the limit of $\kappa\,\rho\rightarrow0$ Eq. (\ref{polar}) becomes
the Liouville differential equation whose solutions are known
\cite{Liouville}.\\ In section 6.1 we have shown that the value of
the main physical observable associated to the solutions of Eq.
(\ref{bogo}) is zero. This statement is true with the exception of
the mass, of the angular momentum and of the electric charge that,
for the vortex solutions, is given by:
\begin{equation}
Q=2\,\pi\,e\int\limits_0\limits^\infty\rho(r)\,r\,dr \ .\label{pp}
\end{equation}
Using Eq. (\ref{polar}), we can write:
\begin{equation}
Q=\frac{\pi\,e}{\alpha}\left[\lim_{r\rightarrow0}\,r\,f(\kappa\,\rho)
-\lim_{r\rightarrow\infty}\,r\,f(\kappa\,\rho)\right] \ , \label{carica}
\end{equation}
with:
\begin{equation}
f(\kappa\,\rho)=\frac{1}{(1+\kappa\,\rho)^{1/2}}\frac{d}{d\,r}\log(\kappa\,\rho)
 \ ,
\end{equation}
where only the asymptotic behavior of $\rho$ is invoked.\\ We
remark that finite energy solutions of the system described by
Hamiltonian (\ref{ee}) with the nonlinear potential given by
(\ref{potential}) must vanish to infinity, while around the origin
become a constant which can also be zero. We can expand the
solution of Eq. (\ref{polar}) in correspondence of the zeros of
$\rho$ in a power series of $\rho$ and, taking only the first
order terms, we can see that the following equation
$\rho^{\prime\prime}-(\rho^\prime)^2/\rho+\rho^\prime/r=0$, where
the prime indicates a derivative with respect to the $r$ variable,
must be satisfied. This equation admits power like solutions, so
that we can write the following asymptotic behaviors of the
solutions of Eq. (\ref{lio}):
\begin{equation}
\rho=C_0\,r^\beta \ ,\hspace{20mm}\beta>0 \ ,\hspace{20mm}r\rightarrow0
\ , \label{asi1}
\end{equation}
\begin{equation}
\rho=C_\infty\,r^{-\gamma} \ ,\hspace{20mm}\gamma>0 \
,\hspace{20mm}r\rightarrow\infty \ ,\label{asi2}
\end{equation}
where $C_0,\,C_\infty,\,\beta$ and $\gamma$ are integration constants.\\
Inserting (\ref{asi1}) and (\ref{asi2}) into Eq. (\ref{carica}) we obtain the expression of the charge:
\begin{equation}
Q=\frac{\pi\,e}{\alpha}\,(\beta+\gamma) \ ,\label{q1}
\end{equation}
holding for solutions vanishing at the origin.\\
Differently, we note from (\ref{carica}) that if the solutions $\rho(r)$ at the origin
are equal to a constant, only their behavior at the infinity determines the electric
charge. In this case the expression of the charge $Q$ becomes:
\begin{equation}
Q=\frac{\pi\,e}{\alpha}\,\gamma \ .\label{q2}
\end{equation}
By comparing Eqs. (\ref{q1}) and (\ref{q2}) we can conclude that the expression of the charge $Q$ is
given by (\ref{q1}) with $\gamma>0$ and $\beta\geq0$. Taking into account Eq. (\ref{chi}), we can deduce the
asymptotic behavior for $r\rightarrow\infty$ of the gauge field $\chi^i$:
\begin{equation}
\chi^i
{\atop\stackrel{\Huge\sim}{\scriptstyle r\rightarrow
\infty}}\pm\frac{\hbar\,c}{2\,e}\,\gamma\,\frac{\epsilon^{ij}\,x_j}{r^2} \ ,\label{chi2}
\end{equation}
and
\begin{equation}
\chi^i{\atop\stackrel{\Huge\sim}{\scriptstyle r\rightarrow
0}}\mp\frac{\hbar\,c}{2\,e}\,\beta\,\frac{\epsilon^{ij}\,x_j}{r^2} \ ,\label{chi1}
\end{equation}
for $r\rightarrow0$ when $\rho\rightarrow0$. The singularity of $\chi^i$
for $r\rightarrow0$ is due to the particular choice of the
gauge and can be eliminated by making a gauge transformation
$U=\exp(i\,e\,\omega/\hbar\,c)$ with:
\begin{equation}
\omega=\mp\frac{\hbar\,c}{2\,e}\,\beta\,\theta
\hspace{5mm};\hspace{10mm}\theta=\arctan(x^2/x^1) \ .\label{phase}
\end{equation}
By using the (\ref{asi1}) and (\ref{phase}) we can write the expression of the field $\psi$ for small values of $r$ as:
\begin{equation}
\psi(r,\,\theta)=C_0\,r^\beta\,e^{\pm i\,\frac{\beta}{2}\,\theta} \ .
\end{equation}
In order to derive the field $\psi$ as a monodrome function,
the positive number $\beta$ should be an even integer:
\begin{equation}
\beta=2\,(\nu-1) \ ,
\end{equation}
where $\nu\in I\!\!N$ is the vorticity of the system. Therefore,
we can write the charge $Q$ in the form:
\begin{equation}
Q=\frac{\pi\,e}{\alpha}\,[2\,\nu+\gamma-2] \ .\label{qq1}
\end{equation}
In section 6.2 we have remarked that when $\kappa\rightarrow0$ the
model studied in this paper reduces to the model of Jackiw and Pi.
Therefore, in this limit, we have $\gamma\rightarrow2\,(\nu+1)$
and, consequently, the discretization of the charge
$Q=4\,\pi\,e\,\nu/\alpha$ \cite{Jackiw,Pi1}. This observation
enables us to write $\gamma$ in the form
$\gamma=2(\xi_{_{\nu,\kappa}}+1)$ with $\xi_{_{\nu,\kappa}}=\nu$
when $\kappa\rightarrow0$. Therefore the electric charge $Q$ can
be written as:
\begin{equation}
Q=\frac{2\,\pi\,e}{\alpha}\,(\nu+\xi_{_{\nu,\kappa}}) \ .
\end{equation}
In the case $\kappa\not=0$ the parameter $\xi_{_{\nu,\kappa}}$ is a continuous function of
the boundary conditions. As a consequence the charge $Q$ loses its
discretization. The behavior of the function $\xi_{_{1,\kappa}}$ for a system
with vorticity $\nu=1$ will be studied in the next section. Finally,
an alternative expression of the charge $Q$ can be obtained from the
integral form of Eq. (\ref{Gauss2}):
\begin{equation}
Q=-g\int\bfm\nabla\wedge{\bfm \chi}\,d^2x=
-g\oint{\bfm \chi}\cdot d{\bfm l} \ ,\label{carica1}
\end{equation}
where the integral is performed both on the boundary at infinity and
on the infinitesimal circles around the poles of the field $\bfm\chi$ (zeros of $\rho$).
By using (\ref{chi2}) and (\ref{chi1}) we obtain again the (\ref{q1}) and (\ref{q2}).
This result means that the poles of the field $\bfm\chi$ (zeros of $\rho$) are
placed at the origin and at the infinity.\\
Finally, from Eq. (\ref{mom}), taking into account the asymptotic behavior of
$\bfm\chi$ and
using (\ref{q2}) we obtain the angular momentum for the 1-vortex solutions:
\begin{equation}
<\,M\,>=\frac{Q^2}{4\,\pi\,c\,g}
 \ .\label{m1}
\end{equation}
Analogously by using (\ref{qq1}) we obtain the following expression of
$<\,M\,>$ for the $\nu$-vortex solution:
\begin{equation}
<\,M\,>=\frac{e\,Q\,(\xi_{_{\nu,\kappa}}+1)}{c\,g\,\alpha}-\frac{Q^2}{4\,\pi\,c\,g} \ .\label{m2}
\end{equation}
We observe that Eq. (\ref{m2}) in the case $\nu=1$ reproduces
exactly the (\ref{m1}) and then it holds for $\forall\,\nu\in
I\!\!N$.  We recall now that the spin of the $\nu$-vortex
solutions is given by $<\,S\,>=-Q^2/4\pi\,c\,g$. From (\ref{m2})
we have immediately the following expression for the orbital
angular momentum $<\,L\,>=<\,M\,>-<\,S\,>$:
\begin{equation}
<\,L\,>=\frac{e\,Q\,(\xi_{_{\nu,\kappa}}+1)}{c\,g\,\alpha} \ .
\end{equation}

\setcounter{equation}{0}
\section{Numerical analysis}
In section 3.2, we have shown that the transformation:
\begin{eqnarray}
\nonumber
\kappa&\rightarrow&\lambda^{-2}\,\kappa \ ,\\
\psi(t,\,{\bfm x})&\rightarrow&\lambda\,\psi(t,\,{\bfm x}) \ , \label{tras}
\end{eqnarray}
can be used to rescale, in the evolution equation of the system,
the value of $\kappa$ to $\pm1$. Introducing now the adimensional
variables:
\begin{eqnarray}
\nonumber &&y=\sqrt{\frac{2\,\alpha}{|\kappa|}}\,r \ ,\\
\nonumber\label{rip}\\
&&n=|\kappa|\,\rho \ ,
\end{eqnarray}
Eq. (\ref{polar}) becomes:
\begin{equation}
\frac{d}{d\,y}\left\{y\,\frac{d}{d\,y}\log
\frac{[1+\sigma\,n(y)]^{1/2}-1}{[1+\sigma\,n(y)]^{1/2}+1}\right\}=-n(y)\,y
\ ,\label{eq1}
\end{equation}
where $\sigma=\kappa/|\kappa|$ takes the value $+1$ when the inclusion principle
holds $(\kappa>0,\, n\geq0)$ and, analogously for the exclusion principle
$(\kappa<0,\,0\leq n\leq1)$, we have $\sigma=-1$. If we take into account Eq.
(\ref{rip}), the expression for the matter field becomes:
\begin{equation}
\rho_\kappa(r)=\frac{1}{|\kappa|}\,n\left(\sqrt
{\frac{2\,\alpha}{|\kappa|}}\,r\right) \ .
\end{equation}
In order to integrate numerically Eq. (\ref{eq1}), let us introduce the auxiliary field $z(y)$:
\begin{equation}
z(y)=y\,\frac{d}{d\,y}\log
\frac{[1+\sigma\,n(y)]^{1/2}-1}{[1+\sigma\,n(y)]^{1/2}+1} \ ,
\end{equation}
so that the second order differential equation (\ref{eq1}) can be
transformed into a system of two first order differential equations:
\begin{eqnarray}
&&\frac{d\,n(y)}{d\,y}=n(y)\,[1+\sigma\,n(y)]^{1/2}\,\frac{z(y)}{y} \ ,\label{sistem1}\\
&&\frac{d\,z(y)}{d\,y}=-y\,n(y) \ .\label{sistem2}
\end{eqnarray}
The system (\ref{sistem1})-(\ref{sistem2}) can be integrated
numerically by using of the Runge-Kutta method, obtaining the
shape of the field $n(y)$.

In the following we discuss a few numerical results. In figure 6.1
are plotted the normalized shapes of a 1-vortex $(\nu=1)$ for the
three cases $\kappa=0,\,\pm1$. Because of the proportionality
between $\rho$ and the magnetic field $B$, figure 6.1 reproduces
also the behavior of the field $B$ having the same polarity in
each point of the space and a toroidal configuration around the
core of the vortex. This is strictly true when the CS coupling
constant $g$ is negative. For positive values of $g$ the behavior
of the magnetic field $B$ have opposed sign to the field $\rho$
and therefore the magnetic flux attached to each particle is
opposed to the case $g<0$. This is in agreement with the following
symmetry that hold in presence of the CS term: $g\rightarrow-g$,
$x^1\leftrightarrow x^2$, $A^1\leftrightarrow A^2$.\\ In figure
6.2, the electric field $E_r$ for the same 1-vortex of figure 6.1,
is plotted as a function of $r$. The polarity of $E_r$ for the
nontopological 1-vortex results to be always positive and directed
radially.\\ In figure 6.3 and 6.4 the shape of the fields $\rho$
and $E_r$, in the case of a 2-vortex $(\nu=2)$ are plotted. Now
the field $B$ has an annular shape around the core of the vortex
and vanishes at the origin and the infinity. As a consequence, the
electric field $E_r$ takes a polarity inversion corresponding to
the points of maximum of the field $B$ and it is distributed in
two annular concentric regions of opposite polarity placed around
the core of the vortex. All the curves show the effect of the EIP
consisting in a main localization ($\kappa>0$) or delocalization
($\kappa<0$) of the vortex. The same effect can be observed also
in the shape of the electric field. Moreover, the intensity of the
maximum of the field $E_r$ is emphasized in the attractive system
respect to the repulsive one, in agreement to the inclusion or
exclusion effect produced by the presence of EIP. Finally, in
figure 6.5 it is shown the behavior of the function
$\xi_{_{\nu,\kappa}}$ for an 1-vortex in the three cases
$\kappa=0,\,\pm1$ as a function of the value of the field
$\rho(r)$ at the origin, here indicated as $\rho_{_{\rm max}}$,
which we assume as an initial condition to integrate Eq.
(\ref{eq1}) (the other condition is given by $d\,\rho/d\,r=0$ for
$r=0$). We note that in the case $\kappa=0$ the function
$\xi_{_{1,0}}$ does not depend on $\rho_{_{\rm max}}$ and its
value is 1. This result, obtained numerically, is equal to the one
derived analytically within the model of Jackiw and Pi, and
implies the discretization of the electric charge. On the
contrary, in the case $\kappa=\pm1$ the quantity $\xi_{_{1,0}}$ is
a continuous increasing $(\nu=-1)$ or decreasing $(\nu=1)$
monotonic function of $\rho_{_{\rm max}}$. \\ This dependence of
$\xi_{_{1,\pm1}}$ on the initial condition implies that now $Q$
and $<M>$ lose their discretization and become continuous
quantities. In fact, the value of the charge and of the other
quantities depend from the asymptotic behavior of the fields
around the zeros of $\rho$ which turn out to be function of the
integration constant. Now, the requirement that the field $\psi$
should be a single value forces the constant $C_0$ to be an
integer, but no condition are met for $C_\infty$.\\ In presence of
the EIP we can observe a new fact. From figure 6.5 we note that
when $\kappa=-1$ the parameter $\xi_{_{1,-1}}$ and than also $Q$
and $<M>$ have an upper bound. So, as for other nonanalitically
integrable models we have obtained continuous quantities but for a
repulsive system these quantities can run in a limited range.
Numerically we obtain $\xi_{_{1,-1}}^{^{\rm max}}\simeq1.156$.

In conclusion, we have studied the static solutions of a model
describing a many body system in the mean field approximation,
obeying to a generalized exclusion-inclusion principle and in the
presence of the Chern-Simons interaction. By selecting the
nonlinear potential $\widetilde U(\rho)$ in the form
(\ref{potential}) we have shown the existence of self-dual static
solutions, satisfying a nonlinear first order differential
equation {\sl \'a la Bogomol'nyi}. The solutions are states with
zero energy and linear momentum. Subsequently, we have considered
the subset of the vortex-like solutions, obtaining the expression
of the electric charge and the angular momentum, while their shape
has been determinate by numerical integration of Eq. (\ref{eq1}).
The model here studied can be considered as a continuum
deformation of the Jackiw and Pi one \cite{Jackiw,Pi1} performed
by the parameter $\kappa$ introduced by the exclusion-inclusion
principle. In the model of Jackiw and Pi the vortex solution takes
discrete values for the charge, magnetic flux and angular momentum
which are proportional to the vorticity number. In the present
model these quantities take values in the continuum and when the
system obeys to an exclusion principle ($\kappa<0$) it has an
upper bound.\\ Finally, we remark that in the model of Jackiw and
Pi, dynamical solutions can be obtained from the static ones by a
boosting, as a consequence of its invariance respect to the
Galilei symmetry. On the other hand, in the frame of our model, as
we have shown in chapter III, the EIP potential breaks this
symmetry, so that we must specifically study nonstatic solutions.


\setcounter{chapter}{7}
\chapter*{Chapter VII\\
\vspace{10mm}Conclusions}
\markright{Chap. VII - Conclusions}

In this work a nonlinear canonical theory describing systems of
identical particles obeying to a generalized exclusion-inclusion
principle (EIP) has been developed. In the mean field
approximation EIP takes into account collective effects of
repulsive (exclusion) or attractive (inclusion) character.\\ The
system here considered is described by a nonlinear Schr\"odinger
equation obtained, in the picture of the canonical quantization,
from a classical model obeying to EIP, requiring that the quantum
current density in the continuity equation takes the expression
${\bfm j}_\kappa={\bfm j}_0\,(1+\kappa\,\rho)$ where ${\bfm j}_0$
is the standard quantum current density of the linear theory.\\ To
this purpose we have introduced in the evolution equation a
complex nonlinear term $\Lambda(\rho,\,{\bfm j})$ deriving from a
nonlinear potential $U(\psi,\,\psi^\ast)$ which simulates a
collective effect among particles. The expression of this
nonlinearity is strongly affected by the quantization method
adopted. In fact, the kinetic approach here used fixes only the
imaginary part of $\Lambda(\rho,\,{\bf j})$. The real part is
determined by the requirement that the system be canonical, which
leads to an expression for Re$\Lambda(\rho,\,{\bfm j})$ defined
modulo of a quantity obtainable from an arbitrary real potential
$U(\rho)$. This arbitrarily allows to consider within EIP other
interactions acting among the particles.\\ After the introduction
of the Lagrangian and Hamiltonian functions, we have taken into
account the Ehrenfest relations. From their analysis we obtain
that the canonical systems obeying EIP, also in the presence of an
arbitrary potential $U(\rho)$ and in the absence of external
forces, are non dissipative processes ($E$, $<P>$ and $<M>$ are
conserved) and obey to a nonlinear kinetic ($d<{\bfm
x}_c>/dt=\int\,{\bfm j}_0\,(1+\kappa\,\rho)\,dx$ where $<{\bfm
x}_c>$ is the mean value of the central mass of the system). Thus,
EIP potential $U_{\rm EIP}$ does not modify the dynamic behavior
which can be affected only by the presence of an external
potential $V$. Conversely, $U_{\rm EIP}$ is responsible for the
formation of localized stationary states (solitons).\\ These
properties of the nonlinear EIP potential are obtained rigorously
studying the symmetries of Schr\"odinger equation. By studying the
space-time translations and using the N\"other theorem, the
expression of the energy-momentum tensor and the related conserved
quantities have been obtained. The main results are:
\begin{enumerate}
\item EIP changes the expression of the energy density $T^{00}$ and of
the fluxes $T^{i0}$ and $T^{ij}$. The energy density is a
semidefinite positive quantity both for positive and negative
value of $\kappa$, consistently with a nonrelativistic theory.
\item The expression of the momentum density is different from that of
the current (${\bfm P}\not=m\,{\bfm j}$) and, as a consequence, the
Galilei invariance is lost.
\item The presence of $U_{\rm EIP}$ introduces in the system a
dimensional coupling constant ($[\kappa]=L^{-D}$) and, as a consequence,
the conformal symmetry of the system is lost.
\item We have found a scaling transformation which permits us to reduce
the coupling constant $\kappa$ to 1 when the inclusion principle holds, or
to -1 when the exclusion principle holds.
\item The discretized symmetries $P$ and $T$ are not lost in the presence of
$U_{\rm EIP}$.
\end{enumerate}
A powerful issue, for its generality, is obtained introducing a
class of nonlinear unitary transformations. In the Schr\"odinger
equation the nonlinear term introduced by EIP is a complex one. By
means of an appropriate transformation, it is possible to make
real this quantity. As a consequence, the transformed system,
which in general is described again by a nonlinear Schr\"odinger
equation, obeys to a linear continuity equation where the quantum
current is the same as in the linear theory. This method,
introduced by us for EIP systems, can be applied to a large class
of nonlinear Schr\"odinger equations obtained from a variational
principle.\\ As an application of the model describing the
dynamics of a collective neutral quantum particles obeying to EIP
we have studied the property of solitary wave solutions. These
results can be used to study Bose-Einstein condensates recently
obtained at low temperature in rarefied vapor of metals like
$^7$Li, $^{23}$Na and $^{87}$Rb. Typically these structures are
studied by means of the cubic nonlinear Schr\"odinger equation. In
place of this nonlinearity, other nonlinear terms can be
considered like, for instance, those introduced by EIP with
positive value of the $\kappa$ parameter to take into account the
attractive effects due to the statistical interaction.\\ We have
studied systems subjected to EIP where the matter field is coupled
to a gauge field of the abelian group $U(1)$.\\ We have also
studied a theoretical model in the presence of gauge fields with a
dynamics described by a Maxwell-Chern-Simons Lagrangian. The most
important properties of the system and its symmetries have been
studied. We have shown that the anyonic behavior due to the
Chern-Simons term is not disturbed by the presence of EIP. Thus
the system describes anyons obeying to EIP in the configuration
space.\\ Applications of this model can be found in the physics of
condensed matter and in the high $T_c$ superconductors models. We
have investigated stationary solutions when the dynamics of the
gauge fields is described by the Chern-Simons term alone.\\ In
presence of the potential $U_{\rm EIP}$ and of a nonlinear
potential, real function of the field $\rho$, we have found
self-dual solutions describing $N$-vortex systems.\\ The principal
properties of these excited states are analytically studied. After
a numerical analysis, we can draw the shape of the vortices and
compare them to the ones already known in literature and deduced
without considering EIP. The conclusive results are:
\begin{enumerate}
\item The values of the charge and angular momentum of the
$N$-vortices are continuous, while in absence of EIP are discrete.
\item In case of systems obeying to an exclusion principle the value
of these observables hase an upper bound and the limits can be
calculated numerically.
\end{enumerate}

The numerical analysis permits us to relate the vortex profiles of
systems with both the exclusive and inclusive effects of different
weight showing how the condensate amplitude changes when the intensity of
EIP changes.


\newpage
\chapter*{Author's Publications}

\begin{enumerate}
\item G. Kaniadakis, P. Quarati and  A. M. Scarfone, Phys. Rev. E {\bf 58},
5574 (1998).

\item G. Kaniadakis, P. Quarati and A. M. Scarfone,  Physica A {\bf 255}, 474 (1998).

\item G. Kaniadakis, P. Quarati and A. M. Scarfone,  Rep. Math. Phys. {\bf44}, 121 (1999).

\item G. Kaniadakis, P. Quarati and A. M. Scarfone,  Rep. Math. Phys. {\bf44}, 127 (1999).

\item G. Kaniadakis, and A. M. Scarfone, {\sl Nonlinear gauge
transformation for a class of Schr\"odinger equations containing complex
nonlinearities}, Rep. Math. Phys. in press.

\item G. Kaniadakis, A. Lavagno, P. Quarati, and A. M. Scarfone, {\sl
Nonlinear gauge transformation of a quantum system obeying an
exclusion-inclusion principle}, J. Nonlin. Math. Phys. in press.

\item G. Kaniadakis, and A. M. Scarfone, {\sl A Maxwell-Chern-Simons model for a quantum
system obeying an exclusion-inclusion principle}, submitted

\item G. Kaniadakis, and A. M. Scarfone, {\sl Chern-Simons
vortices in particle systems obeying an exclusion-inclusion
principle}, submitted

\item G. Kaniadakis, P. Quarati, A. M. Scarfone, \sl Canonical quantum systems obeying an
exclusion-inclusion principle, \rm Electronic Proc. of 11th
International Workshop on NEEDS (Nonlinear Evolution Equations and
Dynamical Systems), June 1997, Kolymbari-Chania, Crete, Greece.
\end{enumerate}
\setcounter{chapter}{1}
\begin{figure}[hb]
\begin{center}
\includegraphics[height=3.8 cm, width=8 cm]{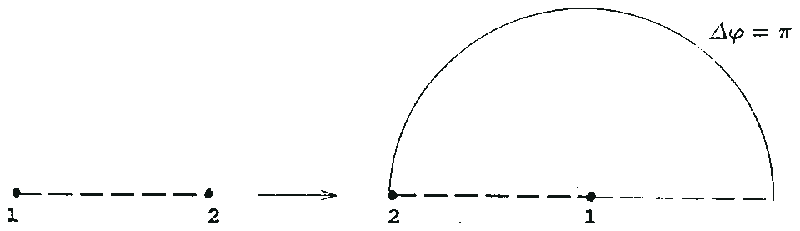}\label{fig1.1}
\includegraphics[height=3 cm, width=7.5 cm]{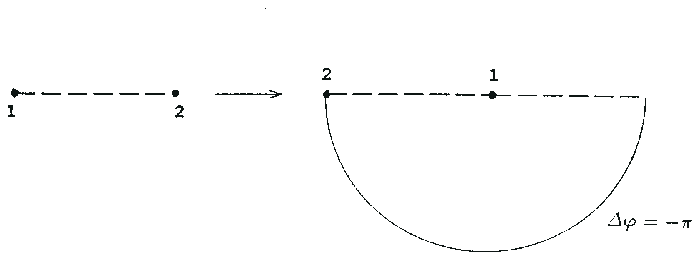}\label{fig1.2}
\caption{The exchange of two particle is realized by means of
rotation of one around the other by an angle of
$\Delta\varphi=\pm\,\pi$}
\end{center}
\end{figure}
\begin{figure}[hb]
\begin{center}
\includegraphics[height=4 cm, width=10 cm]{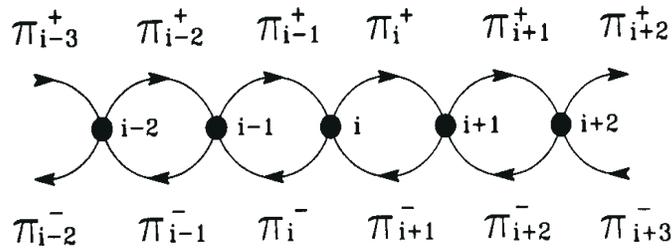}\label{fig3.1}
\caption{One-dimensional discrete Markoffian chain where the
transition probabilities are indicated.}
\end{center}
\end{figure}
\setcounter{chapter}{5} \setcounter{figure}{0}
\begin{figure}[hb]
\begin{center}
\includegraphics[height=8 cm, width=12 cm]{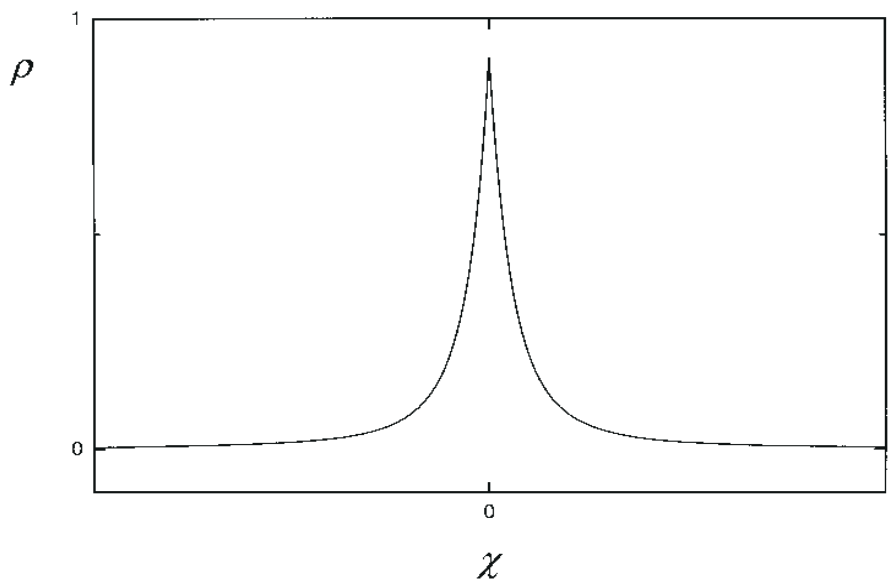}
\caption{Shape of soliton (in arbitrary units) with
$\alpha=0,\,\beta=-1$ and $\sigma=1$.}
\end{center}
\end{figure}
\setcounter{chapter}{6} \setcounter{figure}{0}
\newpage
\begin{figure}[hb]
\begin{center}
\includegraphics[height=7 cm, width=10 cm]{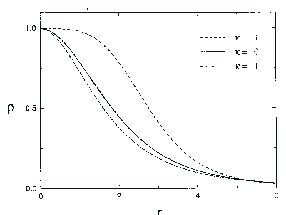}
\caption{Plot of $\rho$ for 1-vortex with $\kappa=0,\,\pm1$ versus
$r$ [in $r_0=(\sqrt{2\,\alpha})^{-1}$ units].}\label{fig6.1}
\includegraphics[height=7 cm, width=10 cm]{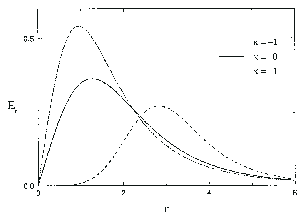}
\caption{Plot of the electric field (in
$E_{0_r}=\hbar^2\,\alpha^{3/2}/e\,m\,\sqrt{2}$ units) for 1-vortex
with $\kappa=0,\,\pm1$ versus $r$ [in
$r_0=(\sqrt{2\,\alpha})^{-1}$ units].}\label{fig6.2}
\end{center}
\end{figure}
\newpage
\begin{figure}[hb]
\begin{center}
\includegraphics[height=7 cm, width=10.5 cm]{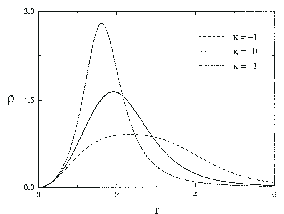}
\caption{Plot of $\rho$ for 2-vortex with $\kappa=0,\,\pm1$ versus
$r$ [in $r_0=(\sqrt{2\,\alpha})^{-1}$ units].}\label{fig6.3}
\includegraphics[height=7 cm, width=10 cm]{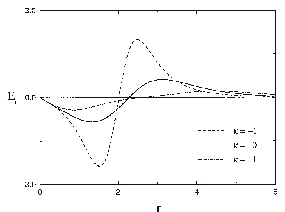}
\caption{Plot of the electric field (in
$E_{0_r}=\hbar^2\,\alpha^{3/2}/e\,m\,\sqrt{2}$ units) for 2-vortex
with $\kappa=0,\,\pm1$ versus $r$ [in
$r_0=(\sqrt{2\,\alpha})^{-1}$ units].}\label{fig6.4}
\end{center}
\end{figure}
\newpage
\begin{figure}[hb]
\begin{center}
\includegraphics[height=7 cm, width=11 cm]{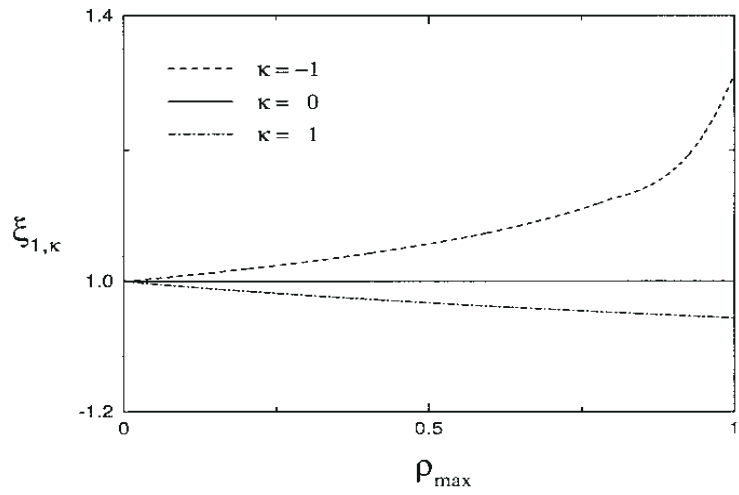}
\caption{Plot of the parameter $\xi_{_{1,\kappa}}$ versus
$\rho_{_{\rm max}}$ with $\kappa=0,\,\pm1$.}\label{fig6.5}
\end{center}
\end{figure}
\vfill \eject
\end{document}